\documentclass[aps,12pt,superscriptaddress,preprintnumbers,amsmath,amssymb,nofootinbib]{revtex4}
\usepackage{epsfig} 
\usepackage{amsmath,amsfonts} 
\usepackage{graphicx}

\large

\newcommand{\nn}{\nonumber \\}
\newcommand{\ov}[1]{\overline{#1}}

\newlength{\nseparation}
\newcommand{\dg}{\ensuremath{\Delta \Gamma}}

\def\bs{B \rightarrow  X_s\, l^+ \, l^-}

\def\v44{V_{4 \times 4}}
\def\v34{V_{3 \times 4}}
\def\bea{\begin{eqnarray}}
\def\eea{\end{eqnarray}}
\def\nn{\nonumber}
\def\barr{\begin{eqnarray}}
\def\earr{\end{eqnarray}}

\newcommand{\ltt}{\lambda_{t t'}^s}
\newcommand{\ltu}{\lambda_{tu}^s}

\newcommand{\re}{{\rm Re}}
\newcommand{\im}{{\rm Im}}

\def\lesssim{\mathrel{\hbox{\rlap{\hbox{\lower4pt\hbox{$\sim$}}}\hbox{$<$}}}} 
\def\gtrsim{\mathrel{\hbox{\rlap{\hbox{\lower4pt\hbox{$\sim$}}}\hbox{$>$}}}}

\def\be{\begin{equation}}
\def\ee{\end{equation}}
\def\beq{\begin{equation}}
\def\eeq{\end{equation}}

\newcommand{\bers}{\begin{eqnarray*}}
\newcommand{\eers}{\end{eqnarray*}}
\def\bra {\langle}
\def\ket {\rangle}

\def\nn{\nonumber}
\def\bbuildrel#1_#2^#3{\mathrel{\mathop{\kern 0pt#1}\limits_{#2}^{#3}}}
\def\slash#1{\setbox0=\hbox{$#1$}#1\hskip-\wd0\dimen0=5pt\advance
       \dimen0 by-\ht0\advance\dimen0 by\dp0\lower0.5\dimen0\hbox
         to\wd0{\hss\sl/\/\hss}}

\newcommand{\gae}{\lower 2pt \hbox{$\, \buildrel {\scriptstyle >}\over {\scriptstyle
\sim}\,$}}
\newcommand{\lae}{\lower 2pt \hbox{$\, \buildrel {\scriptstyle <}\over {\scriptstyle
\sim}\,$}}

\newcommand{\lbar}{\ov{\Lambda}}

\def\lsim{\:\raisebox{-0.5ex}{$\stackrel{\textstyle<}{\sim}$}\:}
\def\gsim{\:\raisebox{-0.5ex}{$\stackrel{\textstyle>}{\sim}$}\:}

\def\issue(#1,#2,#3){{\bf #1}, #2 (#3)}
\def\opcit(#1){ {\em op. cit.}, #1}

\def\APP(#1,#2,#3){Acta Phys.\ Polon.\ \issue(#1,#2,#3)}
\def\ARNPS(#1,#2,#3){Ann.\ Rev.\ Nucl.\ Part.\ Sci.\ \issue(#1,#2,#3)}
\def\CPC(#1,#2,#3){Comp.\ Phys.\ Comm.\ \issue(#1,#2,#3)}
\def\CIP(#1,#2,#3){Comput.\ Phys.\ \issue(#1,#2,#3)}
\def\EPJC(#1,#2,#3){Eur.\ Phys.\ J.\ C\ \issue(#1,#2,#3)}
\def\EPJD(#1,#2,#3){Eur.\ Phys.\ J. Direct\ C\ \issue(#1,#2,#3)}
\def\IEEETNS(#1,#2,#3){IEEE Trans.\ Nucl.\ Sci.\ \issue(#1,#2,#3)}
\def\IJMP(#1,#2,#3){Int.\ J.\ Mod.\ Phys. \issue(#1,#2,#3)}
\def\JHEP(#1,#2,#3){J.\ High Energy Physics \issue(#1,#2,#3)}
\def\JPG(#1,#2,#3){J.\ Phys.\ G \issue(#1,#2,#3)}
\def\MPL(#1,#2,#3){Mod.\ Phys.\ Lett.\ \issue(#1,#2,#3)}
\def\NP(#1,#2,#3){Nucl.\ Phys.\ \issue(#1,#2,#3)}
\def\NIM(#1,#2,#3){Nucl.\ Instrum.\ Meth.\ \issue(#1,#2,#3)}
\def\PL(#1,#2,#3){Phys.\ Lett.\ \issue(#1,#2,#3)}
\def\PRD(#1,#2,#3){Phys.\ Rev.\ D \issue(#1,#2,#3)}
\def\PRL(#1,#2,#3){Phys.\ Rev.\ Lett.\ \issue(#1,#2,#3)}
\def\SJNP(#1,#2,#3){Sov.\ J. Nucl.\ Phys.\ \issue(#1,#2,#3)}
\def\ZPC(#1,#2,#3){Zeit.\ Phys.\ C \issue(#1,#2,#3)}
\begin{document}


\title{SM with four generations: Selected implications for rare B and K decays}

\author{Amarjit Soni}
\affiliation{Physics Department, Brookhaven National Laboratory,
  Upton, NY 11973, USA}
\author{Ashutosh Kumar Alok}
\affiliation{Physique des Particules,
Universit\'e de Montr\'eal, \\ C.P. 6128, succ. centre-ville,
Montr\'eal, QC, Canada H3C 3J7}
\author{Anjan Giri}
  \affiliation{Physics Department, IIT Hyderabad, Andhra Pradesh-502205, 
  India} 
\author{Rukmani Mohanta}
\affiliation{School of Physics, University of
  Hyderabad, Hyderabad - 500046, India}
\author{Soumitra Nandi}
\affiliation{Dipartimento di Fisica 
Teorica, Univ. di Torino and INFN, Sezione di Torino, I-10125 Torino, Italy}

\begin{abstract} 
We extend our recent work and study implications of the Standard Model with four generations (SM4) for rare B and K decays. We again take seriously the several
2-3 $\sigma$ anomalies seen in B, $B_s$ decays and interpret them in the context
of this simple extension of the SM. SM4 is also of course of considerable interest for its potential relevance to dynamical electroweak symmetry breaking
and to baryogenesis. Using experimental information from
processes such as $B \to X_s \gamma$, $B_d$ and  $B_s$ mixings,
indirect CP-violation from $K_L \to \pi \pi$ etc along with oblique corrections,
we constrain the relevant parameter space of the SM4, and find $m_{t'}$ of
about 400-600 GeV with a mixing angle $| V_{t'b}^*V_{t's}|$ in the range of
about (0.05 to 1.4)$\times 10^{-2}$ and with an appreciable CP-odd associated
phase, are favored by the current data. Given the unique role of the CP asymmetry in $B_s \to \psi \phi$ due to its
gold-plated nature, correlation of that with many other interesting observables, including the 
semileptonic asymmetry ($A_{SL}$) are studied in SM4. We also identify several processes,  such 
as $B \to X_s \nu \bar\nu$, $K_L \to \pi^0 \nu \bar \nu$ etc, that are significantly different in SM4 from the SM. Experimentally the very distinctive process 
$B_s\to \mu^+\mu^-$ is also discussed; the branching ratio can be larger or 
smaller than in SM, $(3.2 \to 4.2)\times 10^{-9}$, by a factor of ${\cal{O}}(3)$.
 
\end{abstract} 
\setcounter{footnote}{0}
\renewcommand{\thefootnote}{\arabic{footnote}}

\maketitle 
 
\newpage

\section{Introduction}
Though the CKM paradigm~\cite{NC63, KM73} of CP violation in the Standard Model (SM) has been extremely successful in
describing a multitude of experimental data, in the past few years some indications of
deviations have surfaced, specifically in the flavor sector~\cite{LS07,LS08,LS09,uli_lenz,bona}. An intriguing aspect
of these deviations is that so far they have more prominently,
though not exclusively, 
occurred in CP violating observables
only. While many beyond the standard model (BSM) scenarios
can account for such effects~\cite{APS1, APS2, AJB081, AJB082, MN08,lang,paridi}, a very simple extension of the SM
that can cause these anomalies is the addition of an
extra family as we emphasized in a recent study~\cite{SAGMN08,as_moriond09}. In this paper,
we will extend our previous work and study the implications of the  standard model
with four generations (SM4) in rare B and K decays.

Although our initial motivation for studying SM4 was triggered
by the deviations in the CP violating observables in B, $B_s$ decays,
we want to stress that actually  SM4 is, in fact,  a very simple and interesting
extension of the three generation SM (SM3). The fact that the heavier 
quarks and leptons in this family can play a crucial role in dynamical
electroweak-symmetry breaking (DEWSB) as an economical way to address the hierarchy
puzzle renders this extension of  SM3 especially interesting. In addition,
whereas, as is widely recognized SM3 does not have enough CP
to facilitate baryogenesis, that difficulty is readily and significantly ameliorated in SM4~\cite{Hou08,CJ88, GK08}. Besides, given that three families exist, it is
clearly important to search for the fourth.

That rare B-decays are particularly sensitive to the fourth generation was in fact emphasized 
long ago~\cite{AS_olds1,AS_olds2,AS_olds3,AS_olds4,ge_olds}. The  potential role of heavy quarks in DEWSB was also another reason
for the earlier interest~\cite{norton,symp_SM4_8789,Holdom:1986rn,Hung:2009ia,Hashimoto:2009ty}. LEP/SLC discovery that a fourth family
(essentially) massless neutrino does not exist was one reason that caused 
some pause in the interest
on SM4. A decade later discovery of neutrino oscillations and of neutrino mass
managed to  off-set to some degree this concern about the 4th family's
necessarily involving massive neutrino.
Electroweak precision tests provide a very important constraint on the mass difference  of the 4th family isodoublet. In this context the PDG reviews
for a number of years may have been declaring a ``prematured death" of the fourth
family~\cite{erler}; careful studies show in fact that while mass difference between the isodoublet quarks is constrained to be less than $\approx 75$ GeV, an extra generation of quarks is not excluded by the current data. In fact,
it is also claimed that  for
certain values of particle masses the quality of the fit with four generations
is comparable to that of the SM3~\cite{novikov1,novikov2,Kribs_EWPT,chanowitz}.

The addition of fourth generation to the SM means that the quark
mixing matrix will now become a $4 \times 4$ matrix ($V_{CKM4}$) and the parametrization of
this unitary matrix requires six real parameters and three phases. The two extra phases imply the possibility
of extra sources of CP violation \cite{AS_olds3}.

In \cite{SAGMN08}, it was shown that a fourth family of quarks with $m_{t'}$ in the range of
(400 - 600) 
GeV provides a simple explanation for the several indications of new 
physics that have been observed involving CP asymmetries in the B, $B_s$ 
decays~\cite{LS07,LS08,LS09,uli_lenz,bona}. The 
built-in hierarchy of $V_{CKM4}$ is such that the $t'$ 
readily provides a needed perturbation ($\approx 15\%$) to $\sin 2 \beta$ as measured in
$B \to \psi K_s$ and simultaneously is the
dominant source of CP asymmetry in $B_s \to \psi \phi$. 

While most of the B, $B_s$ CP-anomalies are easily accommodated and explained 
by SM4, we note that, in contrast, EW precision tests constrain the 
mass-splitting between $t'$ and $b'$  to be small,
around $70$ GeV \cite{NEED, novikov1,novikov2, Kribs_EWPT}; so for $m_{t'}$ of O(500 GeV) their masses have to be degenerate
to O(15\%).  As far 
as the lepton sector is concerned, it is clear that the 4th family lepton has
 to be quite different from the previous three families in that the neutral 
lepton has to be rather massive, with mass $> m_Z/2$. This may also be a 
clue that the underlying nature of the 4th family may be quite different
from the previous three families \cite{DM4}.

In this paper we extend our previous work~\cite{SAGMN08} on the implications of SM4, to 
study the direct CP asymmetry in $B\to X_s \gamma$, $B\to X_s\, l^+ \, l^-$ and in 
$B_s\to X_s\ell\nu$, forward-backward (FB) asymmetry in $B\to X_s\ (K^*) l^+ \ l^-$, 
decay rates of $B\to X_s \nu \bar \nu $, $B_s\to \mu^+\mu^-, \tau^+\tau^-$ and 
$K_L \to \pi^0\nu \bar \nu$ and 
CP violation in $B \to \pi K$ and $B^0 \to \pi^0 \pi^0$ modes. We show that
SM4 can ameliorate the difficulty in understanding the large
difference, O(15\%), between the direct CP asymmetries in neutral B decays
to $K^+ \pi^-$ versus that of the charged B-decays to $K^+ \pi^0$ partly due to the
enhanced isospin violation that SM4 causes in flavor-changing penguin
transitions due to the heavy $m_{t'}$~\cite{AS_olds1} originating from the
evasion of the decoupling theorem and partly if the corresponding strong phase(s) are large
in SM4. The enhanced electroweak penguin amplitude
provides a color-allowed ($Z\to \pi^0$) contribution which is not present for
$\pi^{\pm}$ case. However, we want to emphasize that the prediction obtained 
 using the QCD factorization approach~\cite{QCDF1,QCDF,LS07} depends on many 
input parameters therefore it has large theoretical uncertainties. Apart 
from the SM parameters such as CKM matrix, quark masses, the strong 
coupling constant and hadronic parameters there are large 
theoretical uncertainties related to the modeling of power corrections 
corresponding to weak annihilation effects and the chirally-enhanced power 
corrections to hard spectator scattering. Therefore the numerical results for 
the direct CP asymmetries are not reliable.

Several of these observables like FB asymmetry in $B\to  K^* l^+ \ l^-$ 
\cite{Hovhannisyan:2007pb}, CP asymmetry in $B_s\to \psi\phi$ \cite{hou03}
 and the decay rate of $K_L \to \pi^0\nu \bar \nu$ 
\cite{Hou:2005yb} have also been studied before, as well as 
many other interesting aspects of SM4 by Hou and collaborators \cite{Hou:2005hd, 
Arhrib:2006pm, Hou:2006zza, Hou:2006jy}, see also \cite{lenz4}. 
However, their analysis was generally restricted to $m_{t'}$ of $\sim\,$ 300 {\it GeV}. 
On the other hand, our analysis seems to favor $m_{t'}$ in the range of 
(400 - 600) GeV to explain the observed CP asymmetries in the B, $B_s$ decays. 
We note also that recent analysis by Chanowitz seems to disfavor most of the 
parameter space they have used \cite{chanowitz} whereas our parameter space is 
largely unaffected \cite{NTU}.

We identify several processes wherein SM4 causes large deviations from the expectations of SM3; 
for example, $B \to X_s \nu \bar \nu$, $B_s\to \mu^+\mu^-$, $A_{SL}(B_s\to X_s\ell\nu )$, 
$a_{CP} (B \to \pi K)$, $a_{CP} (B \to \pi^0 \pi^0)$, 
$K_L \to \pi^0 \nu \bar \nu$ and of course mixing-induced CP in
$B_s \to \psi \phi$ etc.
These observables will be measured with higher statistics  at the upcoming
high intensity K, B, $B_s$ experiments at CERN, FERMILAB, JPARC facilities etc
and in particular at the  LHCb experiment  and possibly also at the Super-B factories and hence may provide further indirect evidence for an additional family of quarks. 

The paper is arranged as follows. After the introduction, we provide constraints on the 4$\times$4 CKM matrix by incorporating oblique corrections along
with experimental data from important observables
involving Z, B and K decays as well as $B_d$ and $B_s$ mixings etc.
In Sec. \ref{results}, we present the estimates of many useful  observables in the SM4. Finally in Sec. \ref{concl}, we present our summary.

\section{Constraints on the CKM4 matrix elements }
\label{constraint_ckm}

In our previous article \cite{SAGMN08}, to find the limits on $V_{CKM4}$ 
elements, we concentrated mainly on the constraints that will come from  
vertex correction to $Z\to b\bar{b}$, $Br(B\to X_s \gamma)$, 
$Br(B\to X_s \, l^+ \, l^-)$, $B_d - \bar{B_d}$ and
 $B_s - \bar{B_s}$ mixing, $Br(K^+\to \pi^+\nu\nu)$ and  
indirect {\it CP} violation in $K_L \to \pi\pi$ described 
by $|\epsilon_k|$. We did not consider $\epsilon'/\epsilon$ as a constraint 
because of its large hadronic uncertainties. Chanowitz \cite{chanowitz}
has shown that as $m_{t'}$ becomes very large more important constraint is from non decoupling oblique 
corrections rather than the vertex correction to $Z\to b\bar{b}$. In this 
article we have extended  our analysis by including the constraint form 
non decoupling oblique corrections as well; we note that for $m_{t'}\, \lsim 500$ {\it GeV} our 
previous constraints are largely unaffected 
but for $m_{t'}\approx 600$ {\it GeV} the oblique corrections start to have effect. 
With the inputs given in Table. (\ref{tab3})
 we have made the scan over the entire parameter space by a flat 
random number generator and obtained the constraints on various parameters of 
the 4$\times$4 mixing matrix. In the following subsections we briefly 
discuss the various input parameters used in our analysis.

\subsection{Oblique correction} 
The $Z$ pole, $W$ mass, and low-energy data can be used to search for and set limits on 
deviations from the SM. Most of the effects on precision measurements can be described 
by the three gauge self-energy parameters $S$, $T$ and $U$. We assume these parameters to 
be arising from new physics only i.e 
they are equal to zero exactly in SM, and do not include any contributions
from $m_t$ and $M_H$.

The effects of non-degenerate multiplets of chiral fermions can be 
described by just three parameters, $S$, $T$ and $U$ at the one-loop level 
\cite{peskin,novikov1,sher,erler,novikov3}. $T$ is proportional 
to the difference between the $W$ and $Z$ self-energies at $Q^2=0$, while $S$ 
is associated with the difference between the $Z$ self-energy at 
$Q^2=M^2_{Z}$ and $Q^2=0$ and $(S+U)$ is associated with the difference 
between $W$ self-energy at $Q^2=M^2_{W}$ and $Q^2=0$.
A non-degenerate $SU(2)$ doublet $\binom{f1}{f2}$ with masses $m_1$ and $m_2$ 
respectively yields the contributions \cite{peskin}

\barr
S &=&\frac{1}{6\pi}\Big[1- Y \ln({m^2_1/ m^2_2})\Big],\\ \nonumber
T &=& \frac{1}{16 \pi s_W^2 c_W^2 M^2_Z}\Big[m^2_1+m^2_2-\frac{2 m^2_1 m^2_2}
{m^2_1-m^2_2}\ln({m^2_1/ m^2_2})\Big],\\ \nonumber
U &=&\frac{1}{6\pi}\Big[- \frac{5 m^4_1-22 m^2_1 m^2_2 + 5 m^4_2}{3(m^2_1-m^2_2)^2}+ \frac{m^6_1-3 m^4_1 m^2_2 - 3 m^2_1 m^4_2+ m^6_2}{(m^2_1-m^2_2)^3}
\ln({m^2_1/ m^2_2})\Big],
\earr
where $Y$ is the hypercharge of the doublet.
A heavy non-degenerate doublet of fermions contributes positively
to $T$ as 
\be
\rho^{\ast}_0-1= \frac{1}{1-\alpha T}-1 \approx \alpha T,
\ee
where $\rho^{\ast}_0$ denotes the low-energy ratio of neutral to charged current
couplings in neutrino interactions.

The parameter $U$ plays a fairly unimportant role, all the neutral current and
low energy observables depend only on $S$ and $T$ \cite{peskin}. 
In addition $U$ is often predicted to be very small. In most of the models $U$ 
should differ from zero by only a percent of $T$. 

In the case of an extra family with the doublet $\binom{t'}{b'}$, the 
contribution to $T$ and $S$ parameters are given by \cite{chanowitz}
\barr
T_4&=&\frac{1}{8 \pi x_W (1- x_W)}\Big[3 \Big(|V_{t'b'}|^2 \delta{m}_{t'b'}+ 
|V_{t'b}|^2 \delta{m}_{t'b} + |V_{tb'}|^2 \delta{m}_{tb'} - |V_{t'b}|^2 
\delta{m}_{tb} \\ \nonumber
&{}& + |V_{t's}|^2 \delta{m}_{t's}\Big) + \delta{m}_{l_4\nu_4}\Big],\\ 
S_4&=&\frac{3}{6\pi}\Big(1- \frac{1}{3} \ln{\frac{m^2_{t'}}{m^2_{b'}}}\Big),
\earr  
with 
\be
\delta{m}_{12}= \frac{1}{2 M^2_Z}\Big({m^2_1+m^2_2}-\frac{2 m^2_1 m^2_2}
{m^2_1-m^2_2} \ln({m^2_1/ m^2_2})\Big).
\ee

\subsection{Vertex corrections to $Z\to b\bar{b}$} 

Including QCD and QED corrections, the $Z\to b\bar{b}$ decay width can be 
written as \cite{zbb1}
\bea
\Gamma(Z\to q\bar{q}) &=& {\frac{N_c}{48}} {\frac{\alpha} {s^2_W c^2_W}} m_Z\left(|a_q|^2 + |v_q|^2\right) 
\nonumber \\
&{}& \times\big(1 + \delta^{(0)}_b \big) \big(1 + \delta^q_{QED}\big)\big(1 + \delta^q_{QCD}\big)\big(1 + \delta^q_{\mu}\big)\big(1 + \delta^q_{tQCD}\big)\big(1 + \delta_q\big),
\eea
where
\be
v_q = \Big(2 I^q_3 - 4 |Q_q| s^2_W\Big), \hskip 20pt a_q = 2 I^q_3,
\ee
and $\delta$'s are various corrections which are discussed below.

In the decay of the $Z\to b\bar{b}$, the top quark mass
enters in the loop correction to the vertex mediated by the W gauge boson. Due
to spontaneous symmetry breaking effects the top mass can not be neglected in
the calculation. In fact there is a top mass dependence that grows like 
$\frac{m^2_t}{m^2_Z}$ as in many other one-loop weak processes such as $K-\bar{K}$,\, $B-\bar{B}$\, ($\Delta F = 2$ mixings),\, $b\to s\ell^+\ell^-$\, etc. 
The additional contribution to the $Zb\bar{b}$ vertex, 
due to nonzero value of the top quark mass can be written as:
\be
\delta_b \approx  10^{-2}\left(\Big(-\frac{m^2_t}{2 m^2_Z} + 0.5\Big)|V_{tb}|^2
 + \Big(- \frac{m^2_{t'}}{2 m^2_Z} + 0.5\Big)|V_{t'b}|^2\right).
\ee

$\delta^{q}_{QED}$ gives small final-state QED corrections
that depend on the charge of final fermion,
\be
\delta^q_{QED} = \frac{3\alpha}{4\pi} Q^2_q.
\ee
It is very small (0.2\% for charged leptons, 0.8\% for u-type quarks and 0.02\%
for d-type quarks).

$\delta_{QCD}$ gives the QCD corrections common to all quarks and it is given
by
\be
\delta_{QCD} = \frac{\alpha_s}{\pi}  + 1.41 \Big(\frac{\alpha_s}{\pi} \Big)^2.
\ee
$\alpha_s$ is the QCD coupling constant taken at the $m_Z$ scale, i.e. $\alpha_s= \alpha_s(m^2_Z) = 0.12$.

$\delta^q_{\mu}$ contains the kinematical effects of the external fermion masses, including some mass-dependent QCD radiative corrections. It is only important
for the b-quark (0.5\%) and to a lesser extent for the $\tau$-lepton (0.2\%) 
and the c-quark (0.05\%). It is given by
\be
\delta^q_{\mu} = \frac{3 \mu^2_q} {v^2_q + a^2_q}\left(- \frac{1}{2} a^2_q\left(1 + \frac{8 \alpha_s} {3\pi}\right) + v^2_q \frac{\alpha_s}{\pi}\right),
\ee
where $\mu^2_q \equiv 4\bar{m}^2_q(m^2_Z)/ m^2_Z$.

By taking appropriate branching ratios it is possible to isolate the large top
mass dependent $Zb\bar{b}$ vertex $\delta_b$ \cite{zbb1},
 
\be
R_h \equiv \frac{\Gamma(Z\to b\bar{b})} {\Gamma(Z\to hadrons)} = \big( 1 + 2/R_s
+ 1/R_c + 1/R_u \big)^{-1},
\ee
where $R_q \equiv \frac{\Gamma(Z\to b\bar{b})} {\Gamma(Z\to q\bar{q})}$. 

All other corrections cancel exactly in this branching ratio except the 
correction to the $Zb\bar{b}$ vertex which only depends on the top quark mass.

\subsection {$B \to X_s \gamma$ decay}

Radiative B decays have been a topic of great theoretical and experimental interest for long.
Although the inclusive radiative decay $B  \to X_s \gamma$ is
loop suppressed within the SM, it has relatively large
branching ratio making it statistically favorable from the experimental
point of view and hence it serves as an important
probe to test SM and its possible extensions. The present world
average of $Br(B\to X_s \gamma)$ is $(3.55 \pm 0.25)\times 10^{-4}$ \cite{Barberio:2008fa}
which is in good agreement with its SM prediction \cite{Misiak:2006ab,Misiak:2006zs}.
Apart from the branching ratio of $B \to X_s \gamma$,
direct $CP$ violation in $B \to X_s \gamma$, $A_{CP}^{B \to X_s \gamma}$ ,
can serve as an important observable to search physics beyond SM;
therefore we will also study this direct CP asymmetry in this paper
(see Section \ref{dacpbsg}).

The quark level transition $b \to s \gamma$ induces the 
inclusive $B \to X_s \gamma$ decay. The effective
Hamiltonian for $b \to s \gamma$ can be written in the following form
\be
{\cal H}_{eff} =  \frac{4 G_F}{\sqrt{2}} V_{ts}^{*}  V_{tb}
\sum_{i=1}^{8} C_i(\mu) \, Q_i(\mu)\;,
\ee
where the form of operators $O_i(\mu)$ and the expressions for calculating the Wilson
coefficients $C_i(\mu)$ are given in \cite{Buras:1994dj}.
The introduction of fourth generation changes the values of 
Wilson coefficients $C_7$ and $C_8$ via the virtual exchange
of the $t'$-quark and can be written as
\be
C_{7,8}^{\rm tot}(\mu) = C_{7,8}(\mu) + \frac{ V_{t' s}^{*} V_{t' b}}
{ V_{ts}^{*}  V_{tb}} C_{7,8}^{t'} (\mu)\;.
\label{wtot_78}
\ee
The values of $C_{7,8}^{t'}$ can be calculated from the expression 
of $C_{7,8}$ by replacing the mass of $t$-quark by $m_{t'}$.

In order to reduce the uncertainties arising from $b$--quark
mass, we consider the following ratio 
\be
R = \frac{Br(B \to X_s \gamma)}{Br(B \to X_c e \bar \nu_e)}\;. \nonumber
\ee
In leading logarithmic approximation this ratio can be written as \cite{Buras:1997fb}
\be
R = \frac{\left| V_{ts}^{*}   V_{tb} \right|^2}{\left| V_{cb} \right|^2} \,\,
\frac{6 \alpha \left| C_7^{\rm tot}(m_b) \right|^2}{\pi f(\hat m_c) \kappa(\hat m_c)}\;. 
\label{R}
\ee
Here the Wilson coefficient $C_7$ is evaluated at the scale $\mu=m_b$.
The phase space factor $f(\hat{m_c})$ in $Br(B \to X_c e {\bar \nu})$ 
is given by \cite{Nir:1989rm}
\be
f(\hat{m}_c) = 1 - 8\hat{m}^2_c + 8\hat{m}_c^6 - \hat{m}_c^8 - 
24\hat{m}_c^4 \ln \hat{m}_c \;.
\ee
$\kappa(\hat{m_c})$ is the $1$-loop QCD correction factor 
\cite{Nir:1989rm} 
 \be
\kappa(\hat{m_c})=1-\frac{2\alpha_s(m_b)}{3\pi}\left[\left(\pi^2-\frac{31}{4}\right)(1-\hat{m_c})^2+\frac{3}{2}\right]\;.
\ee
Here $\hat{m_c}=m_c/m_b$.

\subsection {$B \to X_s\, l^+ \,l^-$ decay}

The quark level transition $b \to s\, l^+ \, l^-$ is responsible for the inclusive decay $B \to X_s \, l^+ \, l^-$.
We apply the same approach introduced for $b \to s \gamma$.
The effective Hamiltonian for the decay $b \to s\, l^+\, l^-$ is given by 
\be
{\cal H}_{eff} =  \frac{4 G_F}{\sqrt{2}} V_{ts}^{*} V_{tb}
\sum_{i=1}^{10} C_i(\mu) \,  Q_i(\mu)\;.
\ee
In addition to the operators relevant for $b \to s \gamma$, there are two new 
operators:
\be
Q_{9} = (\bar{s}b)_{V-A}(\bar{l}l)_V, \hskip 20pt Q_{10} = (\bar{s}b)_{V-A}(\bar{l}l)_V\;.
\ee

The amplitude for the decay $B \to X_s \, l^+ \, l^-$ 
in SM4 is given by
\bea
M &~=~& \frac{G_F \alpha}{\sqrt{2} \pi}
\, V_{ts}^*V_{tb}  \, \Bigl[ C_9^{\rm tot} \, \bar{s}
\gamma_\mu P_L b \, \bar{l} \gamma_\mu l+ C^{\rm tot}_{10}
\, \bar{s} \gamma_\mu P_L b \, \bar{l} \gamma_\mu \gamma_5
l  \nn \\
&& \hskip2.5truecm
 -~2m_b \, \frac{C^{\rm tot}_7}{q^2} \, \bar{s} i
  \sigma_{\mu\nu} q^\nu P_R b \,
\bar{l} \gamma_\mu l \Bigr] ~,
\label{matrix}
\eea
where $P_{L,R} = (1 \mp \gamma_5)/2$ and $q$ is the sum of
$l^+$ and $l^-$ momenta. Here the Wilson coefficients are evaluated at $\mu$=$m_b$.

The differential branching ratio is given by
\be
\frac{{\rm d}Br(\bs)}{{\rm d}z}
 = \frac{\alpha^2 B(B\rightarrow X_c e {\bar \nu})}
 {4 \pi^2 f(\hat{m_c})\kappa(\hat{m}_c)} (1-z)^2\left(1-\frac{4t^2}{z}\right)^{1/2}
 \frac{|V_{tb}^{*}V_{ts}|^2}{|V_{cb}|^2} D(z)\,,
\label{eq:dbr_bsll}
\ee
where
 \bea
 D(z) &=& |C_9^{\rm tot}|^2\left(1+\frac{2t^2}{z}\right)(1+2z)
      + 4|C_7^{\rm tot}|^2\left(1+ \frac{2t^2}{z}\right)\left(1+\frac{2}{z}\right) \nonumber \\
    & &  + |C_{10}^{\rm tot}|^2 \left[ ( 1 + 2z) + \frac{2t^2}{z}(1-4z)\right]
      +12 {\rm Re}(C_7^{\rm tot} C_{9}^{\rm tot*})\left(1+\frac{2t^2}{z}\right)\;.
\label{bsll_Dz}
\eea
Here $z \equiv q^2/m_b^2$, $t \equiv m_{l}/m_{b}$ and 
$\hat{m}_q=m_q/m_b$ for all quarks $q$.

In the framework of SM4, the Wilson coefficients $C_{7}^{\rm tot }$, 
$C_{9}^{\rm tot}$ and $C_{10}^{\rm tot}$ are given by
\bea
C_{7,10}^{\rm tot }& = & C_{7,10}(m_b)\,+\, \frac{V_{t's}^*V_{t'b}}{V_{ts}^*V_{tb}}\,C_{7,10}^{t' }(m_b)\;, \\
C_{9}^{\rm tot }& = & C_9(m_b) \, +\, Y(z)\,+\, 
\frac{V_{t's}^*V_{t'b}}{V_{ts}^*V_{tb}}\,C_{9}^{t' }(m_b)\;,
\label{wcsm4}
\eea
where the function $Y(z)$ is given in \cite{Buras:1994dj}.

The measurements of the $B\to X_s\ell^+ \ell^-$ in the two regions, so called 
low $q^2$ $(q^2\protect\lsim\, 6 GeV^2)$  and high $q^2$
$(q^2\protect\gsim \,14 {GeV}^2)$, are complementary as they have different
sensitivities to the short distance physics. Compared to small $q^2$, the rate
in the large $q^2$ region has a smaller renormalization scale dependence and
$m_c$ dependence. Although the rate is smaller at large $q^2$, the experimental
efficiency is better. Large $q^2$ constrains the $X_s$ to have small invariant
mass, $m_{X_s}$, which suppresses the background from $B\to X_c \ell^-\bar{\nu}
\to X_s \ell^+\ell^- \nu \bar{\nu}$. To suppress this background at small $q^2$
region an upper cut on $m_{X_s}$ is required, complicating the theoretical 
description due to the dependence of the measured rate on the shape function, 
which is absent at large $q^2$. In the low $q^2$ region the
dominant contribution to $B_s\to X_s\ell^+\ell^-$ comes from virtual photon and 
much less from $Z$. It is the $Z$ that is very sensitive to $m_{t'}$ as
that amplitude grow with $m^2_{t'}$. The photonic contribution cares only
about the electric charge, modulo logarithmic QCD corrections. For these 
reasons we will be using the branching ratio only in the high $q^2$ region to constrain SM4.

The theoretical calculations shown above for the branching ratio of
$B \to X_s\, l^+ \,l^-$ are rather uncertain in the intermediate $q^2$ region
($7$~GeV$^2 < q^2 < 12$~GeV$^2$) owing to the vicinity of
charmed resonances. The predictions are relatively more robust
in the low-$q^2$ ($1 \,{\rm GeV^2} < q^2 < 6\, {\rm GeV^2}$)  
and the high-$q^2$ ($14.4\, {\rm GeV^2} < q^2 < m_b^2$) regions.

For $m_{t'}> 300\, {\rm GeV}$, $Br(B\to X_s \,l^+ \, l^-)$ is completely dominated by the
Wilson coefficient $C_{10}^{\rm tot}$. Hence in our numerical analysis, we neglect the small $z$-dependence in
$C_{9}^{\rm tot}$. 

\subsection {$B_{q}-\bar B_{q}$ mixing}

Within SM, $B_{q}-\bar B_{q}$ mixing ($q=d,s$) proceeds to an excellent approximation only 
 through the box diagrams with internal top quark exchanges. In case of 
four generations there is an additional contribution to  $B_{q}-\bar B_{q}$ mixing
coming from the virtual exchange of the fourth generation up quark $t'$. 
The mass difference $\Delta M_q$ in SM4 is given by
\be
\Delta M_q = 2|M_{12}|\;,
\ee
where
\bea
M_{12} &=& \frac{G_F^2 m_W^2}{12\pi^2} m_{B_q} B_{bq} f_{B_q}^2 \Big\{
\eta_t \left( V_{tq} V_{tb}^{*}  \right)^2 S_0(x_t)   +
\eta_{t'} \left( V_{t' q} V_{t' b}^{*}\right)^2
S_0(x_{t'}) 
\nonumber \\&& 
+ 2\eta_{tt'} \left(V_{tq} V_{tb}^{*} \right) \left( V_{t' q} V_{t' b}^{*}\right) S_0(x_t,x_{t'}) \Big\}~,
\eea
where $x_t=m_t^2/m_W^2$, $x_{t'}=m_{t'}^2/M_W^2$ and
\bea
S_0(x_t) &=& \frac{4 x_t - 11 x_t^2 + x_t^3}{4 (1-x_t)^2}
-\frac{3}{2} \frac{x_t^3 \mbox{\rm ln} x_t}{(1-x_t)^3}~,\\
S_0(x_{t'}) &=& S_0(x_t \to x_{t'})~, \\
S_0(x_t,x_{t'}) &=& x_t x_{t'} \Bigg\{
\frac{\mbox{\rm ln} x_{t'}}{x_{t'}-x_t} \Bigg[
\frac{1}{4} + \frac{3}{2} \frac{1}{1-x_{t'}} -
\frac{3}{4} \frac{1}{(1-x_{t'})^2} \Bigg]  
 \nonumber \\
&& - \frac{\mbox{\rm ln} x_t}{x_{t'}-x_t} \Bigg[
\frac{1}{4} + \frac{3}{2} \frac{1}{1-x_t} -
\frac{3}{4} \frac{1}{(1-x_t)^2} \Bigg]  
 \nonumber \\
&& -\frac{3}{4} \frac{1}{(1-x_t) (1-x_{t^\prime})} \Bigg\}\;.
\eea
Here $\eta_t$ is the QCD correction factor and its value is $0.5765\pm0.0065$ \cite{buras1}. 
The QCD correction factor $\eta_{t'}$ is given by \cite{Hattori:1999ap}
\bea
\eta_{t'} = \Big(\alpha_s(m_t)\Big)^{6/23} 
\left( \frac{\alpha_s(m_{b^\prime})}{\alpha_s(m_t)} \right)^{6/21} 
\left( \frac{\alpha_s(m_{t^\prime})}{\alpha_s(m_{b^\prime})} \right)^{6/19}\;.
\eea
$\alpha_s(\mu)$ is the running coupling constant at the scale $\mu$ at NLO \cite{Buchalla:1995vs}. 
Here we assume $\eta_{t'}=\eta_{tt'}$ for simplicity. The numerical values of 
the structure functions $S_0(x_{t'})$, $S_0(x_t,x_{t'})$ 
and the QCD correction factor $\eta_{t'}$ are given in Table~\ref{tab1} and Table~\ref{tab2} respectively for various $t'$ mass.

\begin{table}[t]
\begin{center}
\begin{tabular}{|c||c|c|}
\hline
$m_{t'}$(GeV)   &400  & 600    \\
\hline
$S_0 (x_{t'})$  &9.225    &17.970   \\
\hline
$S_0 (x_t, x_{t'})$   &4.302    &5.225   \\
\hline
\end{tabular}
\caption{The structure functions $S_0(x_{t'})$ and $S_0(x_t,x_{t'})$.}
\label{tab1}
\end{center}
\end{table}
\begin{table}[t]
\begin{center}
\begin{tabular}{|c||c|c|}
\hline
$m_{t'}$(GeV)   &400   & 600    \\
\hline
$\eta_{t'}$  &0.522  &0.514  \\
\hline
\end{tabular}
\caption{The QCD correction factor  $\eta_{t'}$.}
\label{tab2}
\end{center}
\end{table}

\subsection{Indirect {\it CP} violation in $K_L \to \pi\pi$}

Indirect CP violation in $K_L\to \pi \pi$ is described by the parameter 
$\epsilon_K$, the working formula for it is given by \cite{buras2}
\be
\epsilon_K = {\rm exp}( i \phi_{\epsilon})\sin\phi_{\epsilon} \Big( {{\rm Im} 
{M^{k}_{12}}/\Delta M_k} + \zeta \Big),
\label{epsilon}
\ee
where $\zeta = {{\rm Im}{A_0}\over {\rm Re}{A_0}}$ with $A_0\equiv A\big(K\to (\pi\pi)_{I=0}\big)$ 
and $\Delta M_K$ denoting the $K_L-K_S$ mass difference. The off-diagonal
element $M_{12}$ in the neutral $K$-meson mass matrix represents $K^0-\bar{K^0}$ mixing and
is given by 
\be
M^{*}_{12} = {{\bra\bar{K^0}|{\cal{H}}_{eff}(\Delta S = 2)|K^0\ket} \over 2 m_K}
\ee
The phase $\phi_{\epsilon}$ is given by 
\be
\phi_{\epsilon}=(43.51 \pm 0.05)^{\circ}
\ee

The second term in eq. \ref{epsilon} constitutes a ${\cal{O}}(5)$\% correction 
to $\epsilon_K$. In most of the phenomenological analysis 
$\phi_{\epsilon}$ is taken as $\pi/4$ and $\zeta$ is taken as zero. However 
$\zeta\not= 0 $ and $\phi_{\epsilon} < \pi/4$ results in a suppression effect in 
$\epsilon_k$ relative to the approximate formula with $\zeta = 0 $ and 
$\phi_{\epsilon} = \pi/4$. In order to include these corrections we have used 
the parametrization 
\be
\kappa_{\epsilon}=\sqrt{2}\sin{\phi_{\epsilon}}\bar{\kappa_{\epsilon}},
\label{kappa}
\ee 
where $\bar{\kappa_{\epsilon}}=0.94 \pm 0.02$ and consequently 
${\kappa_{\epsilon}}= 0.92\pm0.02$, $\bar{\kappa_{\epsilon}}$ parameterizing 
the effect of $\zeta\ne 0$ \cite{buras2}.
 
After some calculations it can be shown that \cite{Buras:1997fb}
\bea
M_{12} &=& {G^2_F\over 12\pi^2} f^2_K B_K m_K M^2_W \Big[{\lambda^{*}_c}^2\eta_c S_0(x_c)+ {\lambda^{*}_t}^2\eta_t S_0(x_t) 
+ 2 {\lambda^{*}_c}{\lambda^{*}_t}\eta_{ct} S_0(x_c, x_t)
\nonumber \\ &&
+ {\lambda^{*}_{t'}}^2\eta_{t'} S_0(x_{t'}) 
+ 2 {\lambda^{*}_c}{\lambda^{*}_{t'}}\eta_{c{t'}} S_0(x_c, x_{t'}) 
+ 2 {\lambda^{*}_t}{\lambda^{*}_{t'}}\eta_{t{t'}} S_0(x_t, x_{t'}) \Big]\,,
\label{m12}
\eea 
where $\lambda_i = \lambda^{*}_{is}\lambda_{id}$ and $x_q=(m^2_q/M^2_W)$
for all quarks $q$.

Inserting (\ref{m12}) and (\ref{kappa}) in (\ref{epsilon}) one finds
\bea
\epsilon_K &=& {{G^2_F\over 12\pi^2 \sqrt{2} {\Delta M_K}}} {\kappa_{\epsilon}}
f^2_K B_K m_K M^2_W\, {\rm Im}\Big[{\lambda^{\ast}_c}^2\eta_c S_0(x_c)
+ {\lambda^{\ast}_t}^2\eta_t S_0(x_t) 
\nonumber \\ && +
 2 {\lambda^{\ast}_c}{\lambda^{\ast}_t}\eta_{ct} S_0(x_c, x_t) 
+ {\lambda^{\ast}_{t'}}^2\eta_{t'} S_0(x_{t'}) + 2 {\lambda^{\ast}_c}{\lambda^{\ast}_{t'}}\eta_{c{t'}} S_0(x_c, x_{t'}) 
\nonumber \\&&
+ 2 {\lambda^{\ast}_t}{\lambda^{\ast}_{t'}}\eta_{t{t'}} S_0(x_t, x_{t'}) \Big]\;,
\label{epsilon1}
\eea
where $f_K= 160\, \rm MeV$. The value for $B_K$ has 
been taken from Ref. \cite{latticeold}, in a recent analysis  
\cite{alv,rbc2} the error has been reduced to $\lsim \,4\%$, however, in our analysis 
we use the more conservative value mentioned in Table. \ref{tab3} from \cite{latticeold}.
 
\begin{table}[t]
\begin{center}
\begin{tabular}{|c|c|}
\hline
$B_K = 0.72 \pm 0.05$ \cite{latticeold} & $f_{bs}\sqrt{B_{bs}} = 
0.281 \pm 0.021$ GeV \cite{lattice3}  \\
$\Delta{M_s} = (17.77 \pm 0.12) ps^{-1}$ \cite{cdf}& $\Delta{M_d} = (0.507 \pm 0.005) ps^{-1}$ \\
$\xi_s = 1.2 \pm 0.06$ \cite{lattice3} & $\gamma = (75.0 \pm 22.0)^{\circ} $ \\
$|\epsilon_k|\times 10^{3} = 2.32 \pm 0.007$ & $\sin 2\beta_{\psi K_s} = 0.672 \pm 0.024$\\
$Br(K^+\to \pi^+\nu\nu) = (0.147^{+0.130}_{-0.089})\times 10^{-9}$ & $Br(B\to X_c \ell \nu) = (10.61 \pm 0.17)\times 10^{-2}$\\
$Br(B\to X_s \gamma) = (3.55 \pm 0.25)\times 10^{-4}$ & $Br(B\to X_s \ell^+ \ell^-) = (0.44 \pm 0.12)\times 10^{-6}$ \\
$R_{bb} = 0.216 \pm 0.001$  & ( High $q^2$ region )\\
$|V_{ub}| = (37.2 \pm 2.7)\times 10^{-4}$ & $|V_{cb}| = (40.8 \pm 0.6)\times 10^{-3} $\\
$\eta_c = 1.51\pm 0.24$ \cite{uli1} & $\eta_t = 0.5765\pm 0.0065$ 
\cite{buras1}\\
$\eta_{ct} = 0.47 \pm 0.04$ \cite{uli2} & $m_t = 172.5$ GeV\\
$T_4 = 0.11 \pm 0.14$ & \\
\hline
\end{tabular}
\caption{Inputs that we use in order to constrain the SM4 parameter space,
we have considered the 2$\sigma$ range for $V_{ub}$.}
\label{tab3}
\end{center}
\end{table}


\subsection{$K^+\to \pi^+\nu \bar{\nu}$ decay}

The effective Hamiltonian for $K^+\to \pi^+\nu \bar{\nu}$ can be written as
\bea
{\cal{H}}_{eff} &=& {G_F \over \sqrt{2}}{\alpha\over {2\pi\sin^2{\Theta_w}}} 
\sum_{l=e, \mu, \tau}\Big[ V_{cs}^{\ast}V_{cd} X^l_{NL} + V_{ts}^{\ast}V_{td} X(x_t) 
\nonumber \\ &&
+ V_{t's}^{\ast}V_{t'd} X(x_{t'}) \Big](\bar{s}d)_{V-A}(\bar{\nu}_l\nu_l)_{V-A}\;.
\eea
First term is the contribution from the charm sector.
The function $X(x)$ is relevant for the top part,
\be
X(x) = X_0(x) + {\alpha_s\over 4\pi} X_1(x)\;,
\ee
where $x_q=(m^2_q/M^2_W)$ for all quarks $q$.
Here $X_0(x)$ is the leading contribution given by  
\be
X_0(x) = {x\over 8}\left[- {2 + x \over 1-x} + {3x-6 \over (1-x)^2} \ln{x}\right]\;,
\ee
and $X_1(x)$ is the QCD correction. The expression for $X_1(x)$ is given in \cite{Buras:1997fb}.
The function $X$ can also be written as
\be
X(x_{t/t'}) = \eta_X . X_0(x_{t/t'}), \hskip 20pt \eta_X = 0.994\;.
\ee
Here $\eta_X$ represents the NLO corrections. 

The function $X^l_{NL}$ is the function corresponding to $X(x_t)$ in the charm sector. It results
from the NLO calculations and its explicit form is given in \cite{Buchalla:1995vs,Buchalla:1993wq}.

The branching fraction of $K^+\to \pi^+\nu\bar{\nu}$ can be written as follows
\bea
Br(K^+\to \pi^+\nu\bar{\nu}) &=& \kappa_+ \Big[\left({{\rm Im}\lambda_t \over \lambda^5}X(x_t) 
+ {{\rm Im}\lambda_{t'} \over \lambda^5} X(x_{t'}) \right)^2 
\nonumber \\ &&
 + \Big({{\rm Re}{\lambda_c}\over \lambda}P_0(X) + {{\rm Re}{\lambda_t} \over \lambda^5}X(x_t) 
+ {{\rm Re}{\lambda_{t'}} \over \lambda^5} X(x_{t'})\Big)^2 \Big],
\label{brkpip}
\eea 
where 
\be
\kappa^+ = r_{K+} {{3\alpha^2 \,Br(K^+\to \pi^0 e^+ \nu )}\over {2\pi^2 \sin^4{\Theta_W}}}\lambda^8\;,
\ee
\be
P_0(X) = {1\over\lambda^4}\left[{2\over 3} X^e_{NL} + {1\over 3} X^{\tau}_{NL}\right]\;,
\ee
and $r_{K+} = 0.901$ summarizes the isospin breaking corrections in relating 
the $K^+\to \pi^+\nu\bar{\nu}$ to the well measured leading decay $K^+\to \pi^0 e^+ \nu $.

\begin{table}[t]
\begin{center}
\begin{tabular}{|c||c|c|c|c|}
\hline
$m_{t'}$ ({\it GeV}) & 300 & 400  & 500 & 600  \\
\hline
$\lambda^s_{t'}$& (0.09 - 2.5) & (0.08 - 1.4)& (0.06 - 0.9) &(0.05 - 0.6)  \\
\hline
$\phi_s'$ & 0 $\to$ 80 & 0 $\to$ 80 & 0 $\to$ 80  & 0 $\to$ 80  \\
\hline
\end{tabular}
\caption{Allowed ranges for the parameters, $\lambda^s_{t'}$ ($\times 10^{-2}$)
and phase $\phi_s'$ (in degree) for different masses $m_{t'}$ ( GeV), that has
been obtained from the fitting with the inputs in Table \ref{tab3} and allowed by the 
present experimental bound for {\it CP} asymmetry in $B_s\to J/\psi\phi$  \cite{SAGMN08}.}
\label{tab4}
\end{center}
\end{table}


\section{Predictions in the SM4}
\label{results}
\begin{figure}[htbp]
\includegraphics[width=8cm,height=6cm,clip]{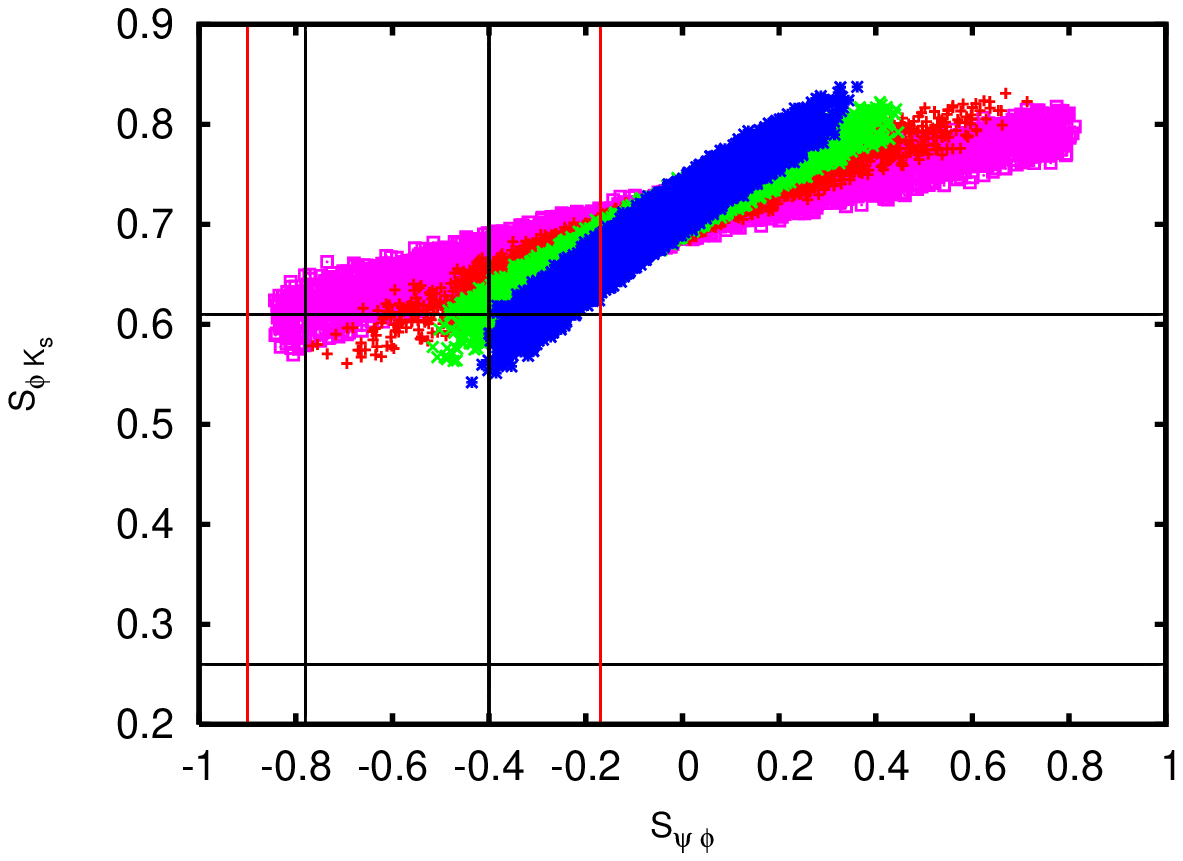}%
\hspace{0.2cm}%
\includegraphics[width=8cm,height=6cm,clip]{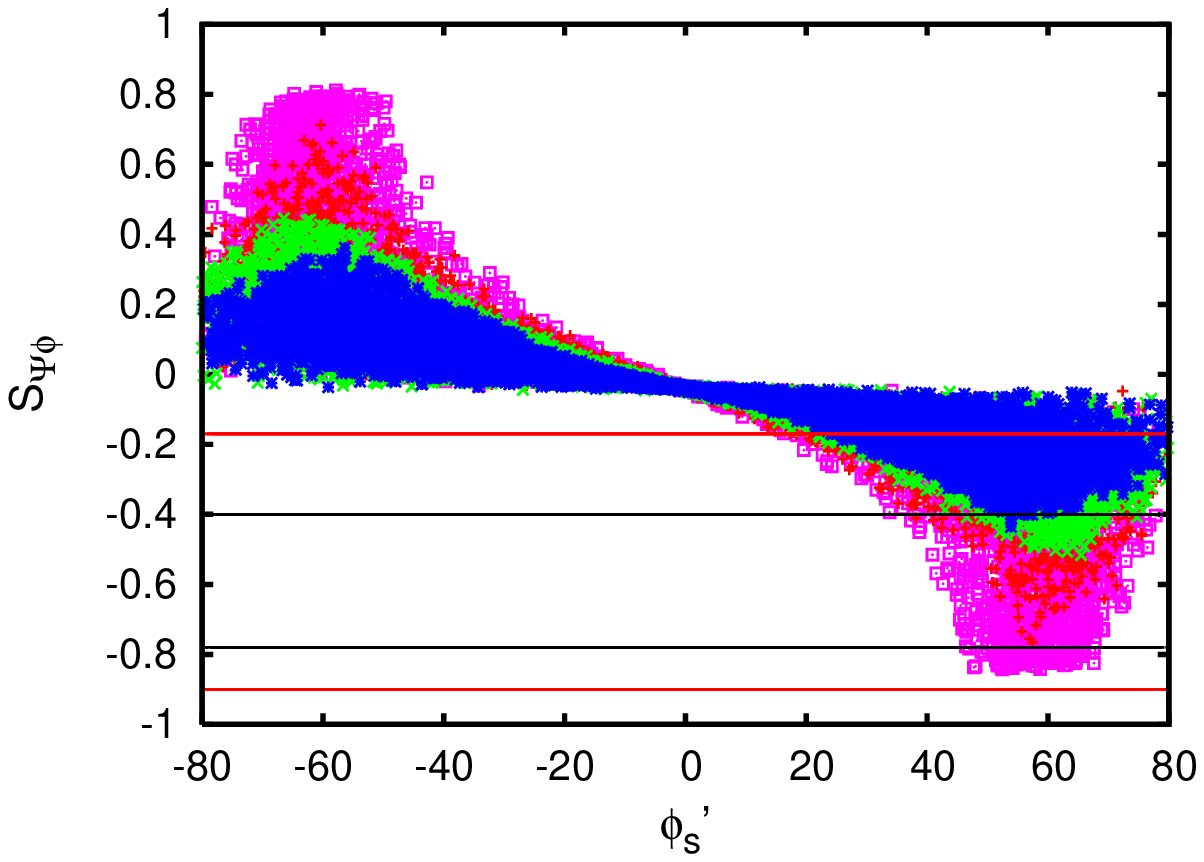}

\caption{(a)   Correlation between $S_{\phi K_s}$ and $S_{\psi\phi}$ 
(left panel) \, and \, (b) Variation of $S_{\psi\phi}$ with the phase $\phi_{s}'$ of 
$\lambda^s_{t'}$ (right panel), for $m_{t'}= 300$ (magenta), $400$ (red), $500$ (green) and 
$600$ (blue) GeV respectively. The horizontal lines (left panel) represent the experimental
$1\sigma$ range for $S_{\phi K_s}$ whereas the vertical lines (black
1-$\sigma$ and red 2-$\sigma$ ) represent that for $S_{\psi\phi}$; in the right panel 
the horizontal lines are for $S_{\psi\phi}$}.
\label{figpap1}
\end{figure}

Fig. \ref{figpap1} (left panel) shows the correlations between the {\it CP} 
asymmetries in $B_d\to \phi K_s$ and $B_s\to \psi \phi$ whereas right panel shows the 
variation $S_{\psi\phi}$ with the new phase $\phi_{s}'$ \footnote{Soon after we posted 
version 1 of our paper , \cite{buras4th} appeared which also discusses about the 
phenomenology of SM4. To facilitate direct comparision with that work we are 
adding few extra figures in this revised version.}; which has already
been shown in our previous article \cite{SAGMN08} for $m_{t'} = 400, \, 500\,$ 
and $600\, {\it GeV}$;  here, we have also included in the plot
$m_{t'} = 300\, {\it GeV}$. This is to clarify the fact 
that the present data on {\it CP} asymmetries tends to favor a fourth family of
quarks with $m_{t'}$ in the range $(400\, - 600\,) {\it GeV}$. 
In this article therefore, we will focus mostly on $m_{t'}\approx 400 - 600$  
{\it GeV} when we provide numerical results for SM4 for some 
interesting observables related to $B$ and $K$ system which could be tested 
experimentally.
  
\subsection{Direct CP asymmetry in  $B \to X_s \gamma$}
\label{dacpbsg}

$A_{CP}$ in $B \to X_s \gamma$ is defined as
\be
 A_{CP}^{B \to X_s \gamma} = \frac{\Gamma(\bar{B}\to X_s \gamma)-\Gamma(B \to X_{\bar{s}} \gamma)}
{\Gamma(\bar{B}\to X_s \gamma)+\Gamma(B \to X_{\bar{s}} \gamma)}
\ee
Within the SM,  $A_{CP}^{B \to X_s \gamma}$ is predicted to be less than $1\%$ \cite{Kagan:1998bh,Kiers:2000xy,Soares:1991te}.
The most recent SM prediction is \cite{Hurth:2003dk} (Here we have calculated the errors by adding all  
errors given in the mentioned reference in quadrature ) 
\be
A_{CP}^{B \to X_s \gamma}|_{E_{\gamma}>1.6\,{\rm GeV}} = \big(0.44^{+0.24}_{
-0.13} \big)\%\;.
\ee

The current world average of $A_{CP}^{B \to X_s \gamma}$ is $(-1.2\pm 2.8)\%$ \cite{Barberio:2008fa},
which is consistent with zero or a very small direct CP asymmetry as we have in the SM.
The present experimental uncertainty is still an order of magnitude 
greater than the theoretical error. However a dramatic improvement in the experimental
sensitivity is possible at the upcoming Super-B factories and sensitivity of 
about
$0.4\%-0.5\%$ can be achieved \cite{Browder:2008em}.

As the CP asymmetry within the SM is less than $1\%$,  
observation of a sizable CP asymmetry would be a clean signal of new physics. 
It is expected that the new physics models with non-standard CP-odd phases 
can enhance $A_{CP}^{B \to X_s \gamma}$ and hence we study $A_{CP}^{B \to X_s \gamma}$
within the framework of SM4.

\begin{figure}[t]
\includegraphics[width= 0.60 \linewidth]{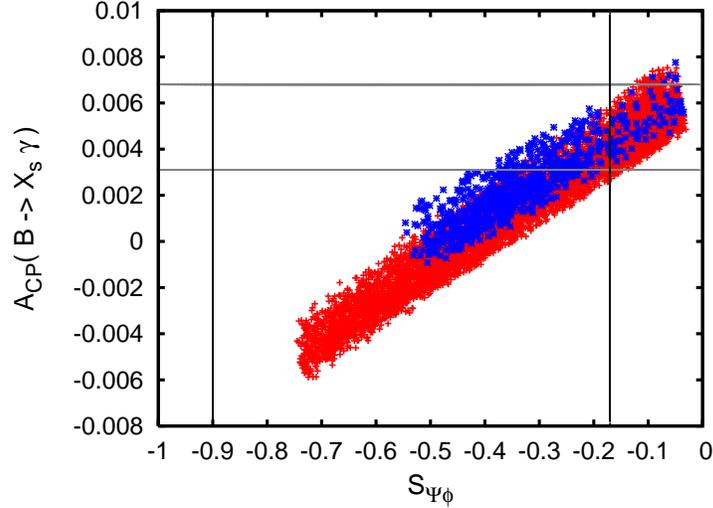}
\caption{Correlation between CP asymmetry in $B\to X_s \gamma$ and 
$S_{\psi\phi}$, the CP asymmetry in $B_s\to J/\psi \phi$; where the red
and blue regions correspond to $m_{t'}$ = 400 and 600 GeV whereas horizontal 
lines represent the SM limit for CP asymmetry and the vertical lines 
represent the $2\sigma$ limit for CP asymmetry in $B_s\to J/\psi \phi$.}
\label{fig:acp_bsg}
\end{figure}

The general expression for the CP asymmetry in $B \to X_s \gamma$ is \cite{Kiers:2000xy}
\bea
A_{CP}^{B \to X_s \gamma} & \simeq & \frac{10^{-2}}{|C^{\rm tot}_7(m_b)|^2}
\Big\{ -1.82\;{\rm Im}\left[C_7^{\rm new}\right] 
+\,1.72\;{\rm Im}\left[C_8^{\rm new}\right] 
-\,4.46\;{\rm Im}\left[C_8^{\rm new}
			C_7^{{\rm new} *}\right] 
\nonumber \\ &&			
+\,3.21\;{\rm Im}\left[\epsilon_{s}\left(
		1-2.18 \; C_7^{{\rm new} *}
		 -0.26\; C_8^{{\rm new} *}\right)\right]\Big\} \;,	
\label{acpgen4}
\eea
where
\be
	\epsilon_s = \frac{V_{us}^* V_{ub}}{V_{ts}^* V_{tb}} \;,
\ee
Here the new physics Wilson coefficients $C_{7,8}^{\rm new}$ are at scale $M_W$.
In SM4, 
\be
C_{7,8}^{\rm new}=\frac{V_{t's}^*V_{t'b}}{V_{ts}^*V_{tb}}C_{7,8}^{t'}(M_W)\;.
\ee

In the Fig. \ref{fig:acp_bsg} we have shown the correlation between CP 
asymmetries in $(B\to X_s \gamma)$ and $B_s\to J/\psi\phi$ 
($S_{\psi\phi}$). The current $2\sigma$ experimental range for $S_{\psi\phi}$ 
is given by $[-0.90,-0.17]$ \cite{cdfd0}. The SM value for $A_{CP}(B\to X_s \gamma)$ 
corresponds to $S_{\psi\phi}\approx 0$ or in other words 
$\phi^{t'}_s\approx 0$. It is easy to understand the nature of the plot i.e 
decrease of $A_{CP}(B\to X_s \gamma)$ 
with increase of $S_{\psi\phi}$. From the expression for 
$A_{CP}(B\to X_s \gamma)$ (eq.~(\ref{acpgen4})), it is clear that in SM the 
only contribution to $A_{CP}$ will come from the first part of the fourth term. 
In the presence of new phase and new coupling, the first two terms 
and the fourth term will contribute to $A_{CP}$. 
Contribution from the first two term is always negative and increases 
(mod value) with the new physics coupling ( within the NP region 
we are interested) whereas the fourth term is always positive and it has
very small increase with the new physics coupling or phase.

\subsection{{\it CP} asymmetry in $B_s\to X_s\ell\nu$ }

In this section we shall concentrate on semileptonic {\it CP} asymmetry
($A_{SL}$) in $B_s$ system \footnote{We were about to post a short
paper reporting our study of $A_{SL}$ in SM4 when the paper \cite{buras4th} appeared wherein
this topic is also discussed-consequently we are making a very breif addition of this in version
\,2 of our paper. Our results agree with Buras {\rm et. al} \cite{buras4th}.}. In general the
{\it CP} asymmetry in semileptonic $B_s$ decays defined as,
\begin{align}
A_{SL} &= \frac{\Gamma[\bar{B}^{phys}_{s}(t)\to \ell^+ X]-\Gamma[B^{phys}_{s}(t)\to \ell^- X]}{\Gamma[\bar{B}^{phys}_{s}(t)\to \ell^+ X]+\Gamma[B^{phys}_{s}(t)\to \ell^- X]},
\end{align}
depends on the relative phase between the absorptive and dispersive parts of
$B_s-\bar{B_s}$ mixing amplitude \cite{hw},
\begin{align}
A_{SL} &={\it Im}\left(\frac{\Gamma_{12}}{M_{12}}\right) =
\frac{|\Gamma^{s}_{12}|}{|M^{SM}_{12}|} \frac{\sin\phi_s}{|\Delta_s|},
\label{asld1}
\end{align}
with $\phi_s = arg\left(-\frac{M^s_{12}}{\Gamma^s_{12}}\right)$, the relative phase
between $B_s-\bar{B_s}$ mixing and the corresponding $b\to c\bar{c} s$ decays and $|\Delta_s|$
parametrises the NP effect in $M^s_{12}$ \cite{uli_lenz}.
$|\Gamma_{12}/M_{12}|=O(m_b^2/M_W^2)$ suppresses $A_{SL}$ to the percent level,
apart from this there is a GIM suppression factor $m_c^2/m_b^2$ reducing
$A_{SL}$ by another order of magnitude. Because of these suppression factors
it is very small in SM, for $B_s$ system it is ${\cal{O}}(10^{-5})$. The GIM
suppression is lifted if new physics contributes to $\arg (M_{12})$. Therefore
$A_{SL}$ is very sensitive to new {\it CP} phases \cite{run2,llnp}.
The situation where new physics could enhance $A_{SL}$ by a factor
$\cal{O}$(10-100) makes this asymmetry a sensitive probe of new physics.

Recently the search for {\it CP} violation in semileptonic $B_s$ decays
achieved a much more improved sensitivity \cite{d07,cdf7}:
\begin{align}
A_{SL} &=(2.45 \pm 1.96)\times 10^{-2} \hskip 50pt {\rm D0}
\nonumber \\
       &=(2.00\pm 2.79)\times 10^{-2} \hskip 50pt {\rm CDF} .
\end{align}
Present world average is given by \cite{hfag},
\begin{align}
A_{SL} &=(-0.37 \pm 0.94)\times 10^{-2} \hskip 50pt {\rm HFAG} .
\end{align}
In near future more precise measurements can exclude SM prediction if it is much
enhanced then the SM prediction.
It is important to note that the scenarios like SM4 can significantly
affect $M_{12}^s$, but not $\Gamma_{12}^s$, which is dominated by the
CKM-favoured $b\to c\ov{c}s$ tree-level decays. The leading contribution to $\Gamma_{12}^s$ was
obtained in \cite{hw,LO}. At present $\Gamma_{12}^s$ is known to
next-to-leading-order (NLO) in both $\lbar/m_b$ \cite{bbd1} and
$\alpha_s(m_b)$ \cite{bbgln1,rome03,bbln}, later in 2006 Nierste and Lenz
\cite{uli_lenz} have improved the NLO calculation
for $\dg_s$ and  updated the value for $\dg_s$.

\begin{figure}[htbp]
\includegraphics[width=8cm,height=6cm,clip]{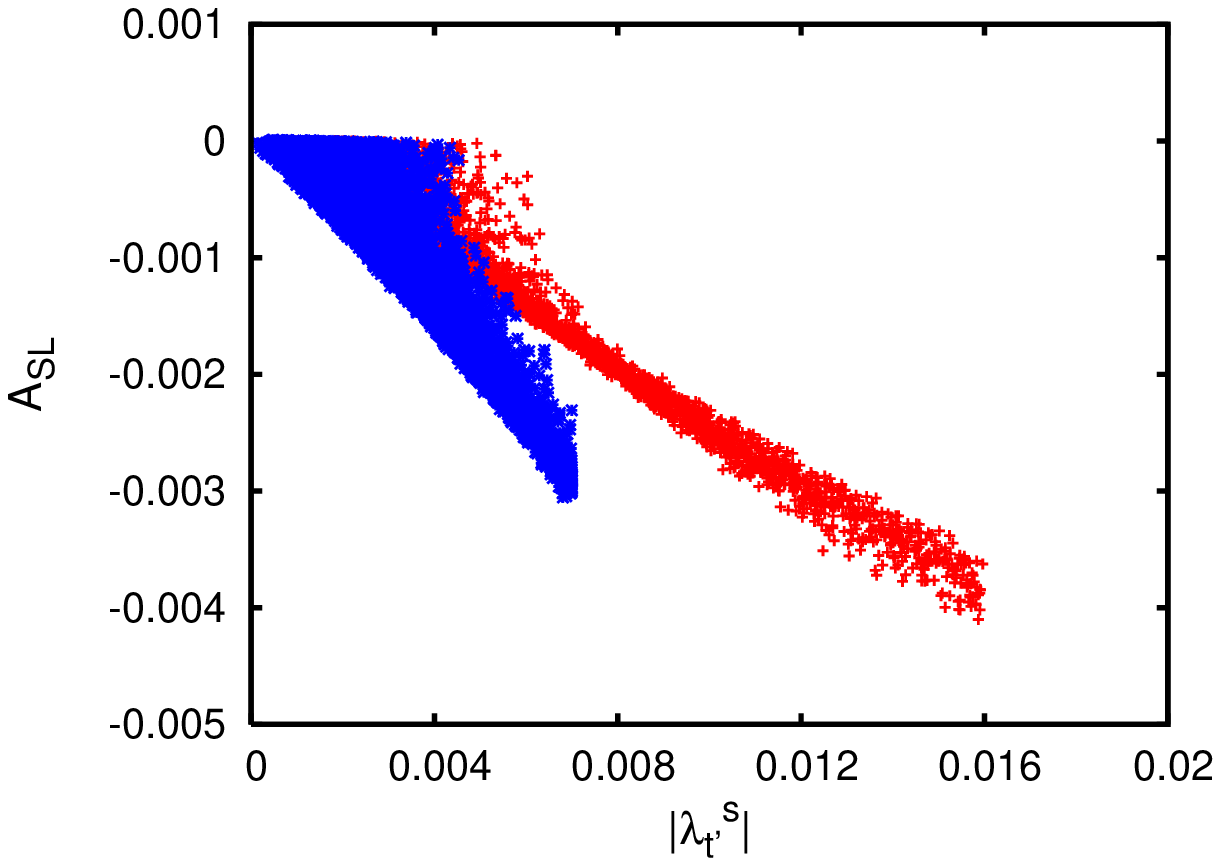}%
\hspace{0.2cm}%
\includegraphics[width=8cm,height=6cm,clip]{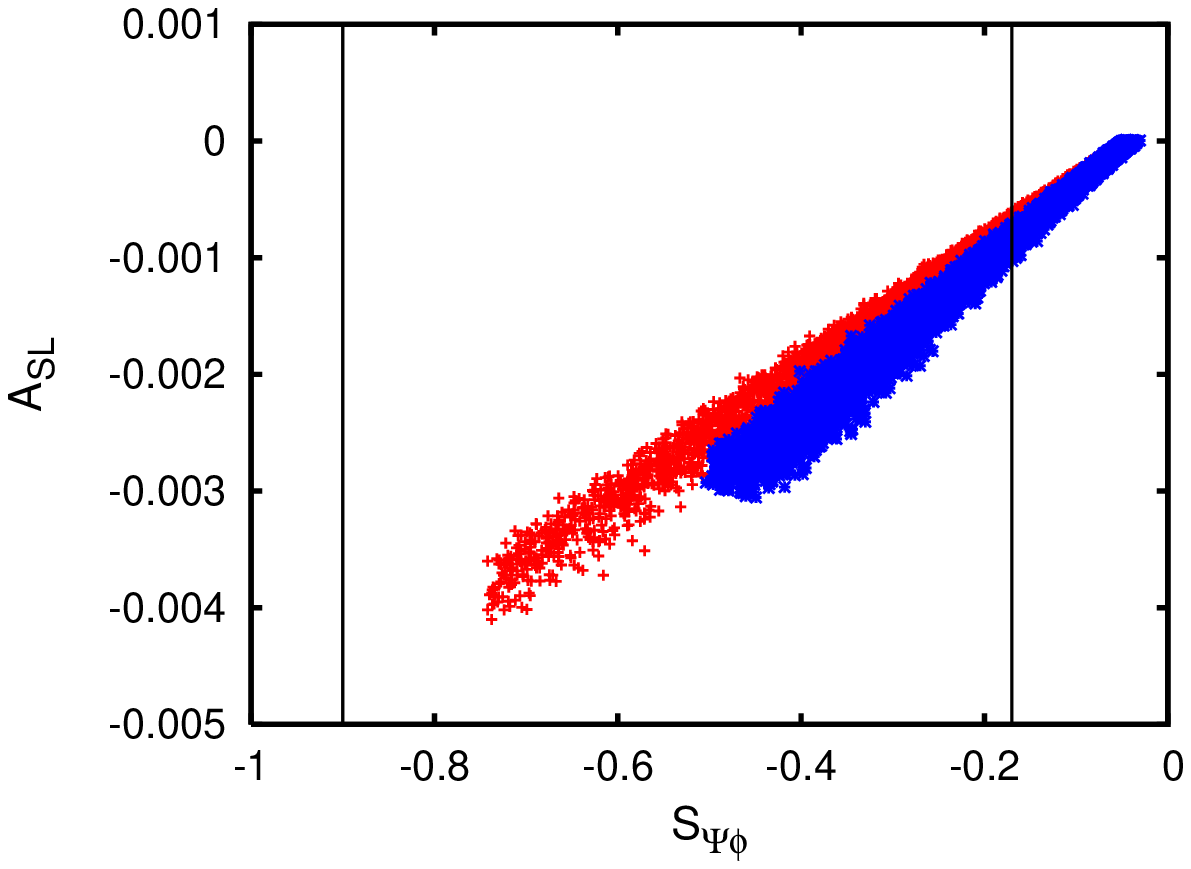}

\caption{Left panel shows the semileptonic {\it CP} asymmetry $A_{SL}$ as a function of 
$|\lambda^s_{t'}|$ whereas in the right panel correlation between
$A_{SL}$ and $S_{\psi\phi}$ is shown; red and blue region corresponds to $m_{t'}$ = 400
and 600 $\rm GeV$ respectively, the SM value of $A_{SL}$ (of order $10^{-5}$) is too close to
zero to be visible in the plot whereas the SM value for $S_{\psi\phi}$ is $-0.04$.}
\label{fig_asl}
\end{figure}
In Fig. \ref{fig_asl} the sensitivity of semileptonic {\it CP} asymmetry
to SM4 is shown and we note an enhancement by a factor of 100 from its SM predicion
of order $10^{-5}$. It could have a value $-0.4\%$ and $-0.3\%$ corresponding to maximum
values of $S_{\psi\phi}$ for $m_{t'} = $ 400 and 600 \,{\it GeV} respectively.

\subsection{CP asymmetry in $\bs$}
\label{acpbq}

It is very useful to consider new physics effects in the 
observables which are either zero or highly suppressed in the SM as they constitute null test of the SM \cite{soni_gers} . 
The reason is that any finite or large measurement of
such an observable may signal the existence of new physics. 
The CP asymmetry in  $B \to X_s \, l^+ \, l^-$ is one such 
observable. 
In the SM, the CP asymmetry in 
$B \to X_s\, l^+ \,l^-$ is $\sim 10^{-3}$ \cite{Du:1995ez,Ali:1998sf}.   
In the SM, 
the only source of CP violation 
is the unique phase in the CKM
quark mixing matrix.
However in many possible extensions of the SM, 
there can be extra phases contributing to the
CP asymmetry. Hence the CP asymmetry in $B \to X_s\, l^+ \,l^-$ is sensitive to SM4.

The CP asymmetry in $\bs$ is defined as
\begin{equation}
A_{\rm CP}(z)=\frac{(dBr/dz)-(d\overline{Br}/dz)}{(dBr/dz)+(d\overline{Br}/dz)}=
\frac{D(z)-\overline{D(z)}}{D(z)+\overline{D(z)}}\;,
\end{equation}
where $Br$  and $\overline{Br}$ represent the branching ratio of $\bar{B} \to  X_{s}l^+ l^-$
and its complex conjugate $B \to  \bar{X_{s}} l^+ l^- $ respectively. $dBr/dz$ is given in eq.~(\ref{eq:dbr_bsll}). 
The Wilson coefficients $C_{7}^{\rm tot }$, $C_{9}^{\rm tot}$, and $C_{10}^{\rm tot}$ can be written as
\begin{eqnarray}
C_{7}^{\rm tot }& = & C_{7}(m_b)\,+\, \lambda_{tt'}^s\,C_{7}^{t' }(m_b)\;, \\
C_{9}^{\rm tot }& = & \xi_1\,+\, \lambda_{tu}^s\xi_2\,+\, 
\lambda_{tt'}^s\,C_{9}^{t' }(m_b)\;, \\
C_{10}^{\rm tot } & = & C_{10}(m_b)+ \lambda_{tt'}^s\,C_{10}^{t' }(m_b)\;,
\label{c10sm4}
\end{eqnarray}
where
\begin{equation}
\lambda_{tu}^s=\frac{\lambda_u^s}{\lambda_t^s}=\frac{V_{ub}^*V_{us}}{V_{tb}^*V_{ts}}\;,
\end{equation}
\begin{equation}
\lambda_{tt'}^s=\frac{\lambda_{t'}^s}{\lambda_t^s}=\frac{V_{t'b}^*V_{t's}}{V_{tb}^*V_{ts}}\;,
\end{equation}
so that all three relevant Wilson coefficients are complex in general.
The parameters $\xi_i$ are given by \cite{Buras:1994dj}
\begin{eqnarray}
\xi_1 & = & C_9(m_b) \, +\, 0.138 \,\omega(z)\,+\,g(\hat{m}_{c},z) (3 C_1 + C_2 + 3 C_3 +
C_4 + 3 C_5 + C_6)\nonumber\\&&- \frac{1}{2}g(\hat{m}_{d},z)
(C_3 + 3C_4) - \frac{1}{2}
   g(\hat{m}_{b},z)(4 C_3 + 4 C_4 + 3C_5 + C_6) \nonumber\\
  & & +\frac{2}{9} (3 C_3 + C_4 + 3C_5 + C_6)\;, \\
\xi_2 & = & [ g(\hat{m}_{c},z)- g(\hat{m}_{u},z)](3C_1 + C_2)\; .
\end{eqnarray}
Here
\bea 
\omega(z) & = & - \frac{2}{9} \pi^2 - \frac{4}{3}\mbox{Li}_2(z) - \frac{2}{3}
\ln z \ln(1-z) - \frac{5+4z}{3(1+2z)}\ln(1-z)  \nonumber \\
& &  - \frac{2 z (1+z) (1-2z)}{3(1-z)^2 (1+2z)} \ln z + \frac{5+9z-6z^2}{6
(1-z) (1+2z)} \; ,
\eea
with
\begin{equation}
\mbox{Li}_2(z)\,=\,-\int_0^z dt\, \frac{{\rm ln}(1-t)}{t}\;.
\end{equation}
The function $g(\hat m,z)$ represents the one loop
corrections to the four-quark operators $O_1-O_6$ and is given by \cite{Buras:1994dj}
\begin{eqnarray}
g(\hat m, z) &  = & -\frac{8}{9}\ln\frac{m_b}{\mu_b} - \frac{8}{9}\ln \hat m +
\frac{8}{27} + \frac{4}{9} x \\
& & - \frac{2}{9} (2+x) |1-x|^{1/2} 
\left\{\begin{array}{ll}
\left( \ln\left| \frac{\sqrt{1-x} + 1}{\sqrt{1-x} - 1}\right| - i\pi \right), &
\mbox{for } x \equiv \frac{4\hat m^2}{z} < 1 \nonumber \\
2 \arctan \frac{1}{\sqrt{x-1}}, & \mbox{for } x \equiv \frac
{4\hat m^2}{z} > 1,
\end{array}
\right. 
\end{eqnarray}
For light quarks, we have $\hat{m}_{u}\simeq \hat{m}_{d}\simeq0$. 
In this limit,
\be
g(0, z) =  \frac{8}{27} -\frac{8}{9} \ln\frac{m_b}{\mu_b} - \frac{4}{9} \ln
z + \frac{4}{9} i\pi\;.
\ee
We compute $g(\hat{m},z)$ at $\mu_b = m_b$.

$d\overline{Br}/dz$ can be obtained from $dBr/dz$
by making the following replacements: 
\begin{eqnarray}
C_{7}^{\rm tot } =C_{7}(m_b)\,+\, \lambda_{tt'}^s\,C_{7}^{t' }(m_b)
& \to & \overline {C_{7}^{\rm tot }}=C_{7}(m_b)\,+\, 
\lambda_{tt'}^{s*}\,C_{7}^{t' }(m_b)\;, \\
C_{9}^{\rm tot }=\xi_1\,+\, \lambda_{tu}^s\xi_2\,+\, \lambda_{tt'}^s\,C_{9}^{t' }(m_b)
& \to &
\overline {C_{9}^{\rm tot }}=\xi_1\,+\, \lambda_{tu}^{s*}\xi_2\,+\, \lambda_{tt'}^{s*}\,C_{9}^{t' }(m_b)\;, \\
C_{10}^{\rm tot }=C_{10}(m_b)+ \lambda_{tt'}^s\,C_{10}^{t' }(m_b)
& \to & 
\overline {C_{10}^{\rm tot }}=C_{10}(m_b)+ \lambda_{tt'}^{s*}\,C_{10}^{t' }(m_b)\;.
\end{eqnarray}

Then we get \cite{alok}
\begin{eqnarray}
D(z) - \overline{D(z)} 
&=& 2 \left( 1 + \frac{2 t^2}{z} \right) 
\bigg[ \im(\ltu) \left\{ 2(1+2z) \im(\xi_1 \xi_2^\ast) - 12 C_7 \im(\xi_2) \right\}  \nonumber \\
&& \phantom{2 \left( 1 + \frac{2 t^2}{z} \right) } 
+  X_{im} \left\{ (1+2 z) C_9^{t'} + 6 C_{7}^{t'} \right\} \bigg] \; ,
\label{acp_num} \\
D(z) + \overline{D(z)} 
&=& \left( 1 + \frac{2 t^2}{z} \right) 
\Bigl[ (1 + 2z) \left\{ B_1 + 2 C_9^{t'} \left( | \ltt |^2 C_9^{t'} + X_{re} \right) \right\} \Bigr. \nonumber \\
&& \Bigl.  + 12 \left\{ B_2 + 2 C_7 C_9^{t'} \re(\ltt) 
+ C_{7}^{t'} \left(  2 | \ltt |^2 C_9^{t'}  + X_{re} \right)  \right\}\Bigr] \nonumber \\
&&+  8  \left( 1 + \frac{2 t^2}{z} \right)  \left( 1 + \frac{2}{z} \right) |C_7^{\rm tot}|^2   \nonumber \\
&&+2 \left[ \left( 1 + 2 z \right) +  \frac{2 t^2}{z} \left( 1 - 4 z \right) \right] |C_{10}^{\rm tot}|^2 \; , 
\label{acp_den}
\end{eqnarray}
where
\begin{eqnarray}
X_{re} &=& 2  \left\{ \re\left( \ltt \right) \re\left( \xi_1 \right) 
+ \re\left( \ltt {\ltu}^\ast \right) \re\left( \xi_2 \right)\right\} \; , \\
X_{im} &=& 2  \left\{ \im\left( \ltt \right) \im\left( \xi_1 \right) 
+ \im\left( \ltt {\ltu}^\ast \right) \im\left( \xi_2 \right)\right\} \; , 
\label{xim}\\
B_1 &=& 2 \left\{  |\xi_1|^2 + {|\ltu \xi_2|}^2 + 2 \re\left( \ltu \right) \re\left( \xi_1 \xi_2^\ast \right) \right\} \; , \\
B_2 &=& 2 C_{7} \left\{ \re(\xi_1) + \re(\ltu) \re(\xi_2)  \right\} \; , \\
|C_{10}^{\rm tot}|^2 
&=& {\left( C_{10} \right)}^2
+ |\ltt |^2 {\left( C_{10}^{t'} \right)}^2 + 2 C_{10} C_{10}^{t'} \re\left( \ltt \right) \; , \\
|C_{7}^{\rm tot}|^2 
&=&{\left( C_{7} \right)}^2 
+ |\ltt |^2 { \left( C_{7}^{t'} \right)}^2 
+ 2 C_{7}  C_{7}^{t'}  \re\left( \ltt \right) \; .
\end{eqnarray}

\begin{figure}
\includegraphics[width= 0.60 \linewidth]{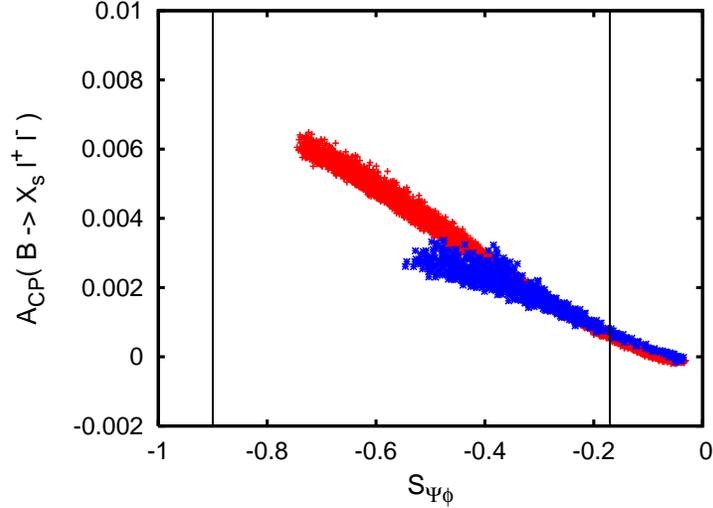}
\caption{Correlation between CP asymmetry in $B\to X_s \ell^+\ell^-$ 
(high-$q^2$ region) and $S_{\psi\phi}$. In the SM both the values are very 
small and in the plot they correspond to the point $[-0.04,0.0]$ .
The red and blue regions correspond to $m_{t'}$ = 400 and 600 GeV whereas the 
vertical lines represent $2\sigma$ experimental range for $S_{\psi\phi}$. }
\label{fig:acp_bsll}
\end{figure}

From the expression for $g(\hat{m}, z)$ it is clear that the strong phase in 
$g(\hat{m}_{u/d}, z)$ and $g(\hat{m}_{c}, z)$ is responsible for CP 
asymmetry in $B\to X_s \ell^+\ell^-$ within the SM. $g(\hat{m}_{u/d}, z)$ is 
complex in both high and low-$q^2$ region whereas $g(\hat{m}_{c}, z)$ is 
complex only in the high-$q^2$ region. On the other hand $g(\hat{m}_{b}, z)$ 
is always real. The SM CP asymmetry in high-$q^2$ region is almost zero 
since Im($\xi_2$) is very small, almost one order in magnitude relative to 
its value in low-$q^2$ region, due to the relative cancellations of strong 
phases in $\xi_2$. In the presence of new physics $\xi_2$ is unaffected but
 $\xi_1$ increases with the new physics coupling . On the other hand we have 
contribution from the second term of eq.~(\ref{acp_den}) as a whole the CP
 asymmetry will increase with $S_{\psi\phi}$, as shown in the figure \ref{fig:acp_bsll}.  

\subsection{FB asymmetry in  $B \to X_s\, l^+ \, l^-$}

The quark level transition $b \to s\, l^+\, l^-$ is forbidden at the
tree level within the SM and can occur only via one or more loops. Hence it has the potential
to test higher order corrections to the SM and also to constrain many of its
possible extensions. It gives rise to the inclusive decay $B \to X_s\, l^+ \, l^-$
which has been experimentally observed \cite{Aubert:2004it,Iwasaki:2005sy}
with a branching ratio close to its
SM predictions, $Br(B\to X_s \ell^+ \ell^-)(1 < q^2 < 6$ ${\rm GeV}^2)$=
$(1.63 \pm 0.20)\times 10^{-6}$ and $Br(B\to X_s \ell^+ \ell^-)(q^2 > 14.4$
${\rm GeV}^2)$=
$(3.84 \pm 0.75) \times 10^{-7}$ \cite{Ghinculov:2003bx,Ali:2002jg,Ali:2002ik}.

Apart from the branching ratio of semi-leptonic decay, there are other
observables
which are sensitive to new physics contribution to $b \to s$ transition.
One such observable is forward-backward (FB) asymmetry of leptons in
$B \to X_s\, l^+ \, l^-$. The FB asymmetry of leptons in
$B(p_b) \to X_s(p_s)\, l^+(p_{l^+})\, l^-(p_{l^-})$ is obtained by
integrating the double differential branching ratio
($d^2 Br/dz dcos\theta$)
with respect to the angular variable $cos\theta$ \cite{Ali:1991is}
\bea
A_{FB}(z)= \frac{\int_0^{1}dcos\theta \frac{d^2Br}{dz \
dcos\theta}-\int_{-1}^{0}dcos\theta \frac{d^2Br}{dz\ dcos\theta}}
{\int_0^{1}dcos\theta \frac{d^2Br}{dz\ dcos\theta}+\int_{-1}^{0}dcos\theta \frac{d^2Br}{dz
\ d\cos\theta}}\;,
\eea
where $z \equiv q^2/m_{b}^{2}\equiv (p_{l^+}+p_{l^-})^2/m_{b}^{2} $ and
$\theta$
is the angle between the momentum of the $B$-meson (or the outgoing $s$-quark)
and that of $l^+$ in the center of mass frame of the dileptons $l^+ l^-$.
FB asymmetry
measures the difference in the right-chiral and left-chiral couplings of the
leptonic current. FB asymmetry is driven by the top quark \cite{Ali:1991is}
and hence it is sensitive to the fourth generation up type quark $t'$.

\begin{figure}[t]
\includegraphics[width= 0.60 \linewidth]{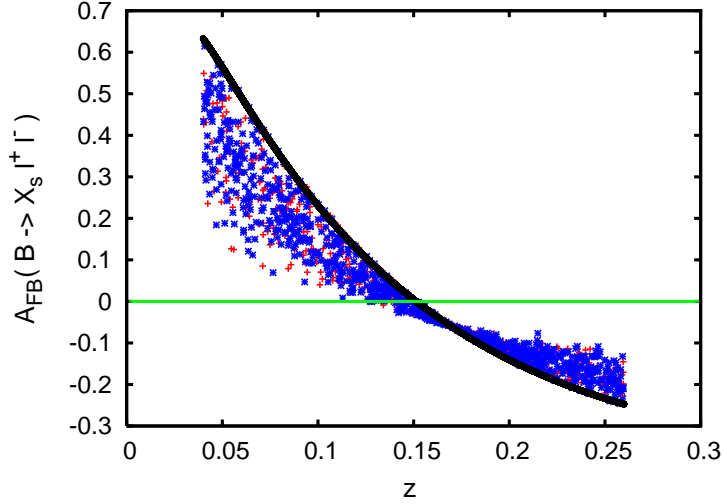}
\caption{Forward backward (FB) asymmetry in $B\to X_s \ell^+\ell^-$ has been
plotted with $z={q^2 \over m^2_b}$ , the red and blue regions correspond to
$m_{t'}$ = 400 and 600 GeV respectively and the black thick line represents
that for SM and the green line represents the zero of $A_{FB}$.}
\label{fig:fb}
\end{figure}
Within the framework of SM4, the FB asymmetry in $B \to X_s\, l^+ \, l^-$ is given by
\be
A_{FB}(z) =- 3 \left(1-\frac{4t^2}{z}\right)^{1/2}\,\frac {E(z)}{D(z)}\;,
 \ee
where
\be
E(z)= {\rm Re}(C_9^{\rm tot}C_{10}^{\rm tot*})z
  + 2{\rm Re}(C_7^{\rm tot}C_{10}^{\rm tot*})\;,
 \ee
and $D(z)$ is given in eq.~(\ref{bsll_Dz}).

The FB asymmetry in $B \to X_s\, l^+ \, l^-$ becomes zero for a particular
value of the dilepton invariant mass.
Within SM, the zero of $A_{FB}(q^2)$ appears in the low $q^2$ region, sufficiently away
from the charm resonance region to allow the precise prediction of
its position in perturbation theory. The value of the zero of the FB asymmetry
is one of the most precisely calculated observables in flavor physics
with a theoretical error of order $5\%$. The NNLO prediction for the
zero of FB asymmetry is with $m_b=4.8$ $\rm GeV$\cite{Huber:2007vv}
\begin{equation}
(q^2)_0= (3.5 \pm 0.12)\, {\rm GeV}^2 \,.
\end{equation}
This zero varies from model to model.
Thus it can serve as an important probe
to test SM4 experimentally.

As far as experiments are concerned, this quantity has not been
measured as yet. But estimates show that a precision of about $5\%$ could
be obtained at Super-B factories \cite{Browder:2008em}.

From Fig. \ref{fig:fb} one can see that the value of $z = \frac{q^2}{m^2_b}$,
for which $A_{FB}(z)$-asymmetry is zero, could be shifted to a lower value
than its SM value (although it is consistent with the SM within the
uncertainty). For $m_{t'}$ = 400 and 600 GeV, one could have the value for
$(q^2)_0$ ranging between $(3.09 \to 3.57)$ ${\rm GeV}^2$ for
$m_b = 4.8$ $\rm GeV$.

\subsection{FB asymmetry in  $B\to K^* \ell^+\ell^-$}

The quark level transition $b \to s \ell^+\ell^-$ is responsible for the exclusive decay $B\to K^* \ell^+\ell^-$.
The exclusive decay $B\to K^* \ell^+\ell^-$ has relatively large theoretical errors as compared to the inclusive decay $b \to s \ell^+\ell^-$
due to the uncertainty in the determination of the hadronic form factors appearing in the transition amplitude $B\to K^*$. However the exclusive decays are more readily accessible in the experiments. Therefore despite the large theoretical errors, the precise measurement of the exclusive decays could provide hints for possible deviations from the SM. 
The decay  $B\to K^* \ell^+\ell^-$ has been observed at the Babar and Belle experiments \cite{babar-03,babar-06,belle-03}. Within the present experimental
and theoretical precisions, the measured branching ratio is in agreement with the SM prediction \cite{Ali:2002jg,lunghi}. 
However the measurements of the invariant dilepton mass is sparse. It is expected that the precise measurements 
of the Dalitz distributions in $B\to K^* \ell^+\ell^-$
is possible at the LHCb and at the Super B factories. 
In particular, the measurement of FB asymmetry in $B\to K^* \ell^+\ell^-$ is of great importance. This is because the uncertainty due to the form factors is minimal \cite{ali-00}.

\begin{figure}[t]
\includegraphics[width=8cm,height=6cm,clip]{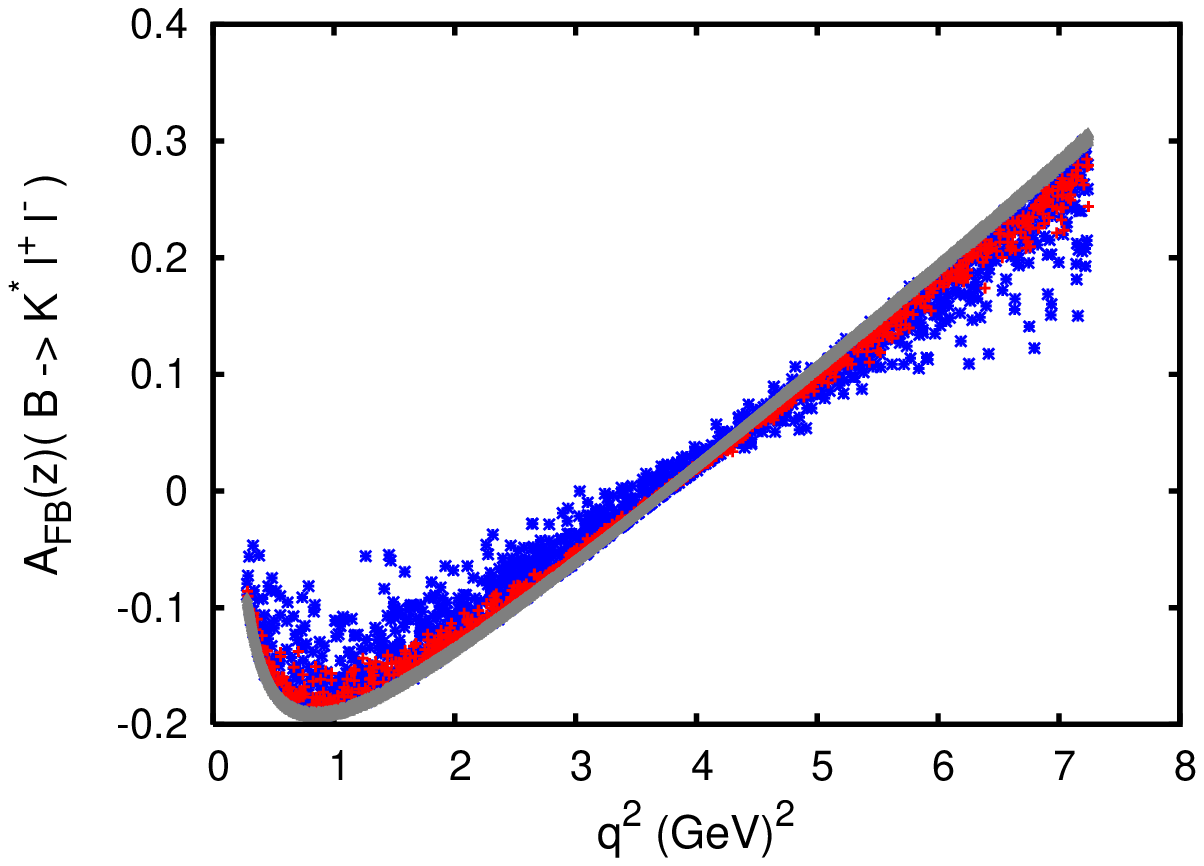}%
\hspace{0.2cm}%
\includegraphics[width=8cm,height=6cm,clip]{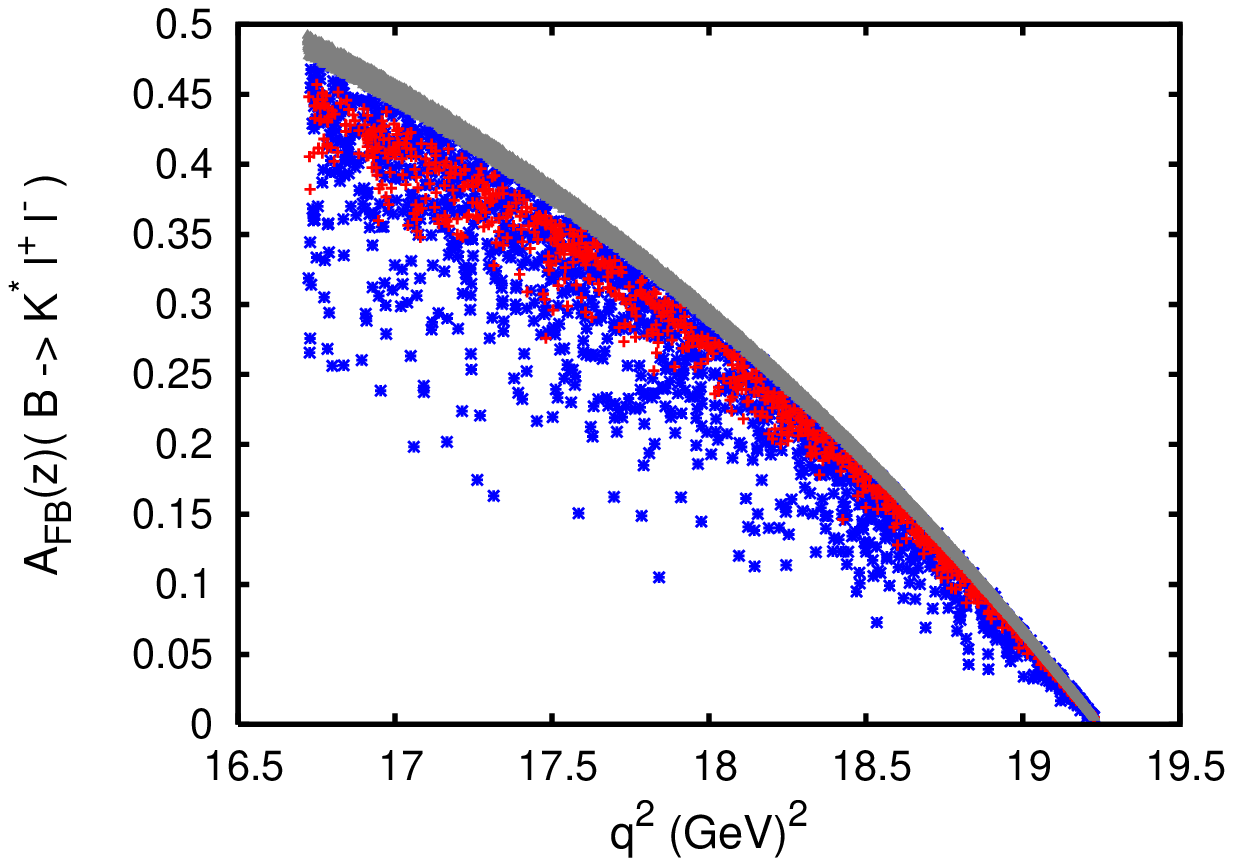}
\caption{FB asymmetry in $B\to K^* \ell^+\ell^-$
in the low-$q^2$  (left panel) and the high-$q^2$ region (right panel).  The 
red and the blue regions correspond to 
$m_{t'}$ = 400 and 600 GeV respectively and the grey region represents the SM 
prediction.}
\label{fb_excld}
\end{figure}

Within the SM4, the normalized FB-asymmetry in $B\to K^* \ell^+\ell^-$ is 
given by \cite{ali-00}
\begin{eqnarray}
A_{FB}(z) &=&- \frac{G^2_F\alpha^2m^4_B}{2^8 \pi^5 (d\Gamma/dz)}|V^*_{ts}V_{tb}|^2 z \lambda \left(1-\frac{4\hat{m}^2_l}{z}\right)
\times \Bigg[{\rm Re}(C^{\rm tot}_9 C^{\rm tot *}_{10})VA_1 
\nonumber\\ &&
+ \frac{\hat{m}_b}{z} {\rm Re}(C^{\rm tot}_7 C^{\rm tot *}_{10})\Big\{VT_2(1-\hat{m}_{K^*})+A_1T_1(1+\hat{m}_{K^*})\Big\}   \Bigg]\;,
\end{eqnarray}
where
\begin{eqnarray}
 \lambda &=&1+\hat{m}^4_{K^*} + z^2-2 z - 2\hat{m}^2_{K^*} (1+z) \; ,\\
z&=&\frac{q^2}{m^2_B}\;, \\
\hat{m}_{K^*} &=& \frac{m_{K^*}}{m_B}\;.
\end{eqnarray}
Here $(d\Gamma/dz)$ is the $B\to K^* \ell^+\ell^-$ differential decay 
distributions and its detailed expression can be seen from Ref. \cite{ali-00}. 
The form factors $A_i, \, V,\, T_i$ are calculated in the light cone QCD approach and their values are given in \cite{ali-00}.

The zero of FB-asymmetry is determined by the equation,
\begin{equation}
Re\Big(C^{eff}_9({z}_0)\Big)= - 2 \frac{\hat{m}_b}{{z}_0} C^{eff}_7 
\frac{1 - {z}_0}{1 + m^2_{K^*}-{z}_0},
\end{equation}
where ${z}_0$ corresponds to the value of  ${z}$ for which FB-asymmetry is 
zero, within SM the value of $(q^2)_0$ for $m_b= 4.8$ $\rm GeV$ is given by 
\cite{ali-00} 
\begin{equation}
(q^2)_0 = z_0 M^2_B = 2.88^{+0.44}_{-0.28} \hskip 5pt {\rm GeV}^2.
\end{equation}

\begin{table}[t]
\begin{center}
\begin{tabular}{|c|c|c|c|c|}
\hline
$q^2 ({\rm GeV}^2/c^2)$ & \multicolumn{1}{c}{} & \multicolumn{1}{c}{} {$A_{FB}$}&
\multicolumn{1}{c}{} &   \\
\cline{2-5}
  & exp & SM & $m_t'=400\, {\rm GeV}$ & $m_t'=600\, {\rm GeV}$ \\
\hline
$0.6 - 1.0$  & $0.47^{+0.26}_{-0.33}$ & $(-0.18 \to -0.19)$ &
$(-0.13 \to -0.19)$ & $(-0.08 \to -0.19 )$ \\
\hline
$1.0 - 6.0$  & $0.26^{+0.28}_{-0.31}$ & $(-0.2 \to 0.2)$ &
$(-0.2 \to 0.2)$ & $(-0.2 \to 0.2 )$ \\
\hline
$6.0 - 8.0$ & $0.45^{+0.21}_{-0.26}$ & $(0.19 \to 0.30)$ &
$(0.17 \to 0.28)$ & $(0.11 \to 0.30 )$ \\
\hline
$16.5 - 18.0$ & $0.66^{+0.12}_{-0.16}$ & $(0.28 \to 0.49)$
 & $(0.25\to 0.45)$ & $0.15 \to 0.47$     \\
\cline{3-5}
$18.0 - 19.5$ & For $ (q^2 > 16)$ & $(0.003 \to 0.30)$ &
$(0.003\to 0.27)$ & $0.003 \to 0.28$     \\
\hline
\end{tabular}
\caption{Values of FB-asymmetry in different $q^2$ region.}
\label{tabfb}
\end{center}
\end{table}

From the left panel of Fig. \ref{fb_excld}, it is clear that within the 
uncertainty, the zero of the FB asymmetry in the SM4 
is consistent with the SM prediction.


In Table \ref{tabfb} we have made a comparative study between SM, SM4 and 
experimental ranges for $A_{FB}(q^2)$ in different $q^2$ region and one could 
see that the SM and SM4 predictions are within the present experimental bound. 
One interesting feature of data is that for low $q^2$ (first two bins), the 
central value (with appreciable errors) of $A_{FB}$ is positive whereas SM 
predicts negative $A_{FB}$ for these bins. Note also that there are deviations between 
SM and SM4 predicted FB-asymmetries in some regions of $q^2$, for example 
$q^2$ (${ \rm GeV}^2$) with values in between $(0.6\to 1.0)$, $(6.0 \to 8.0)$ 
and $(16.5 \to 18.0)$ the lower limit of SM4 predicted values are lower in 
magnitude than that for SM predictions; these differences are more prominent for 
$m_{t'} = 600$ GeV (see Table. \ref{tabfb}).

\subsection {$B_s \to l^+ l^-$ decay}
The purely leptonic decays   $B_s \to l^+ l^-$, where $l=e,\,\mu,\,\tau$, are chirally suppressed within the SM and hence
have appreciably smaller branching ratios as compared to that of the semi-
leptonic decays. The helicity suppression is more dominant in the case of 
$B_s \to e^+ e^-$ and $B_s \to \mu^+ \mu^-$ which have branching ratio of 
$\sim\,(7.7\pm 0.74)\times 10^{-14}$ and $\sim\,(3.35 \pm 0.32)\times 10^{-9}$ respectively \cite{buras03}, within the SM. 
However the suppression is evaded to some extent in the case of 
$B_s \to \tau^+ \tau^-$ due to the large $m_{\tau}$, which has a branching ratio of $\sim 10^{-7}$. These decays are yet to be observed experimentally. The present upper bound on $B_s \to e^+ e^-$ and $B_s \to \mu^+ \mu^-$ are \cite{Barberio:2008fa}
\bea
Br(B_s \to e^+ e^-)  &<& 0.28 \times 10^{-6}\;,
\nonumber \\
Br(B_s \to \mu^+ \mu^-)  &<& 3.60 \times 10^{-8}\;.
\eea
As far as the $\tau$ channel is concerned, the current experimental information is rather poor. Using the LEP data on $B \to \tau \nu$ decays, the indirect bound on $Br(B_s \to \tau^+ \tau^-)$ is obtained to be \cite{Grossman:1996qj}
\beq
Br(B_s \to \tau^+ \tau^-)  <  5\% \;.
\eeq

Though the decay $B_s \to \tau^+ \tau^-$ has relatively larger branching ratio 
compared to $B_s \to e^+ e^-$ and $B_s \to \mu^+ \mu^-$, its 
observation will also be extremely difficult as the reconstruction of $\tau$ 
is a very challenging task. However, the upcoming experiments at the LHC can 
reach the SM sensitivity of $B_s \to \mu^+ \mu^-$ and hence it can serve as an 
important probe to test the SM and constrain many new physics models. The LHCb 
will be able to probe the SM predictions for $B_s \to \mu^+ \mu^-$ at 
$3 \sigma$ with $2\,fb^{-1}$ of data \cite{Lenzi:2007nq} whereas the ATLAS and 
CMS will be able to reconstruct the $B_s \to \mu^+ \mu^-$ signal at 
$3 \sigma$ with $30\,fb^{-1}$ of data collection \cite{Smizanska:2008qm}. 

Here we study the decay $B_s \to \mu^+ \mu^-$ and $B_s \to \tau^+ \tau^-$ in 
the context of SM4. Within the SM4, the branching ratio of $B_s \to l^+ l^-$ 
is given by
\begin{equation}
  Br(B_s \to l^+ l^-)  = \frac{G^2_F \alpha^2 m_{B_s }m_l^2 f_{B_s}^2 \tau_{B_s}}{16 \pi^3}
    |V_{tb}^{}V_{ts}^{\ast}|^2 \sqrt{1 - \frac{4 m_l^2}{m_{B_s}^2}}\, \Big| C_{10}^{\rm tot}\Big|^2\;.
\end{equation}

The branching ratio of $B_s \to l^+ l^-$ can be predicted with higher accuracy 
by correlating it with the $B_{s}-\bar B_{s}$ mixing and then considerable 
uncertainty due to mixing angle and $f_{B_s}$ gets removed. We have
\beq
 Br(B_s \to l^+ l^-)  = \frac{3\alpha^2\tau_{B_s}m_l^2}{8\pi B_{bs}m_W^2}\, \sqrt{1 - \frac{4 m_l^2}{m_{B_s}^2}}\,
\frac{\Big| C_{10}^{\rm tot}\Big|^2}{|\Delta'|}\Delta M_s\;,
\label{bsmu1}
\eeq
where $B_{bs}$ is the ``Bag-parameter" for $B_s$ mesons for which lattice 
result is given by \cite{latticebag},
\beq
B_{bs} = 1.33 \pm 0.06, 
\eeq
however, in order to be conservative we use the value $1.33 \pm 0.15$ .
In eq. \ref{bsmu1} the parameter $\Delta'$ is defined as,
\beq
\Delta' = \Big[\eta_t S_0(x_t) + \eta_{t'} \frac{\left( V_{t' s} V_{t' b}^{*}\right)^2}{\left( V_{ts} V_{tb}^{*}  \right)^2}S_0(x_{t'}) +  2\eta_{tt'} \frac{\left( V_{t' s} V_{t' b}^{*}\right)}{\left( V_{ts} V_{tb}^{*}  \right)}S_0(x_t,x_{t'})\Big]\;.
\label{del2}
\eeq

\begin{figure}[t]
\includegraphics[width=8cm,height=6cm,clip]{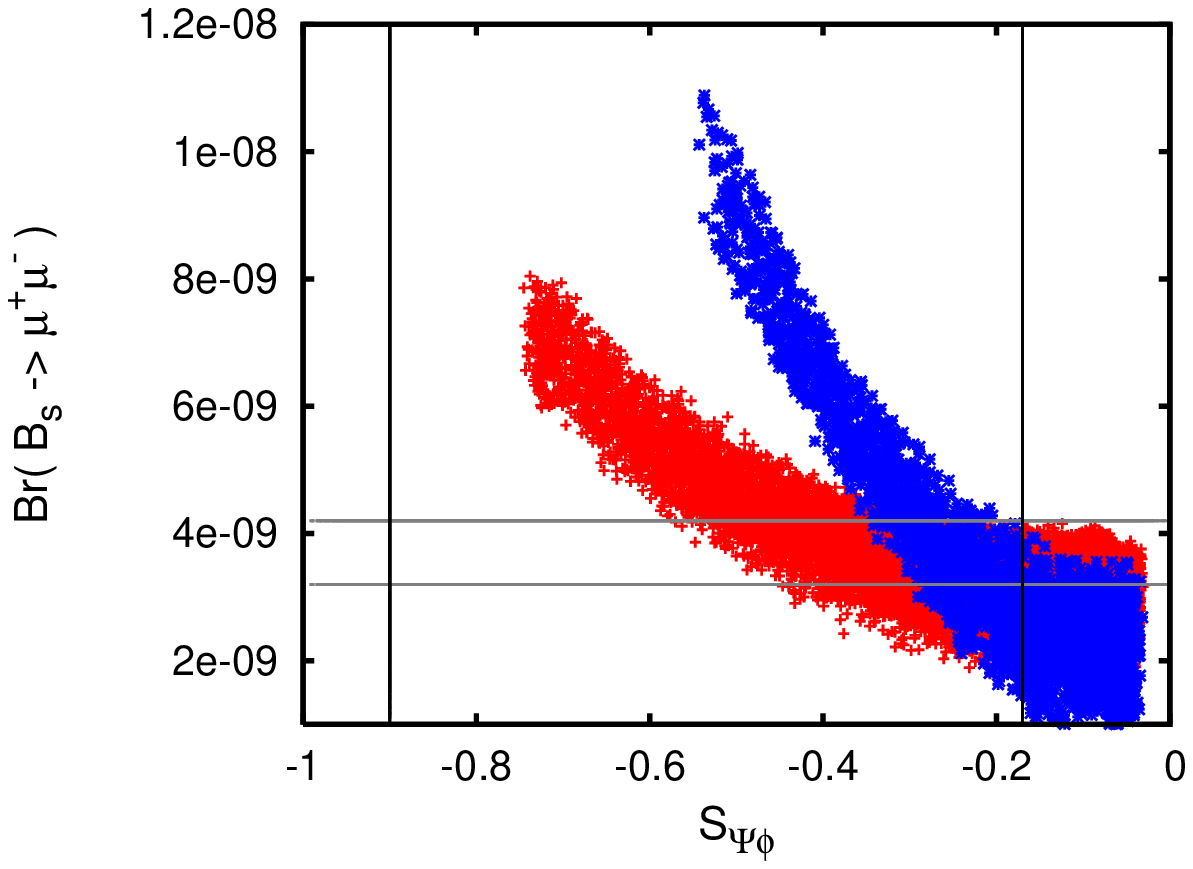}%
\hspace{0.2cm}%
\includegraphics[width=8cm,height=6cm,clip]{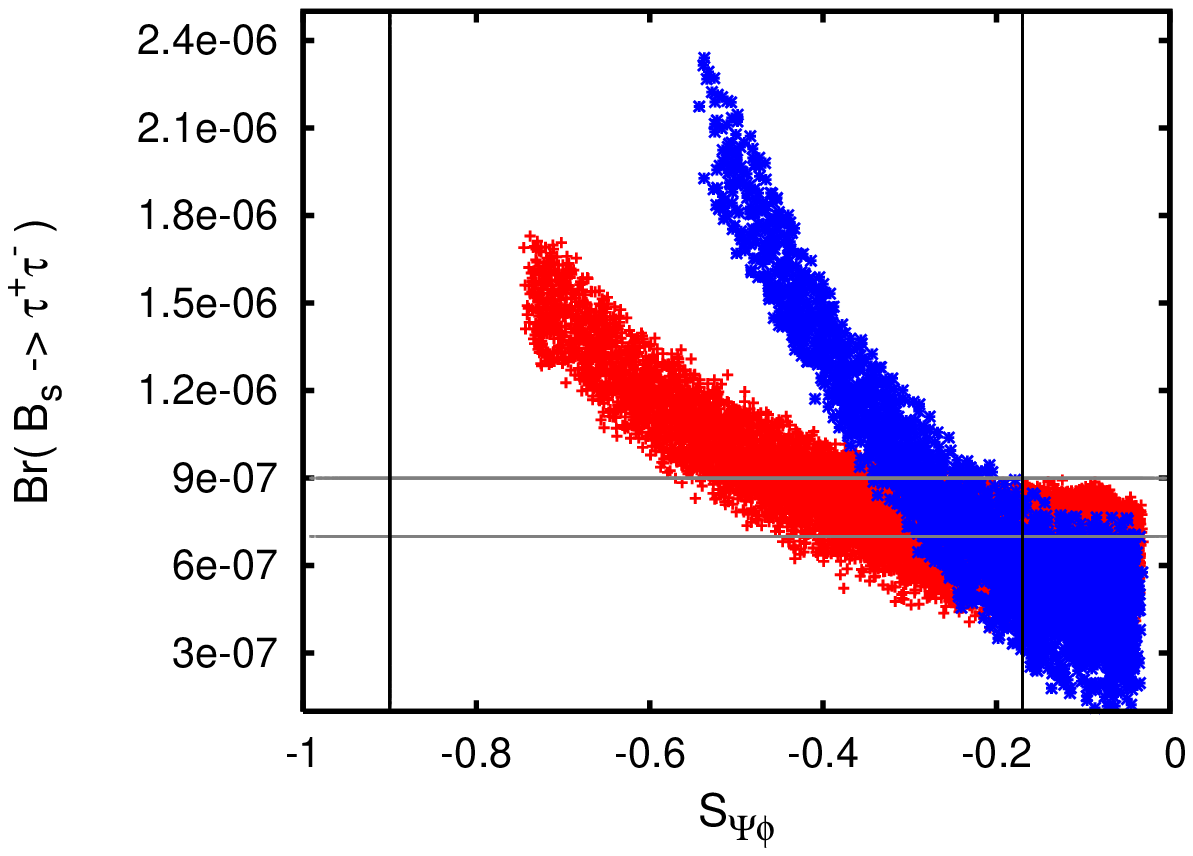}
\caption{Correlation between branching fraction in $B_s\to \mu^+\mu^- $ 
(left panel) and $B_s\to \tau^+\tau^- $ (right panel) with $S_{\psi\phi}$, 
where the red and blue regions correspond to $m_{t'}$ = 400 and 600 GeV 
respectively, the horizontal lines represent the SM limit for 
$Br(B_s \to \ell^+\ell^-)$ whereas the vertical lines represent the 
$2\sigma$ experimental range for $S_{\psi\phi}$.}
\label{fig_brbsmu}
\end{figure}

In fig. \ref{fig_brbsmu} we have shown the correlation between the branching 
fraction $Br(B_s\to \ell^+\ell^-)$ and {\it CP} asymmetry in $B_s\to \psi\phi$,
it is clear that there are possibilities for appreciably different predictions 
in SM4 compared to SM, enhanced or diminished by a factor of ${\cal{O}}(3)$.
Note also that enhanced branching fractions correspond to a large {\it CP} 
asymmetry in $B_s\to \psi\phi$ and smaller branching fractions correspond to
smaller asymmetry. The corresponding upper limit on the branching fractions 
are given by, 
\begin{align}
Br(B_s\to \mu^+\mu^-) & <\, 8.0\times 10^{-9} \hskip 20pt m_{t'} = 400\, 
{\it GeV}, \nonumber \\
& < 1.2\, \times 10^{-8}, \hskip 20pt m_{t'} = 600\,
{\it GeV},\nonumber \\
Br(B_s\to \tau^+\tau^-) & <\, 1.8\times 10^{-6} \hskip 20pt m_{t'} = 400\,
{\it GeV}, \nonumber \\
& < 2.4\, \times 10^{-6}, \hskip 20pt m_{t'} = 600\,
{\it GeV}.
\end{align} 
However, when $S_{\psi\phi}$ is close to its SM value i.e when the {\it CP} 
violating phase, $\phi^s_{t'}$, of $V_{t's}$ is close to zero, the branching 
fractions reduce from their SM value since $|C_{10}^{\rm tot}|$ and $\delta'$ in 
eq. \ref{del2} are reduced from its SM value due to destructive interference with 
SM4 counterpart. 

\subsection{Branching fraction $B\to X_s \nu \bar{\nu}$}

The decays $B\to X_s \nu \bar{\nu}$ are the theoretically cleanest decays in
the field of rare $B$-decays. They are dominated by the same $Z^0$-penguin and
box diagrams involving top quark exchanges which we encounter in the
case of $K_L\to \pi^0 \nu \bar{\nu}$ ,
since the change of the external quark flavors has no impact on the
$m_{t/t'}$ dependence, the later is fully described by the function
$X(x_{t/t'})$ which includes the NLO corrections. The charm contribution is 
negligible here. The effective Hamiltonian for the decay 
$B\to X_s\nu\bar{\nu}$ is given by
\beq
{\cal{H}}_{eff} = {G_F \over \sqrt{2}}{\alpha\over {2\pi\sin^2{\Theta_w}}}\left( V^{\ast}_{tb}V_{ts} X(x_t) + V^{\ast}_{t's}V_{t'd} X(x_{t'}) \right)(\bar{b}s)_{V-A}(\bar{\nu}\nu)_{V-A} + h.c.
\eeq
with
\bea
X(x)={x\over8}\Big[{{2+x}\over{x-1}}+{{3 x-6}\over(x-1)^2}\ln x\Big]
\eea
\begin{figure}
\includegraphics[width= 0.60 \linewidth]{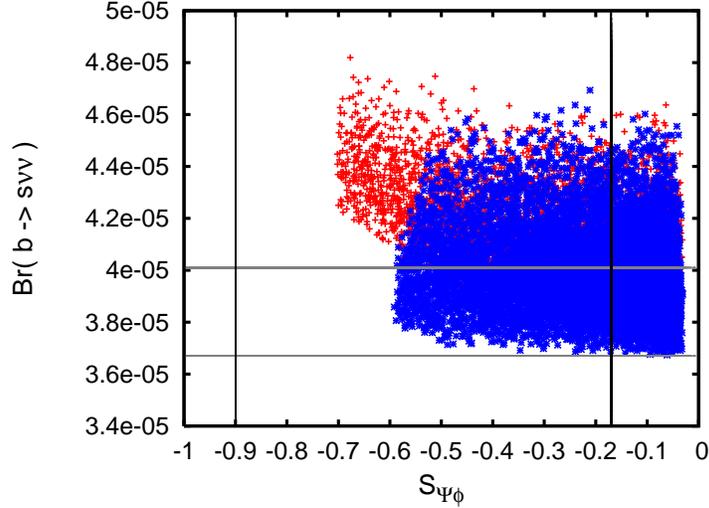}
\caption{Correlation between branching fraction in $B\to X_s \nu \bar{\nu}$
and $S_{\psi\phi}$, where the red and blue regions correspond to $m_{t'}$ = 400
 and 600 GeV respectively, the horizontal lines represent the SM limit 
for $Br(B\to X_s \nu \bar{\nu})$ whereas the vertical lines represent the 
$2\sigma$ experimental range for $S_{\psi\phi}$.}
\label{fig_brbsnu}
\end{figure}

The calculation of the branching fractions for $B\to X_s \nu \bar{\nu}$ can be
done in the spectator model corrected for short distance QCD effects.
Normalizing it to $Br\big(B\to X_c \nu \bar{\nu}\big)$ and summing over
three neutrino flavors one finds \cite{Buras:1997fb,grossman}
\bea
Br\big(B\to X_s \nu \bar{\nu}\big)\over {Br\big(B\to X_c e 
\bar{\nu}\big)} &=& {3 \alpha^2 \over 4\pi^2 \sin^4\Theta_W} {\bar{\eta}\over 
f(z)\kappa(z)}{1\over |V_{cb}|^2} \Big|\lambda_t X(x_t) + 
\lambda_{t'} X(x_{t'})\Big|^2 \nonumber \\
{} &=& {{\tilde{C}^2 \bar{\eta}}\over{|V_{cb}|^2 f(z) \kappa(z)}},
\label{brbsnu1}
\eea
where 
\beq
{\tilde{C}}^2=({\tilde{C}}^{SM})^2\Big|1+ {V^{\ast}_{t'b}V_{t's}\over 
V^{\ast}_{tb}V_{ts}}{X_0(x_{t'})\over X_0(x_{t})}\Big|^2,
\label{brbsnu2}
\eeq
with 
\beq
({\tilde{C}}^{SM})^2={{\alpha}^2\over{2 \pi^2\sin^4\Theta_W}}
\Big|V^{\ast}_{tb}V_{ts}X_0(x_{t})\Big|^2.
\label{brbsnu3}
\eeq
The factor $\bar\eta$ represents the QCD correction to the matrix element of
the $b\to s\nu{\bar{\nu}}$ transition due to virtual and bremsstrahlung
contributions and is given by the well known expression
\beq
{\bar\eta} = \kappa(0)= 1 + {2\alpha_s(m_b) \over 3\pi}\Big({25\over 4} - \pi^2\Big) \approx 0.83.
\eeq

The SM4 predicted branching fraction $Br(B\to X_s \nu \bar{\nu})$ could be 
sufficiently larger than its SM limit, $(3.66 \to 4.01)\times 10^{-5}$ 
\cite{Buras:1997fb} within the uncertainties, for values of $S_{\psi\phi}$ 
sufficiently away from its SM predictions. We are constraining 
$\lambda^{s}_t = V_{tb} V^{\ast}_{ts}$ using CKM4 unitarity with $\lambda^{s}_{t'} = 
V_{t'b} V^{\ast}_{t's}$ as free parameter, with the change
of phase and amplitude of $\lambda^{s}_{t'}$, $|\lambda^s_t|$ 
increases from its SM value resulting an overall enhancement of $Br(B\to X_s \nu \bar{\nu})$ from
its SM prediction. 
For values of $\phi^s_{t'}$ close to $80^{\circ}$, the terms within modulus in 
eq. \ref{brbsnu2} and eq. \ref{brbsnu3} have their maximum values and so the 
branching fraction is sufficiently larger than its SM prediction and 
reach its maximum value $4.8\times 10^{-5}$. In passing, we note incidently that the upper 
limit that we have obtained for SM4 is consistent with that obtained in Ref. \cite{burasewerth}, 
in models with minimal flavor violation (MFV), and with the present 
experimental bound $6.4\times 10^{-4}$ \cite{expbsnu}.

\subsection{Branching fraction $K^+ \to \pi^+ \nu \bar{\nu}$}

\begin{figure}
\includegraphics[width=8cm,height=6cm,clip]{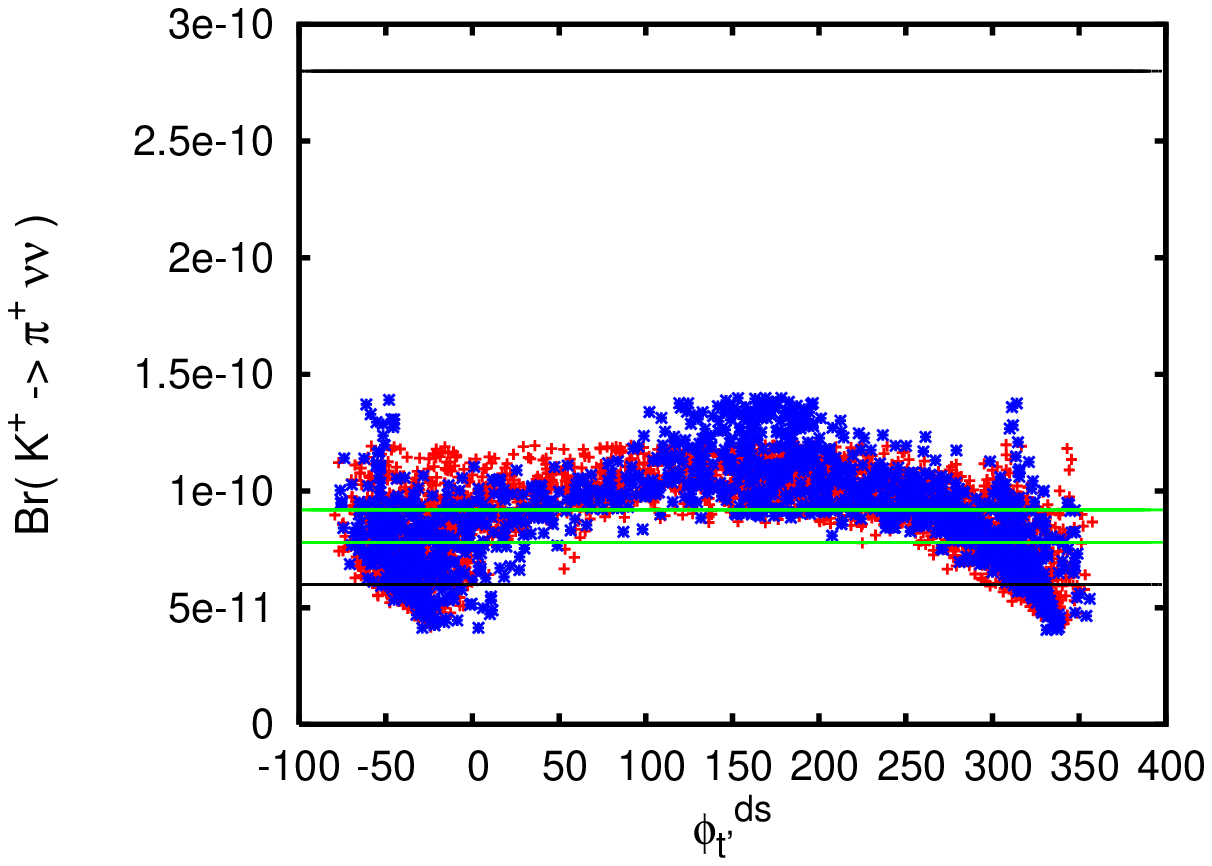}
\hspace{0.2cm}%
\includegraphics[width=8cm,height=6cm,clip]{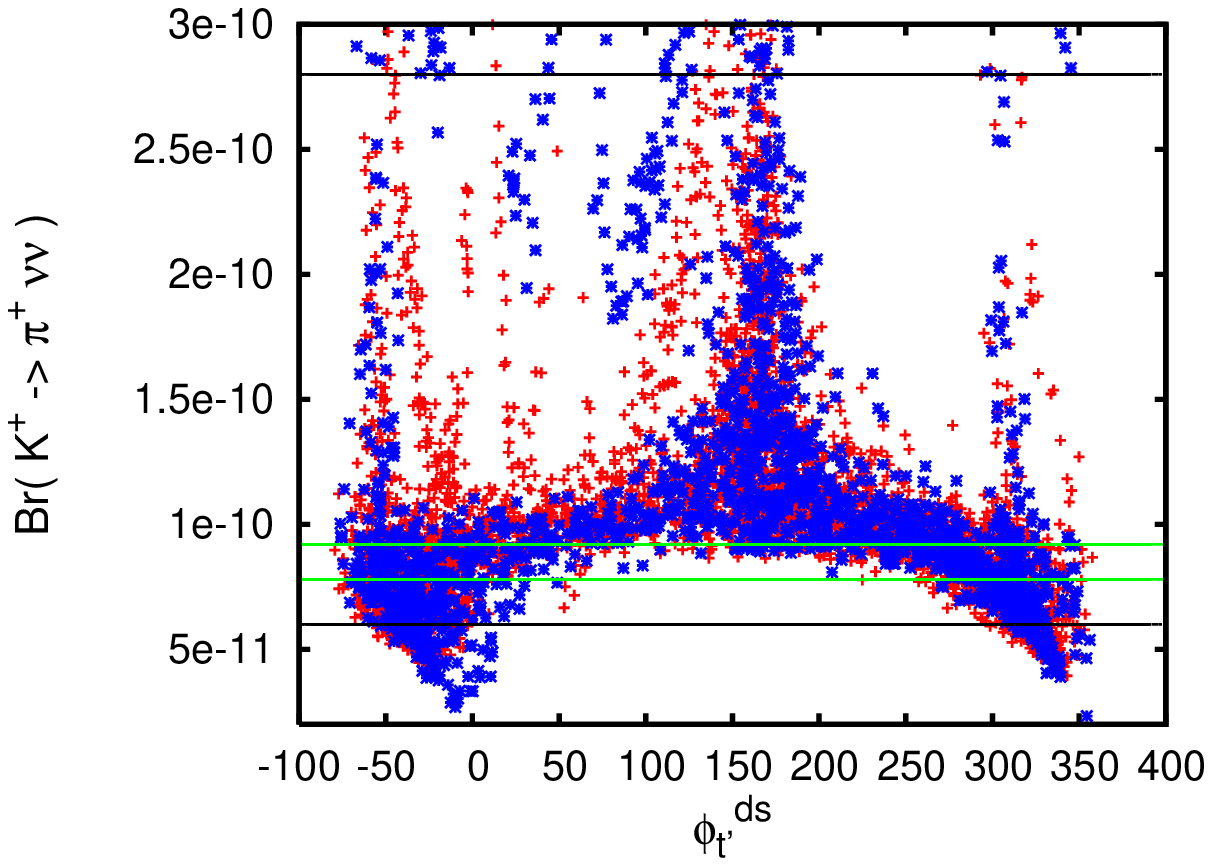}
\caption{Plot between the branching fraction of $K^+ \to \pi^+ \nu \bar{\nu}$ 
with $\phi^{ds}_{t'} = \phi^{d}_{t'}-\phi^{s}_{t'}$ bounded by the present experimental limit,
 red and blue region corresponds to $m_{t'}$ = 400 and 600 $\rm GeV$ 
respectively, the green and black horizontal lines represent $1\sigma$ limit 
for SM and experimental value respectively. Left panel shows only 1\,$\sigma$ range expected 
in SM4; full range is shown in the right panel.}
\label{fig_ktopip}
\end{figure}

Although we have taken branching fraction for  $K^+ \to \pi^+ \nu \bar{\nu}$ as
a constrain to fit $V_{CKM4}$, in Fig. (\ref{fig_ktopip}) we show the effect of SM4;
 note that in the left panel only the $1\sigma$ range 
for the branching fraction using the constraints given in the Table. \ref{tab3} 
(except $Br(K^+ \to \pi^+ \nu \bar{\nu})$) is shown\footnote{Right panel is added in our version
 2 to facilitate direct comparision with \cite{buras4th}.}. 

From Fig. \ref{fig_ktopip} one could see that the 
$Br(K^+ \to \pi^+ \nu \bar{\nu})$ could be enhanced to its present experimental upper limit. 
In order to understand the nature of the plot one needs to concentrate on eq. (\ref{brkpip}), 
and it is important to note that  
$Br(K^+ \to \pi^+ \nu \bar{\nu})$ is dominated by the second term of 
the expression i.e the term proportional to $Re(\lambda_q)$ it should also be 
noted that the SM and SM4 part for each term has a relative sign difference. 
When $\phi^{ds}_{t'}$ is negative (i.e when $\phi^{d}_{t'}$ has values in 
between $(0-80)^{\circ}$) and $\phi^{ds}_{t'} > 270^{\circ}$ the 
branching fraction will decrease because of the destructive interference 
between SM and SM4 part in the second term of eq. (\ref{brkpip}). For 
$\phi^{ds}_{t'}$ in between $(90 - 180)^{\circ}$ the branching fraction have 
values above the SM value it is due to constructive interference between SM 
and SM4 in the second term of eq. (\ref{brkpip}). 

Present NNLO predictions for branching fraction for 
$K^+ \to \pi^+ \nu \bar{\nu}$ within SM is given by \cite{uliburas}
\begin{equation}
Br(K^+ \to \pi^+ \nu \bar{\nu}) = (8.5 \pm 0.7)\times 10^{-11},
\end{equation}
and the SM4 $1\sigma$ limit on $Br(K^+\to \pi^+ \nu \bar{\nu})$ is given by
\bea
Br(K^+\to \pi^+ \nu \bar{\nu}) = (4.0 \to 12.0)\times 10^{-11}; \hskip 20pt  
m_{t'}= 400 \hskip 5pt {\rm GeV}, \nonumber \\
Br(K^+\to \pi^+ \nu \bar{\nu}) = (4.0 \to 13.0)\times 10^{-11}; \hskip 20pt  
m_{t'}= 600 \hskip 5pt {\rm GeV}.
\eea
Again these upper limits are consistent with the 95\% confidence level limit obtained 
in Ref. \cite{burasewerth} calculated in MFV model.

\subsection{Branching fraction $K_L\to \pi^0\nu \bar{\nu}$}
\begin{figure}
\includegraphics[width=8cm,height=6cm,clip]{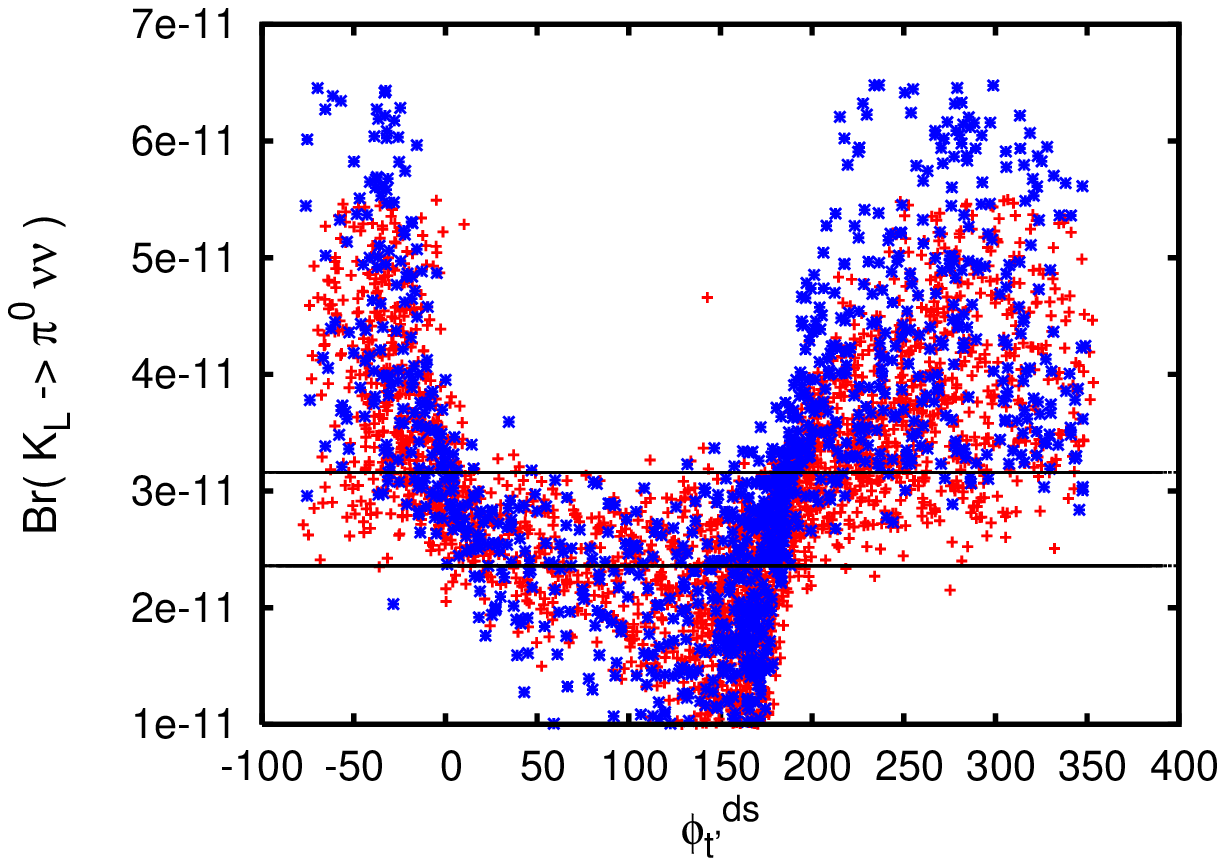}%
\hspace{0.2cm}%
\includegraphics[width=8cm,height=6cm,clip]{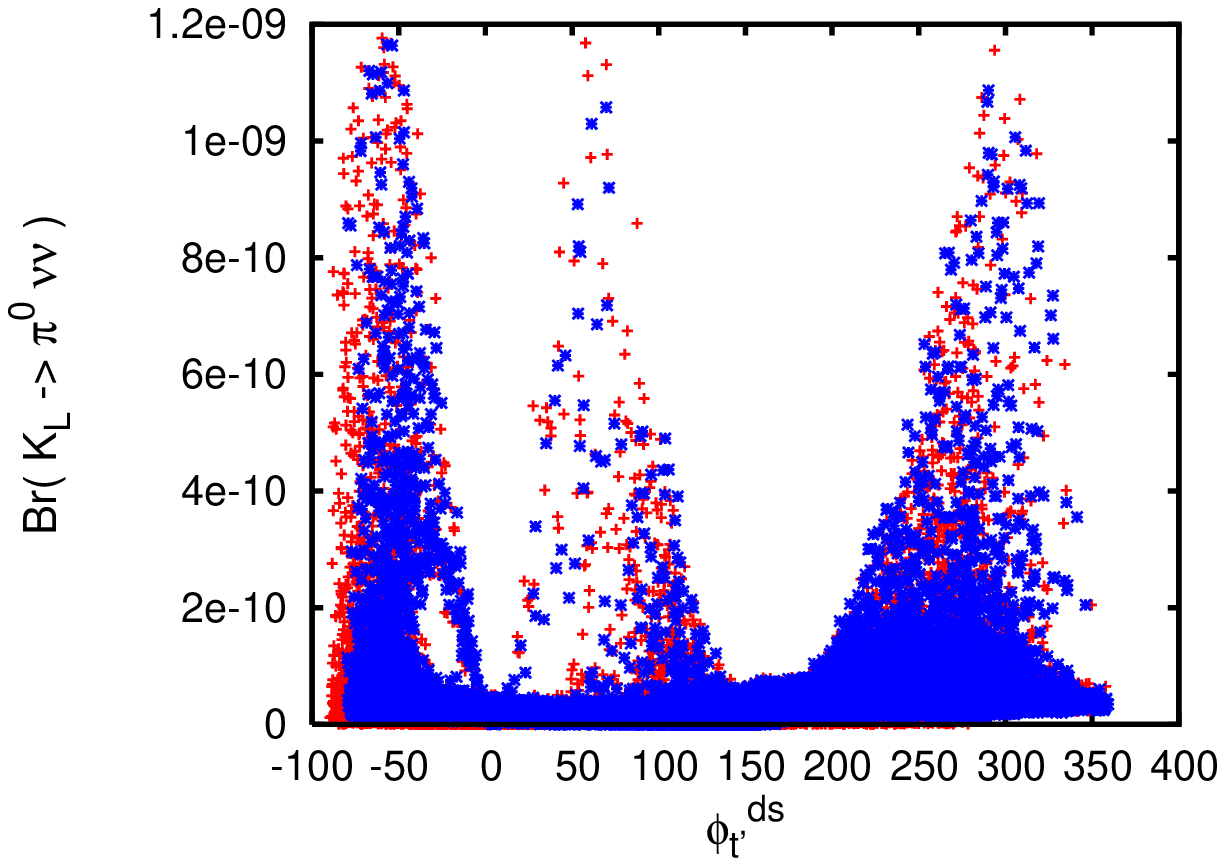}
\caption{The branching fraction of $K_L \to \pi^0 \nu \bar{\nu}$ versus 
 $\phi^{ds}_{t'} = \phi^{d}_{t'}-\phi^{s}_{t'}$ in SM4,  
red and blue region corresponds to $m_{t'}$ = 400 and 600 $\rm GeV$ 
respectively, the black horizontal lines represent $1\sigma$ SM limit; left panel 
shows only $1\sigma$ range expected in SM4, full range for SM4 is shown in the right panel.}
\label{fig_kltopi}
\end{figure}

The effective Hamiltonian for $K_L\to \pi^0 \nu \bar{\nu}$ can be written as
\beq
{\cal{H}}_{eff} = {G_F \over \sqrt{2}}{\alpha\over {2\pi\sin^2{\Theta_w}}}\left( V^{\ast}_{ts}V_{td} X(x_t) + V^{\ast}_{t's}V_{t'd} X(x_{t'}) \right)(\bar{s}d)_{V-A}(\bar{\nu}\nu)_{V-A} + h.c.
\eeq

Within SM $K_L\to \pi^0\nu \bar{\nu}$ decay, proceeds almost entirely through
{\it CP} violation, is completely dominated by short-distance loop diagrams with top quark exchanges, here the charm contribution can be fully neglected.

The branching fraction of $K_L\to \pi^0\nu\bar{\nu}$ can be written as follows
\beq
Br(K_L\to \pi^0\nu\bar{\nu}) = \kappa_L .\left[\left({Im\lambda_t \over \lambda^5} X(x_t) + {Im\lambda_{t'} \over \lambda^5} X(x_{t'}) \right)^2 \right]
,
\label{brkpi0}
\eeq
with
\beq
\kappa_L = {r_{K_L}\over r_{K^+}}{\tau(K_L)\over\tau(K^+)}\kappa_+ = 1.80 \times 10^{-10},
\eeq
$\kappa_+$ and $r_{K_L}=0.944$ summarizing isospin breaking corrections in
relating $K_L\to \pi^0\nu\bar{\nu}$ to $K^+\to \pi^0 e^+ \nu$. The current
value of branching fraction for  $K_L\to \pi^0\nu\bar{\nu}$ with SM is given 
by \cite{uliburas}
\begin{equation}
Br(K_L \to \pi^{0} \nu \bar{\nu}) = (2.76 \pm 0.40)\times 10^{-11}.
\end{equation}

In Fig. (\ref{fig_kltopi}) the variation of branching fraction 
$Br(K_L\to \pi^0\nu\bar{\nu})$ with the phase $\phi^{ds}_{t'}$ is shown\footnote{Right panel is 
added in our revised version to facilitate direct comparision with \cite{buras4th}.}.
We note that with the constraint on  $Br(K^+ \to \pi^+ \nu \bar{\nu})$ (Table. \ref{tab3}), 
while, in principle $Br(K_L\to \pi^0\nu\bar{\nu})$ could be enhanced as much as 
$1.2\times 10^{-9}$ (right panel Fig. \ref{fig_kltopi}), the expected $1\, \sigma$ range in SM4 
(left panel Fig. \ref{fig_kltopi}) is only to $7\times 10^{-11}$, however, at 95\% CL the value
could be enhanced to $8\times 10^{-10}$. The branching fraction has its 
maximum value when the phase $\phi^{ds}_{t'}$ has the value $\pm 90^{\circ}$ and 
$270^{\circ}$ since SM4 contribution  picks up its maximum value at those points 
(eq. \ref{brkpi0}).

The SM4 $1\sigma$ limit on $Br(K_L\to \pi^0\nu\bar{\nu})$ is given by
\bea
Br(K_L\to \pi^0\nu\bar{\nu}) = (1.0 \to 5.2)\times 10^{-11};  \hskip 20pt  
m_{t'}= 400  \hskip 5pt {\rm GeV}, \nonumber \\
Br(K_L\to \pi^0\nu\bar{\nu}) = (1.0 \to 6.2)\times 10^{-11}; \hskip 20pt  
m_{t'}= 600 \hskip 5pt {\rm GeV},
\eea
 the upper limits are consistent with the limit calculated in 
Ref. \cite{burasewerth}.

\subsection{CP violation in  $B \to \pi K$ modes}
The observed data from the currently running two asymmetric $B$ 
factories are almost consistent with the SM
predictions and till now there is no compelling evidence for new physics.
However there are some interesting deviations from the SM
associated with the $b \to s$ transitions, which
provide us with possible indication of new physics.
For example the mixing induced CP asymmetries in many
$b \to s \bar q q$ penguin dominated modes do not seem to agree with
the SM expectations. The measured values in such modes
follow the trend $S_{s
\bar q q} < \sin 2 \beta $ \cite{Barberio:2008fa,LS09}, whereas in the SM they  are
expected to be similar \cite{Grossman:1996ke,London:1997zk}. In this context $B \to  \pi K$
decay modes, which receive dominant contributions from $b \to s$
mediated QCD penguins in the SM, provide another testing
ground to look for new physics.

The first one is the difference in direct CP asymmetries in $B^- \to
\pi^0 K^-$ and $\bar B^0 \to \pi^+ K^-$ modes. These two modes
receive similar dominating contributions from tree and penguin
diagrams and hence one would naively expect that these two channels
will have the same direct CP asymmetries i.e., ${\cal A}_{\pi^0
K^-}= {\cal A}_{\pi^+ K^-}$. In the QCD factorization approach, the
difference between these asymmetries  is found to be \cite{LS07}
 \be \Delta A_{CP} ={\cal A}_{ K^- \pi^0} -{\cal A}_
{ K^- \pi^+} =(2.5 \pm
 1.5)\%
 \ee
 whereas the corresponding experimental value \cite{Barberio:2008fa} is
\be \Delta A_{CP} =(14.8 \pm
 2.8)\%\;,
 \ee
which yields nearly $4 \sigma$ deviation.

The second anomaly is associated with the mixing induced
CP asymmetry in $B^0 \to \pi^0 K^0$ mode.
 The time dependent CP asymmetry in this mode is defined
as \bea \frac{\Gamma(\bar B^0(t) \to \pi^0 K_s)-\Gamma(B^0(t) \to
\pi^0 K_s)}{\Gamma(\bar B^0(t) \to \pi^0 K_s)+\Gamma(B^0(t) \to
\pi^0K_s)}= A_{\pi^0 K_s} \cos(\Delta M_d t)+S_{\pi^0 K_s}
\sin(\Delta M_d t)\;, \eea and in the pure QCD penguin limit one
expects $A_{\pi^0 K_s} \approx 0$ and $S_{\pi^0 K_s} \approx \sin(2
\beta)$. Small non-penguin contributions do provide some corrections
to these asymmetry parameters and it has been shown in Ref.
\cite{Beneke:2005pu,Cheng:2006dk,Buchalla:2005us} that these corrections
generally tend to increase $S_{K
\pi^0}$ from its pure penguin limit of ($\sin 2 \beta$) by a
modest  amount i.e., $ S_{\pi^0 K_s} \approx 0.8$.
Recently, using isospin symmetry it has been shown in \cite{rf, mg1,beak}
that the standard model favors a large $S_{\pi^0 K_s} \approx 0.99$.

However, the  recent results from Belle
\cite{belle} and Babar \cite{babar}  are \bea A_{\pi^0 K_s}&=&
0.14 \pm 0.13 \pm
0.06,~~~~~S_{\pi^0 K_s}=0.67 \pm 0.31 \pm 0.08~~~~({\rm
Belle})\nn\\
A_{\pi^0 K_s}&=& -0.13 \pm 0.13 \pm 0.03,~~~~~S_{\pi^0 K_s}=0.55 \pm
0.20 \pm 0.03~~~~({\rm Babar}) \eea with average \be A_{\pi^0 K_s}=
-0.01 \pm 0.10,~~~~~S_{\pi^0 K_s}=0.57 \pm 0.17\;.\label{av} \ee
As seen from (\ref{av}), the observed value of
$S_{\pi^0 K_s}$ is found to be
smaller than the present  world average value of $\sin 2 \beta=0.672
\pm 0.024 $ measured in $b \to c \bar c s$ transitions
\cite{Barberio:2008fa} by nearly $1\sigma$ and the deviation from the 
SM expectation given above is possibly even larger.
This deviation which is  opposite to the SM expectation, implies the
possible presence of new physics in the $ B^0 \to K^0 \pi^0 $ decay
amplitude. In the SM, this decay mode receives contributions from
QCD penguin ($P$), electroweak penguin ($P_{EW})$ and color
suppressed tree ($C$) diagrams, which follow the hierarchical
pattern $P:P_{EW}:C= 1:\lambda : \lambda^2$, where $\lambda \approx
0.2257$ is the Wolfenstein expansion parameter. Thus, accepting the
above discrepancy seriously one can see that the electroweak penguin
sector is the best place to search for new physics.

To account for these discrepancies here we consider the effect of 
sequential fourth generation quarks \cite{Hou:2006zza,Hou:2005hd,Arhrib:2006pm,Hou:2006jy,AS_olds1}. 
In the SM, the relevant
effective Hamiltonian describing the decay modes $B \to \pi K$
  is given by
\begin{equation}
{\cal H}_{eff}^{SM} = \frac{G_F}{\sqrt{2}}\left[  V_{ub}
V_{us}^*(C_1O_1+C_2 O_2)- V_{tb}V_{ts}^*\sum_{i=3}^{10} C_i O_i
\right].
\end{equation}
With a sequential fourth generation, the Wilson coefficients $C_i$'s
will be modified due to the new contributions from $t^\prime$ quark
in the loop. Furthermore, due to the presence of the $t'$ quark the
unitarity condition becomes $\lambda_u+\lambda_c+\lambda_t+
\lambda_{t'}=0$, where $\lambda_q= V_{qb}V_{qs}^*$.

Thus, including the fourth generation and replacing $\lambda_t
=-(\lambda_u+\lambda_c +\lambda_{t'})$, the modified Hamiltonian
becomes \bea {\cal H}_{eff}& = &\frac{G_F}{\sqrt{2}}\left[
\lambda_u(C_1O_1+C_2 O_2)- \lambda_t \sum_{i=3}^{10} C_i O_i
-\lambda_{t'}\sum_{i=3}^{10} C_i^{t'} O_i\right]\nn\\
&=&\frac{G_F}{\sqrt{2}}\left[  \lambda_u(C_1O_1+C_2 O_2
+\sum_{i=3}^{10} C_i O_i)+
\lambda_c\sum_{i=3}^{10} C_i O_i -\lambda_{t'}\sum_{i=3}^{10} \Delta
C_i O_i\right]\;, \eea where $\Delta C_i$'s  are the effective (t
subtracted) $t'$ contributions.

Thus, one can obtain the transition amplitudes in the QCD
factorization approach as \cite{QCDF1,QCDF}
 \bea \sqrt2 A(B^- \to
\pi^0 K^-)&=&\lambda_u \Big(A_{\pi \bar K}(\alpha_1+\beta_2)+A_{\bar
K \pi}\alpha_2 \Big)\nn\\
&+&\sum_{p=u,c}\lambda_p\Big(A_{\pi \bar
K}(\alpha_4^p+\alpha_{4,EW}^p+\beta_3^p+\beta_{3,EW}^p)+\frac{3}{2}A_{\bar
K \pi}\alpha_{3,EW}^p \Big)\nn\\
&-& \lambda_{t'}\Big(A_{\pi \bar K}(\Delta \alpha_4+\Delta
\alpha_{4,EW}+\Delta \beta_3+\Delta \beta_{3,EW})+\frac{3}{2}A_{\bar
K \pi}\Delta\alpha_{3,EW} \Big),\nn\\ A(\bar B^0 \to  \pi^+
K^-)&=&\lambda_u \Big(A_{\pi \bar K}~\alpha_1 \Big) +
\sum_{p=u,c}\lambda_p A_{\pi
\bar
K}\Big(\alpha_4^p+\alpha_{4,EW}^p+\beta_3^p-\frac{1}{2}
\beta_{3,EW}^p \Big)\nn\\
&-& \lambda_{t'}A_{\pi \bar K}\Big(\Delta \alpha_4+\Delta
\alpha_{4,EW}+\Delta \beta_3-\frac{1}{2}\Delta \beta_{3,EW} \Big),\nn\\
\sqrt 2 A(\bar B^0 \to \pi^0
\bar K^0)&=&\lambda_u A_{\bar
K \pi}\alpha_2 +\sum_{p=u,c}\lambda_p\Big[A_{\pi \bar
K}\Big(-\alpha_4^p+\frac{1}{2}\alpha_{4,EW}^p-
\beta_3^p+\frac{1}{2}\beta_{3,EW}^p \Big)\nn\\
&+&\frac{3}{2}A_{\bar K \pi}\alpha_{3,EW}^p \Big]
-\lambda_{t'}\Big[A_{\pi \bar
K}\Big(-\Delta \alpha_4+\frac{1}{2}\Delta \alpha_{4,EW}-
\Delta \beta_3+\frac{1}{2}\Delta \beta_{3,EW} \Big)\nn\\
&+&\frac{3}{2}A_{\bar K \pi}\Delta \alpha_{3,EW} \Big],
\eea where \be  A_{\pi \bar K}= i\frac{G_F}{\sqrt 2}M_B^2 F_0^{B\to
\pi}f_K~~~~~{\rm and}~~~~A_{ \bar K \pi}= i\frac{G_F}{\sqrt 2}M_B^2
F_0^{B\to K}f_{\pi}\;. \ee
\begin{table}[t]
\begin{tabular}{|ccccccc|}
\hline $m_{t'}$ (in GeV) &&& 400  &&& 600 \\
\hline
$\Delta C_3(m_b) $ &&&  0.628  &&&1.471  \\
$\Delta C_4(m_b) $ &&& $-0.274$  &&& $-0.578$  \\
$\Delta C_5(m_b) $ &&&  0.042  &&& 0.086  \\
$\Delta C_6(m_b) $ &&&  $-0.206$  &&& $-0.362$   \\
$\Delta C_7(m_b) $ &&&  0.443  &&& 1.072   \\
$\Delta C_8(m_b) $ &&& $0.168$  &&& 0.407  \\
$\Delta C_9(m_b) $ &&& $-1.926$  &&& $-4.465$   \\
$\Delta C_{10}(m_b) $ &&&  0.433  &&& 1.005  \\
$\Delta C_{7 \gamma}^{eff}(m_b) $ &&&  $-5.667$  &&& $-7.239$  \\
$\Delta C_{8g}^{eff}(m_b) $ &&&  $-1.452$  &&& $-1.728$  \\
\hline
\end{tabular}
\caption{Values of the Wilson coefficients $\Delta C_i$'s
at different $b$-mass scale.}
\label{tab5}
\end{table}
These amplitudes can be symbolically represented as \bea Amp=
\lambda_u A_u +\lambda_c A_c -\lambda_{t'}A_{t'} .\eea $\lambda$'s
contain the weak phase information and $A_i$'s are associated with
the strong phases. Thus one can explicitly separate the strong and
weak phases and write the amplitudes as \bea Amp=\lambda_c
A_c\Big[1+ r a e^{i(\delta_1-\gamma)}-r' b e^{i(\delta_2+\phi_s)}],
\eea where $a=|\lambda_u/\lambda_c|$, $b=|\lambda_{t'}/\lambda_c|$,
$-\gamma$ is the weak phase of $V_{ub}$ and $\phi_s$ is the weak
phase of $\lambda_{t'}$. $r=|A_u/A_c|$, $r'=|A_{t'}/A_c|$, and
$\delta_1$ ($\delta_2$) is the relative strong phases between $A_u$
and $A_c$ ($A_{t'}$ and $A_c$). From these  amplitudes one
can obtain the direct and mixing induced CP asymmetry parameters as
\bea
A_{\pi K}&=&\frac{2\Big[ra \sin
\delta_1 \sin \gamma +r' b \sin \delta_2 \sin \phi_s + r r' a b
\sin(\delta_2-\delta_1) \sin
(\gamma+\phi_s)\Big]}{\Big[{\cal{R}}+2ra \cos \delta_1 \cos
\gamma-2r' b \cos \delta_2\cos \phi_s  - 2 r r' a b
 \cos(\delta_2-\delta_1)\cos(\gamma+\phi_s)\Big]}\;,\nn\\
S_{\pi K}&= &\frac{X}{{\cal R}+2ra \cos \delta_1 \cos \gamma-2r'
b  \cos \delta_2 \cos \phi_s- 2 r r' a b\cos(\delta_2-\delta_1)
 \cos(\gamma+\phi_s)
}\;,\eea where
${\cal{R}}=1+(ra)^2+(r'b)^2$ and
\bea X &=& \sin 2 \beta+ 2 r a \cos
\delta_1 \sin(2 \beta+\gamma)- 2 r' b \cos \delta_2
\sin(2 \beta- \phi_s) +(r a)^2\sin(2 \beta +2 \gamma) \nn\\
&+&(r' b)^2\sin(2 \beta -2 \phi_s)-2 rr' a b \cos(\delta_2-
\delta_1) \sin(2 \beta+\gamma -\phi_s). \eea

To find out the new contributions due to the fourth generation
effect, first we have to evaluate the new Wilson coefficients
$C_i^{t'}$.  The values of these coefficients at the $M_W$ scale can
be obtained from the corresponding contributions from the $t$ quark
by replacing the mass of $t$  quark in the Inami-Lim functions
\cite{lim}  by $t'$ mass. These values  can then  be evolved to the
$m_b$ scale using the renormalization group equation \cite{Buchalla:1995vs}
\begin{equation}
{\vec{C}} (m_b) = U_{5} (m_{b},M_{W}, \alpha) {\vec C} (M_{W})
\end{equation}
where $C$ is the $10 \times 1$ column vector of the Wilson
coefficients and $U_{5}$ is the five flavor $10 \times 10$ evolution
matrix. The explicit forms of ${\vec C}(M_W)$ and $U_{5} (m_b, M_W,
\alpha)$ are given in \cite{Buchalla:1995vs}. The values of $\Delta
C_{i=1-10} (m_b)$ in the NLO approximation and the coefficients of
the dipole operators $C_{7 \gamma}^{eff}$ and $C_{8g}^{eff}$ in the
LO  for different $m_{t'}$ values are presented in Table \ref{tab5}.

\begin{figure}[t]
\centerline{\epsfysize 2.5 truein \epsfbox{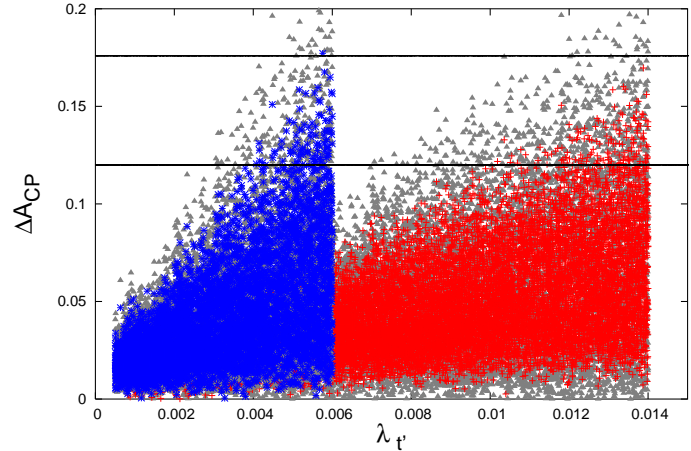}} \caption{The
allowed range of the CP asymmetry difference ($\Delta A_{CP}$) in
the ($\Delta A_{CP}-\lambda_t'$) plane, where the red and blue
regions correspond to $m_{t'}$ = 400 and 600 GeV; grey shaded regions correspond
to the uncertainties due to hadronic parameters.} \label{kpi2}
\end{figure}
%
%
\begin{figure}[t]
\includegraphics[width=8cm,height=6cm,clip]{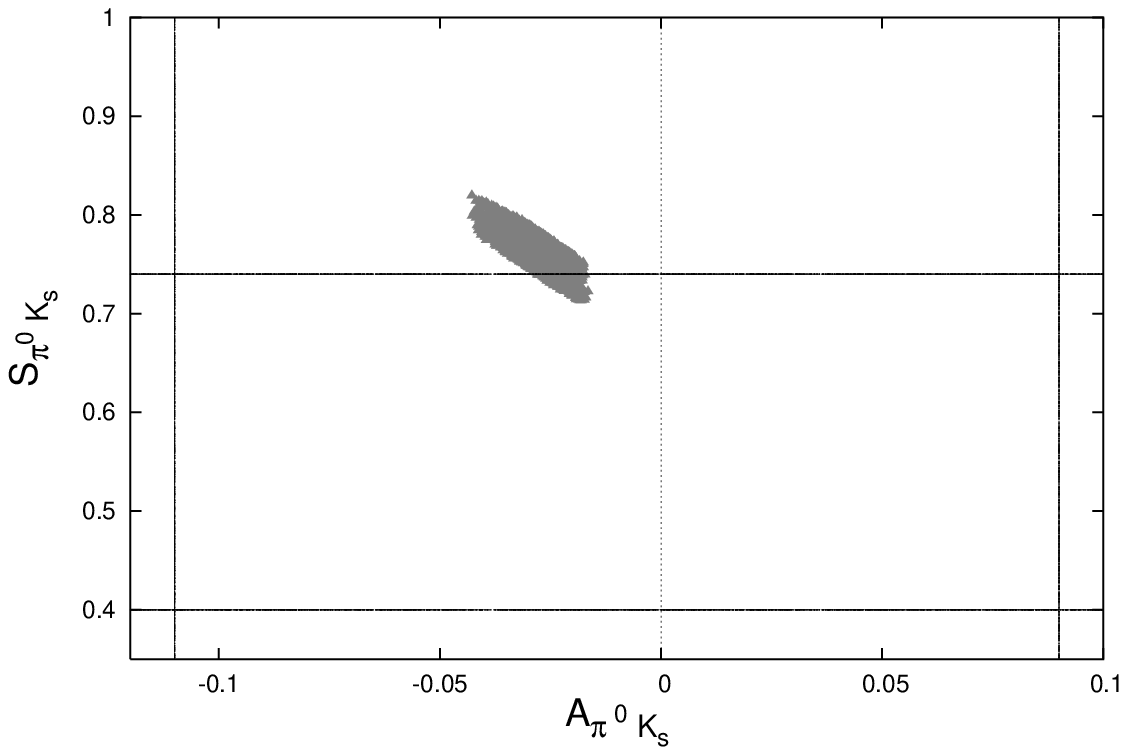}%
\hspace{0.2cm}%
\includegraphics[width=8cm,height=6cm,clip]{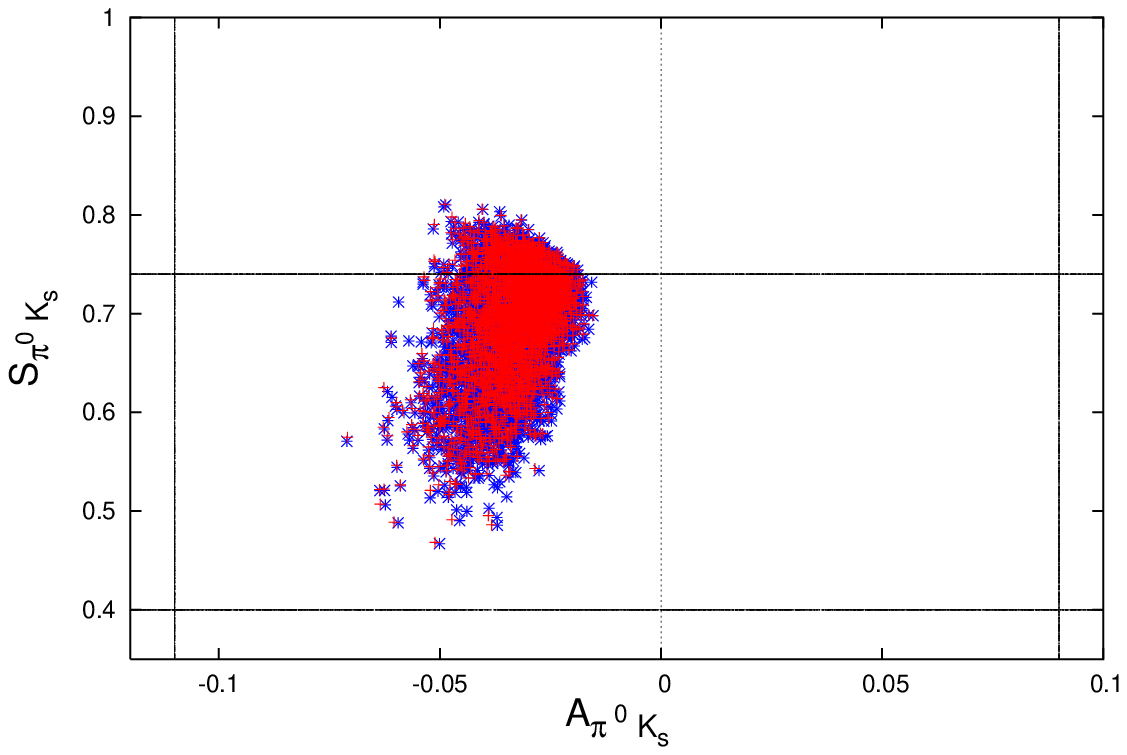}
\caption{Correlation plots between the mixing induced CP asymmetry
$S_{\pi^0 K_s}$ and the direct CP asymmetry $A_{\pi^0 K_s}$ in the
SM (left panel) and in the fourth generation model (right panel)
where the  red and blue regions correspond to $m_{t'}=$400 and 600
GeV . The horizontal and vertical lines represent $1\sigma$ experimental 
allowed ranges.}
\label{correlation2}
\end{figure}

For numerical evaluation,  we use input parameters as follows. For
the form factors and decay constants we use $F_0^{B \to K}(0)=0.34
\pm 0.05$,  $F_0^{B \to \pi}(0)=0.28 \pm 0.05$, $f_{\pi}=0.131$ GeV,
$f_K=0.16$ GeV and for Gegenbauer moments we use $\lambda_B = 350 \pm 150$\, MeV
\cite{QCDF}. We varied the hard spectator and
annihilation phases $\phi_{A,H}$ in the entire range i.e., between
$[-\pi, \pi]$, imposing the constraint that the corresponding
branching ratios should be within the three sigma experimental
range. Also we have included $20\%$ uncertainty in $\Lambda_{QCD}$ i.e we varied 
$\Lambda_{QCD}= 225$\, MeV from its nominal value in SM3 \cite{QCDF} by $\pm 45$\, MeV, 
which enters in the hard spectator contribution \footnote{The corresponding 
choices in the scenario S4 of \cite{QCDF} are given by $F_0^{B \to K}(0)=0.31$, 
$F_0^{B \to \pi}(0)=0.25$, $f_{\pi}=0.131$ GeV, $f_K=0.16$ GeV, $\lambda_B = 200$\, MeV,  
$\phi_{A,H} = - 55^{\circ}$ and $\Lambda_{QCD} = 225$ MeV}. Since $\lambda_B$ and 
$\Lambda_{QCD}$ were previously fixed to 200\, MeV and 225\, MeV respectively to 
fit the data interpreted in SM3, it may not be unreasonable to assume small 
changes for SM4.  For the CKM matrix elements we use values as given in the 
Table \ref{tab1}. We have also used the  range of $\lambda_{t'}$ and $\phi_s$ as obtained 
from the fit for different $m_{t'}$.

Using these values we show the allowed regions in the 
$\Delta A_{CP}-\lambda_{t'}$ plane for different values of $m_{t'}$  in
figure \ref{kpi2} and we note that an enhancement in $\Delta A_{CP}$ upto 
the current $1\,\sigma$ experimental upper bound ($\approx 17.6 \%$) is possible 
for largish strong phases, $\phi_{A,H} \sim (-45\, \to \,- 90)^{\circ}$.
The correlation plots between mixing induced and
direct CP asymmetry parameters in $B^0 \to \pi^0 K^0$ modes are
shown in figure \ref{correlation2}.

\subsection{CP violation in  $B^0 \to \pi^0 \pi^0$ modes}
As discussed earlier there exists several hints for the
possible existence of  new physics in the
$b \to s$ sector. So the next obvious question is:
Do the $b \to d$ penguin amplitudes also have significant new physics
contribution? The present data does not provide any conclusive answer
to it. The obvious example is the $B \to \pi \pi $ processes, which receive
dominant contribution from $b \to u$ tree and from $b \to d$ penguin diagrams.
The present data \cite{Barberio:2008fa} are presented in Table \ref{tab6}.
\begin{table}[htbp]
\begin{center}
\begin{tabular}{|c|c|c|}
\hline
Decay mode & HFAG Average  \\
\hline
 $10^6 \times {\rm Br}(B^0 \to \pi^+ \pi^-)$ & $5.16 \pm 0.22 $\\
 $10^6 \times {\rm Br}(B^- \to \pi^- \pi^0)$ & $5.59 \pm 0.41 $\\
$10^6 \times {\rm Br}(B^0 \to \pi^0 \pi^0)$ & $1.55  \pm 0.19
$ \\
$S_{\pi^+ \pi^-}$& $-0.65 \pm 0.07 $\\
{ $A_{\pi^+ \pi^-}$ } &  { $ 0.38 \pm 0.06 $ }\\
$A_{\pi^- \pi^0}$ & $0.06 \pm 0.05 $ \\
{ $A_{\pi^0 \pi^0}$} & { $0.43_{-0.24}^{+0.25} $ }\\
\hline
\end{tabular}
\end{center}
\caption{Experimental  results for $B \to \pi \pi$ processes}
\label{tab6}
\end{table}
Thus, it can be seen that  the measured value of Br$(B^0
\to \pi^0 \pi^0)$ is nearly two times larger than the corresponding
theoretical predictions \cite{QCDF,lisanda}. 
Also the measured values of  direct CP asymmetry parameters
$A_{\pi^+ \pi^-}$ and $A_{\pi^0 \pi^0}$ are higher than the corresponding
SM predictions \cite{QCDF}.
Thus, the discrepancy between the theoretical and the measured quantities
imply that there may also be some new physics effect in the $b \to
d$ penguins as speculated in $b \to s$ penguins.

Let us first write down the most general topological amplitudes
for $B \to \pi \pi$ modes as
\bea
\sqrt 2 A(B^+ \to \pi^+ \pi^0)& =& -(T +C +P_{ew}),\nn\\
A(B^0 \to \pi^+ \pi^-)& =&-(T+P),\nn\\
\sqrt 2 (B^0\to \pi^0 \pi^0) &=& -(C-(P-P_{ew})).
\eea
From the above relations it can be seen that if there will be additional
new contribution to the penguin sector
with other amplitudes as expected in SM4 then that may explain $B \to \pi \pi$
observations.

As discussed earlier, due to the presence of the additional
generation of quarks the unitarity condition becomes
$\lambda_u + \lambda_c
+\lambda_t +\lambda_{t'}=0$.
Thus, including the new contributions one can symbolically represent
these amplitudes as \bea Amp = \lambda_u^d A_u^d +\lambda_c^d A_c^d
-\lambda_{t'}^d~A_{new} =\lambda_u^d A_u^d\Big[1- r_1~ a_1~
e^{i(\delta_1^d+\gamma)}-r'_1~ b_1~ e^{i(\delta_2^d+\phi_d)}\Big],
\eea where
$b_1=|\lambda_{t'}^d/\lambda_u^d|$, $\phi_d$ is the weak phase of
$\lambda_t'^d$.
$r'_1=|A_{new}/A_u^d|$, and $\delta_2^d$ is
the relative strong phases between $A_{new}$ and
$A_u^d$.
Thus from the
above amplitude one can obtain the CP averaged branching ratio,
direct and mixing induced CP asymmetry parameters as
\bea
{\rm Br}&=& \frac{|p_{c.m}| \tau_B}{8 \pi M_B^2}
\Big[{{\cal{R}}_1}-2r_1a_1 \cos \delta_1^d \cos
\gamma-2r'_1 b_1 \cos \delta_2^d \cos (\phi_d+\gamma) 
\nonumber\\&&
+2 r_1 r'_1 a_1 b_1
 \cos(\delta_2^d-\delta_1^d)\cos\phi_d\Big]\;,\nn\\
A_{\pi \pi}&=&\frac{2\Big[r_1a_1 \sin
\delta_1^d \sin \gamma +r'_1 b_1 \sin \delta_2^d \sin (\phi_d+\gamma)
+ r_1 r'_1 a_1 b_1
\sin(\delta_1^d-\delta_2^d) \sin
\phi_d \Big]}{\Big[{\cal{R}}_1-2r_1a_1 \cos \delta_1^d \cos
\gamma-2r'_1 b_1 \cos \delta_2^d \cos (\phi_d+\gamma) +
2 r_1 r'_1 a_1 b_1
 \cos(\delta_2^d-\delta_1^d)\cos\phi_d\Big]}\;,\nn\\
S_{\pi \pi}&=&\frac{X_1}{\Big[{\cal R}_1
-2r_1a_1 \cos \delta_1^d \cos
\gamma-2r'_1 b_1 \cos \delta_2^d \cos (\phi_d+\gamma) +
2 r_1 r'_1 a_1 b_1
 \cos(\delta_2^d-\delta_1^d)\cos\phi_d\Big]
}\;,
 \eea
where
\bea X_1 &=&-\Big[ \sin (2 \beta+2\gamma)- 2 r_1 a_1 \cos
\delta_1^d \sin(2 \beta+\gamma)+ 2 r'_1 b_1 \cos \delta_2^d
\sin(\phi_d-(2 \beta+ \gamma))\nn\\ &+&(r_1 a_1)^2\sin(2 \beta )
+(r'_1 b_1)^2\sin(2\phi_d-2 \beta)-2 r_1r'_1 a_1 b_1 \cos(\delta_1^d-
\delta_2^d) \sin(\phi_d-2 \beta)\Big].
\eea
and ${\cal{R}}_1=1+(r_1a_1)^2+(r'_1b_1)^2$.
\begin{figure}[htb]
\centerline{\epsfysize 2.5 truein \epsfbox{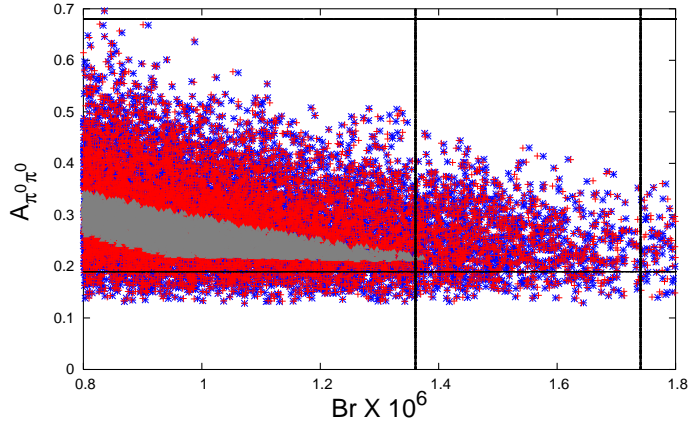}} \caption{The
correlation plot between the direct CP asymmetry and the CP-averaged
branching ratio for the $B^0 \to \pi^0 \pi^0 $ process where the
grey region corresponds to the SM result and 
the red and blue regions correspond to
$m_{t'}=$400 and 600 GeV respectively. The horizontal and vertical lines represent the
1-$\sigma$ experimental range of the corresponding observables.}
\label{pi02}
\end{figure}
Now varying $\lambda_{t'}^d$ between 0 and $1.5\times 10^{-4}$ and $\phi_d$
between $(0-360)^\circ$ we present the correlation plot between the
direct CP asymmetry parameter and branching ratio in Fig. \ref{pi02}. 
From the figure one can see that the observed data could be
accommodated in the SM with four generations.

\section{Summary and Outlook}

\label{concl}
Standard Model with four generations should be considered seriously.
We do not have a good understanding of fermion generations. We have 
already seen three; why not the fourth? Electroweak precision tests 
do not rule out the existence of a fourth family, though they  do require 
that the mass difference between the $t^\prime$ and the $b^\prime$ be less than about 75 GeV.
This degeneracy amounting to O(10\%) for $\approx 500$ GeV masses does not
seem so serious. Of course, the electroweak precision tests suggest then
a possible heavy Higgs particle but this  actually may be hinting at a very
interesting resolution to the hierarchy puzzle. This is because
heavier quarks of the 4th generation can play a significant role in dynamical 
electroweak-symmetry breaking, i.e. a composite Higgs particle .

Another extremely interesting implication of a 4th family is the gigantic improvement over the three generation case in the context of baryogenesis,
as in particular emphasized by Hou~\cite{Hou08}.

These two implications of a 4th family are in themselves so interesting, if not
profound, that even though at this time the repercussions for dark matter and/or unification are not quite clear, the idea should be given a serious consideration.

Although one of us (A.S.) had gotten already interested and involved in the    physics of the 4th generation over twenty years ago, our recent interest 
was instigated by the fact that this obvious extension of the Standard Model
offers a simple solution to many of the anomalies that have been seen in B,
$B_s$ decays. For one thing the predicted value of $\sin 2 \beta$ in the SM
is coming out to be too high from the one directly measured via the gold-plated
$\psi K_s$ mode. Besides, the value of $\sin 2 \beta$ measured via many of the penguin-dominated modes is systematically coming out to be smaller than the
predicted value. Then there is the very large difference in the direct CP
asymmetry between $K^+ \pi^-$ and $K^+ \pi^0$ decays of the $B^0$ and $B^+$.
Finally, there is the fact that both CDF and D0 find that $B_s \to \psi \phi$
decays are exhibiting O(2$\sigma$) non-vanishing CP asymmetries whereas
SM predicts vanishing small asymmetry. 

The effect seen in $B_s \to \psi \phi$ at Fermilab is doubly significant.
First of all two of the anomalies discussed above that were seen at B-factories
taken seriously suggest a non-standard CP-odd phase in $b \to s$ transitions. That then makes it extremely difficult, if not impossible, for new physics 
not to show up as well in $B_s$ mixing; thus the B-factory anomalies basically
imply  non-standard CP effects in mixing induced CP-asymmetry in $B_s \to \psi \phi$. The second crucial aspect of the CP asymmetry in $B_s \to \psi \phi$ 
is that it is a gold-plated effect; that is the fact that in the SM CP asymmetry in that mode 
should be vanishingly small is a very clean prediction with  no serious hadronic uncertainty. 
Therefore it is extremely important that Fermilab gives very high
priority to confirming or refuting this effect. In fact very soon the LHCb
experiment at CERN should also be able to study this mode and clarify this issue.

In an earlier paper we had focused on studying the CP anomalies seen in B, $B_s$
decays in SM4 mentioned above; we found that the SM4 offers a simple explanation for most of the 
anomalies with the heavy quarks of mass around 400 - 600 GeV.
This paper is a follow-up wherein we further explore the implications
of SM4 for K and B, $B_s$ decays. By using a host of measurements in K, B, $B_s$
decays such as indirect CP violation parameter $\epsilon_K$, $K^+ \to \pi^+ \nu \bar \nu$, mixing induced CP asymmetry in $B \to \psi K_s$, Br ($B \to X_s \gamma$), semi-leptonic decays of B etc along with oblique parameters and Br( $Z \to b \bar b$), we first constrained the enlarged 4$\times$4 CKM-matrix.
We then explored the implications of the SM4 for a variety of processes such
as $a_{CP}(B \to X_s \gamma)$, $Br(B_s\to \mu^+\mu^-)$, $a_{CP} (B \to X_s l^+ l^-)$, $A_{SL}(B_s\to X_s \ell\nu)$, $A_{FB}(B \to X_s l^+ l^-)$, $A_{FB}(B \to K^* l^+ l^-)$, 
$Br(B \to X_s \nu \bar \nu)$, CP asymmetries in $B\to \pi^0 K_s$ and in $B\to \pi^0 \pi^0$ etc.
 We identified many processes wherein SM4 predicts significant differences from 
SM3, {\rm e.g} $S(B_s\to \psi\phi)$, $a_{CP}(B \to X_s \gamma)$, $a_{CP} (B \to X_s l^+ l^-)$, 
$A_{SL}(B_s\to X_s \ell\nu)$, $Br(B \to X_s \nu \bar \nu)$, $Br(B_s\to \mu^+\mu^-)$, 
$Br(K_L\to \pi^0\nu\bar{\nu})$ {\rm etc};  thus studies therein should especially 
provide further understanding of the parameter space of SM4.

One of the most interesting aspect of the 4th generation hypothesis is that 
it is testable relatively easily in the LHC experiments where in fact it has 
distinctive 
signatures \cite{Hou08}. In the coming few years not only we should be able to learn about the existence or lack thereof of quarks and leptons of the 4th family, the heavier Higgs that is also favored
in SM4 scenario should be easier to search for in the LHC experiments via
the gold-plated mode: $H \to Z Z$. Also the heavy Higgs has interesting implications for 
flavour-diagonal and flavour-changing final states involving $t'$ and/or $b'$ \cite{shaouly}. 
Therefore,  LHC should shed significant light on the question of SM4 in the next few years.

\begin{acknowledgments} 
We want to thank Andrzej Buras, Martin Beneke, Thorsten Feldmann, Tillmann Heidsieck, 
Alexander Lenz and Giovanni Punzi for many discussions.
SN would also like to thank Carlo Giunti for discussion regarding numerical 
analysis and the theory division of Saha Institute of Nuclear Physics (SINP) 
, in particular to Gautam Bhattacharyya, for hospitality. The work of AKA is 
financially supported by 
NSERC of Canada. The work of AS is suppported in part by the US DOE grant 
\# DE-AC02-98CH10886(BNL). The work of AG is supported in part by CSIR and DST, Govt. of India and the work of RM is supported in part by DST, Govt. of India. 
SN's work is supported in part by MIUR under contract 2008H8F9RA$\_$002 and by the European 
Community’s Marie-Curie Research Training Network under contract MRTN-CT-2006-035505 ‘Tools and Precision Calculations for Physics Discoveries at Colliders’.

\end{acknowledgments}


\begin{thebibliography}{99}

\bibitem{NC63}
  N.~Cabibbo,
  Phys.\ Rev.\ Lett.\  {\bf 10}, 531 (1963).

\bibitem{KM73}
  M.~Kobayashi and T.~Maskawa,
  Prog.\ Theor.\ Phys.\  {\bf 49}, 652 (1973).


\bibitem{LS07}
  E.~Lunghi and A.~Soni,
  JHEP {\bf 0709}, 053 (2007)
  [arXiv:0707.0212 [hep-ph]].
  
  \bibitem{LS08}
 E.~Lunghi and A.~Soni,
  Phys.\ Lett.\  B {\bf 666}, 162 (2008)
  [arXiv:0803.4340 [hep-ph]].


\bibitem{LS09}
E.~Lunghi and A.~Soni,
  JHEP {\bf 0908}, 051 (2009)
  [arXiv:0903.5059 [hep-ph]].

\bibitem{uli_lenz}
 A.~Lenz and U.~Nierste,
  JHEP {\bf 0706}, 072 (2007)
  [arXiv:hep-ph/0612167].

\bibitem{bona}
  M.~Bona {\it et al.}  [UTfit Collaboration],
  arXiv:0803.0659 [hep-ph].
  
\bibitem{APS1}
 K.~Agashe, G.~Perez and A.~Soni,
  Phys.\ Rev.\ Lett.\  {\bf 93}, 201804 (2004)
  [arXiv:hep-ph/0406101].


\bibitem{APS2}
 K.~Agashe, G.~Perez and A.~Soni,
  Phys.\ Rev.\  D {\bf 71}, 016002 (2005)
  [arXiv:hep-ph/0408134].

\bibitem{AJB081}
 M.~Blanke, A.~J.~Buras, B.~Duling, S.~Gori and A.~Weiler,
  JHEP {\bf 0903}, 001 (2009)
  [arXiv:0809.1073 [hep-ph]].

\bibitem{AJB082}
 M.~Blanke, A.~J.~Buras, B.~Duling, K.~Gemmler and S.~Gori,
  JHEP {\bf 0903}, 108 (2009)
  [arXiv:0812.3803 [hep-ph]].

\bibitem{MN08}
 S.~Casagrande, F.~Goertz, U.~Haisch, M.~Neubert and T.~Pfoh,
  JHEP {\bf 0810}, 094 (2008)
  [arXiv:0807.4937 [hep-ph]].


\bibitem{lang}
V.~Barger, L.~L.~Everett, J.~Jiang, P.~Langacker, T.~Liu and C.~E.~M.~Wagner,
  JHEP {\bf 0912}, 048 (2009)
  [arXiv:0906.3745 [hep-ph]].

\bibitem{paridi}
W.~Altmannshofer, A.~J.~Buras, S.~Gori, P.~Paradisi and D.~M.~Straub,
  Nucl.\ Phys.\  B {\bf 830}, 17 (2010)
  [arXiv:0909.1333 [hep-ph]].


\bibitem{SAGMN08}
  A.~Soni, A.~K.~Alok, A.~Giri, R.~Mohanta and S.~Nandi,
  Phys.\ Lett.\  B {\bf 683}, 302 (2010)
  [arXiv:0807.1971 [hep-ph]].


  
\bibitem{as_moriond09}
A.~Soni,
  arXiv:0907.2057 [hep-ph].


\bibitem{Hou08}
W. S. Hou, arxiv:0803.1234. 

\bibitem{CJ88}

For earlier related works see,
C. Jarlskog and R. Stora, \PL(B208,288,1988);
F. del Aguila and J. A. Aguilar-Saavedra, \PL(B386,241,1996);
F. del Aguila and J. A. Aguilar-Saavedra and G. C. Branco,Nucl. Phys. {\bf B510},39,1998.


\bibitem{GK08}
See also,
R.~Fok and G.~D.~Kribs,
        arXiv:0803.4207 .

\bibitem{AS_olds1}
The importance of rare B-decays for searching for the 4th family was
emphasized long ago in
W. -S. Hou,
R. Willey and A.Soni, Phys. Rev. Lett. {\bf 58}, 1608 (1987);
see also~\cite{AS_olds2,AS_olds3,AS_olds4,ge_olds}.
\bibitem{AS_olds2}
W. -S. Hou,
A. Soni and H. Steger, Phys. Rev. Lett. {\bf 59}, 1521 (1987).
\bibitem{AS_olds3}
W. -S. Hou,
A. Soni and H. Steger, \PL(B192,441,1987).
\bibitem{AS_olds4}
 C.~Hamzaoui, A.~I.~Sanda and A.~Soni,
  Nucl.\ Phys.\ Proc.\ Suppl.\  {\bf 13}, 494 (1990).

\bibitem{ge_olds}
G.~Eilam, J.~L.~Hewett and T.~G.~Rizzo,
  Phys.\ Rev.\  D {\bf 34}, 2773 (1986), 
  Phys.\ Lett.\  B {\bf 193} (1987) 533,

\bibitem{norton}
J. Carpenter, R. Norton, S. Siegemund-Broka and A. Soni, Phys. Rev. Lett. {\bf 65}, 153 (1990). 
In passing, we note that the quark mass needed for
dynamical electroweak symmetry breaking in this work, translated to the fourth family 
quasi-degenerate doublet, gives $m_{t'}\sim\, 500$\, {\it GeV} and
$m_H \approx \,\sqrt{2} \,m_{t'}\, \sim\, 700$\, {\it GeV}.

\bibitem{symp_SM4_8789}
See the proceedings of the First International Symposium on the fourth family of 
			  quarks and leptons, Santa Monica, CA, Feb 1987, published
			  by the NY Academy of Sciences; eds D. Cline and A. Soni; see also the proceedings of the Second  International Symposium on the fourth family of 
			  quarks and leptons, Santa Monica, CA, Feb 1989, published
			  by the NY Academy of Sciences; eds D. Cline and A. Soni.
	

\bibitem{Holdom:1986rn}
  B.~Holdom,
  Phys.\ Rev.\ Lett.\  {\bf 57}, 2496 (1986)
  [Erratum-ibid.\  {\bf 58}, 177 (1987)];
 W.~A.~Bardeen, C.~T.~Hill and M.~Lindner,
  Phys.\ Rev.\  D {\bf 41}, 1647 (1990);
 C.~T.~Hill, M.~A.~Luty and E.~A.~Paschos,
  Phys.\ Rev.\  D {\bf 43}, 3011 (1991);
P.~Q.~Hung and G.~Isidori,
  Phys.\ Lett.\  B {\bf 402}, 122 (1997)
  [arXiv:hep-ph/9609518].


\bibitem{Hung:2009ia}
  P.~Q.~Hung and C.~Xiong,
  arXiv:0911.3892 .

\bibitem{Hashimoto:2009ty}
  M.~Hashimoto and V.~A.~Miransky,
  arXiv:0912.4453 .

\bibitem{erler}
 J.~Erler and P.~Langacker,
  Acta Phys.\ Polon.\  B {\bf 39}, 2595 (2008)
  [arXiv:0807.3023 [hep-ph]].
  
  \bibitem{novikov1}
  V.~A.~Novikov, L.~B.~Okun, A.~N.~Rozanov and M.~I.~Vysotsky,
  Phys.\ Lett.\  B {\bf 529}, 111 (2002)
  [arXiv:hep-ph/0111028].
  
  \bibitem{novikov2}
V.~A.~Novikov, L.~B.~Okun, A.~N.~Rozanov and M.~I.~Vysotsky,
  JETP Lett.\  {\bf 76}, 127 (2002)
  [Pisma Zh.\ Eksp.\ Teor.\ Fiz.\  {\bf 76}, 158 (2002)]
  [arXiv:hep-ph/0203132].

\bibitem{Kribs_EWPT}
 G.~D.~Kribs, T.~Plehn, M.~Spannowsky and T.~M.~P.~Tait,
 Phys.\ Rev.\  D {\bf 76}, 075016 (2007)
 [arXiv:0706.3718 [hep-ph]].
 
\bibitem{chanowitz}
M.~S.~Chanowitz,
  Phys.\ Rev.\  D {\bf 79}, 113008 (2009)
  [arXiv:0904.3570 [hep-ph]].
 
 











\bibitem{NEED}
Particle Data Group ( D. E. Groom {\it et al}), \EPJC(15,1,2000);
M. Maltoni {\it et al.}, \PL(B476,107,2000) [hep-ph/9911535];


\bibitem{DM4}
See, {\it e.g.}
  G.~E.~Volovik,
      Pisma Zh.\ Eksp.\ Teor.\ Fiz.\  {\bf 78}, 1203 (2003)
        [JETP Lett.\  {\bf 78}, 691 (2003)]
	  [hep-ph/0310006].
	


\bibitem{QCDF1}
 M.~Beneke, G.~Buchalla, M.~Neubert and C.~T.~Sachrajda,
  Nucl.\ Phys.\  B {\bf 606}, 245 (2001)
  [arXiv:hep-ph/0104110].

\bibitem{QCDF}
  M.~Beneke and M.~Neubert,
  Nucl.\ Phys.\  B {\bf 675}, 333 (2003)
  [arXiv:hep-ph/0308039].


\bibitem{Hovhannisyan:2007pb}
 A.~Hovhannisyan, W.~S.~Hou and N.~Mahajan,
 Phys.\ Rev.\  D {\bf 77}, 014016 (2008)
 [arXiv:hep-ph/0701046].

\bibitem{hou03}
A.~Arhrib and W.~S.~Hou,
  Eur.\ Phys.\ J.\  C {\bf 27}, 555 (2003)
  [arXiv:hep-ph/0211267].

\bibitem{Hou:2005yb}
 W.~S.~Hou, M.~Nagashima and A.~Soddu,
 Phys.\ Rev.\  D {\bf 72}, 115007 (2005)
 [arXiv:hep-ph/0508237].


\bibitem{Hou:2005hd}
  W.~S.~Hou, M.~Nagashima and A.~Soddu,
  Phys.\ Rev.\ Lett.\  {\bf 95}, 141601 (2005)
  [arXiv:hep-ph/0503072].

\bibitem{Arhrib:2006pm}
  A.~Arhrib and W.~S.~Hou,
  JHEP {\bf 0607}, 009 (2006)
  [arXiv:hep-ph/0602035].

\bibitem{Hou:2006zza}
  W.~S.~Hou, M.~Nagashima, G.~Raz and A.~Soddu,
  JHEP {\bf 0609}, 012 (2006)
  [arXiv:hep-ph/0603097].



\bibitem{Hou:2006jy}
  W.~S.~Hou, H.~n.~Li, S.~Mishima and M.~Nagashima,
  Phys.\ Rev.\ Lett.\  {\bf 98}, 131801 (2007)
  [arXiv:hep-ph/0611107].

\bibitem{lenz4}
M.~Bobrowski, A.~Lenz, J.~Riedl and J.~Rohrwild,
  Phys.\ Rev.\  D {\bf 79}, 113006 (2009)
  [arXiv:0902.4883 [hep-ph]];
G.~Eilam, B.~Melic and J.~Trampetic,
  Phys.\ Rev.\  D {\bf 80}, 116003 (2009)
  [arXiv:0909.3227 [hep-ph]].

\bibitem{NTU}
W.~S.~Hou and C.~Y.~Ma have recently updated their analysis; see the talk by 
C.~Y.~Ma at the second workshop on ``Beyond the three generation Standard 
Model", Jan\, 14-16, 2010, The National Taiwan University
, Taipei, Taiwan.
 


\bibitem{peskin}
  M.~E.~Peskin and T.~Takeuchi,
  Phys.\ Rev.\  D {\bf 46}, 381 (1992).



\bibitem{sher}
 P.~Q.~Hung and M.~Sher,
  Phys.\ Rev.\  D {\bf 77}, 037302 (2008)
  [arXiv:0711.4353 [hep-ph]].



\bibitem{novikov3}
V.~A.~Novikov, A.~N.~Rozanov and M.~I.~Vysotsky,
  arXiv:0904.4570 [hep-ph].

\bibitem{zbb1}
 J.~Bernabeu, A.~Pich and A.~Santamaria,
  Nucl.\ Phys.\  B {\bf 363}, 326 (1991).

\bibitem{Barberio:2008fa}
  E.~Barberio {\it et al.}  [Heavy Flavor Averaging Group],
  arXiv:0808.1297 [hep-ex].


\bibitem{Misiak:2006ab}
  M.~Misiak and M.~Steinhauser,
  Nucl.\ Phys.\  B {\bf 764}, 62 (2007)
  [arXiv:hep-ph/0609241].

\bibitem{Misiak:2006zs}
  M.~Misiak {\it et al.},
  Phys.\ Rev.\ Lett.\  {\bf 98}, 022002 (2007)
  [arXiv:hep-ph/0609232].
See also, 
T.~Becher and M.~Neubert,
  Phys.\ Rev.\ Lett.\  {\bf 98}, 022003 (2007)
  [arXiv:hep-ph/0610067].

\bibitem{Buras:1994dj}
A.~J.~Buras and M.~Munz, \PRD(52,186,1995)  [arXiv:hep-ph/9501281].

\bibitem{Buras:1997fb}
  A.~J.~Buras and R.~Fleischer,
  Adv.\ Ser.\ Direct.\ High Energy Phys.\  {\bf 15}, 65 (1998);
  [arXiv:hep-ph/9704376].

\bibitem{Nir:1989rm}
Y.~Nir, Phys.\ Lett.\  B {\bf 221}, 184 (1989).


\bibitem{buras1}
  A.~J.~Buras, M.~Jamin and P.~H.~Weisz,
  Nucl.\ Phys.\  B {\bf 347}, 491 (1990).

\bibitem{Hattori:1999ap}
T.~Hattori, T.~Hasuike and S.~Wakaizumi, Phys.\ Rev.\  D {\bf 60}, 113008 (1999)
[arXiv:hep-ph/9908447].

\bibitem{Buchalla:1995vs}
G.~Buchalla, A.~J.~Buras and M.~E.~Lautenbacher, Rev.\ Mod.\ Phys.\  {\bf 68}, 1125 (1996);
[arXiv:hep-ph/9512380].


\bibitem{buras2} 
 A.~J.~Buras and D.~Guadagnoli,
  Phys.\ Rev.\  D {\bf 78}, 033005 (2008);
  [arXiv:0805.3887 [hep-ph]].

\bibitem{latticeold}
 D.~J.~Antonio {\it et al.}  [RBC Collaboration and UKQCD Collaboration],
  Phys.\ Rev.\ Lett.\  {\bf 100}, 032001 (2008)
  [arXiv:hep-ph/0702042].


\bibitem{alv}
 C.~Aubin, J.~Laiho and R.~S.~Van de Water,
  arXiv:0905.3947 [hep-lat].
  
  \bibitem{rbc2}
  See also C. Kelly talk (for the RBC-UKQCD Collaborations) at the Lattice 2009 Symposium, Beijing, China.

\bibitem{lattice3}
D.~Becirevic, 
{\it In the Proceedings of 2nd Workshop on the CKM Unitarity Triangle, Durham, England, 5-9 Apr 2003, pp WG202} [arXiv:hep-ph/0310072];
  N.~Tantalo,
  arXiv:hep-ph/0703241;
E.~Gamiz, C.~T.~H.~Davies, G.~P.~Lepage, J.~Shigemitsu and M.~Wingate,
  PoS {\bf LAT2007}, 349 (2007)
  [arXiv:0710.0646 [hep-lat]].

\bibitem{cdf}
H.~G.~Evans  [CDF Collaboration and D0 Collaboration],
  Frascati Phys.\ Ser.\  {\bf 44}, 421 (2007)
  [arXiv:0705.4598 [hep-ex]].



\bibitem{uli1}
S. Herrlich and U. Nierste, \NP(B419,292,1994) [hep-ph/9310311].

\bibitem{uli2}
S. Herrlich and U. Nierste, \PRD(52,6505,1995) [hep-ph/9507262].


\bibitem{Buchalla:1993wq}
  G.~Buchalla and A.~J.~Buras,
  Nucl.\ Phys.\  B {\bf 412}, 106 (1994);
  [arXiv:hep-ph/9308272].

\bibitem{buras4th}
 A.~J.~Buras, B.~Duling, T.~Feldmann, T.~Heidsieck, C.~Promberger and S.~Recksiegel,
  arXiv:1002.2126 [hep-ph].


\bibitem{Kagan:1998bh}
  A.~L.~Kagan and M.~Neubert,
  Phys.\ Rev.\  D {\bf 58}, 094012 (1998);
  [arXiv:hep-ph/9803368].


\bibitem{Kiers:2000xy}
  K.~Kiers, A.~Soni and G.~H.~Wu,
  Phys.\ Rev.\  D {\bf 62}, 116004 (2000);
  [arXiv:hep-ph/0006280].

\bibitem{Soares:1991te}
  J.~M.~Soares,
  Nucl.\ Phys.\  B {\bf 367}, 575 (1991).

\bibitem{Hurth:2003dk}
  T.~Hurth, E.~Lunghi and W.~Porod,
  Nucl.\ Phys.\  B {\bf 704}, 56 (2005);
  [arXiv:hep-ph/0312260].

\bibitem{Browder:2008em}
  T.~E.~Browder, T.~Gershon, D.~Pirjol, A.~Soni and J.~Zupan,
  arXiv:0802.3201 [hep-ph].


\bibitem{cdfd0}
T.~Aaltonen {\it et al.}  [CDF Collaboration],
  Phys.\ Rev.\ Lett.\  {\bf 100}, 161802 (2008);
  [arXiv:0712.2397 [hep-ex]];
V.~M.~Abazov {\it et al.}  [D0 Collaboration],
  Phys.\ Rev.\ Lett.\  {\bf 101}, 241801 (2008)
  [arXiv:0802.2255 [hep-ex]].

\bibitem{hw}
E.~H.~Thorndike,
Ann.\ Rev.\ Nucl.\ Part.\ Sci.\ 35 (1985) 195;
J.~S.~Hagelin and M.~B.~Wise,
Nucl.\ Phys.\ B {\bf 189} (1981) 87;
J.~S.~Hagelin,
Nucl.\ Phys.\ B {\bf 193} (1981) 123;
A.~J.~Buras, W.~Slominski and H.~Steger,
Nucl.\ Phys.\ B {\bf 245} (1984) 369.
R.~N.~Cahn and M.~P.~Worah,
Phys.\ Rev.\ D {\bf 60} (1999) 076006;
\bibitem{run2}
K.~Anikeev {\it et al.},
\emph{$B$ physics at the Tevatron: Run II and beyond},
[hep-ph/0201071], Chapters 1.3 and 8.3.
\bibitem{llnp}
S.~Laplace, Z.~Ligeti, Y.~Nir and G.~Perez,
Phys.\ Rev.\ D {\bf 65} (2002) 094040.

\bibitem{d07}
V.~M.~Abazov {\it et al.}  [D0 Collaboration],
  Phys.\ Rev.\ Lett.\  {\bf 98}, 151801 (2007)
  [arXiv:hep-ex/0701007].
\bibitem{cdf7}
CDF Collaboration, "Measurement of {\it CP} Asymmetry in Semileptonic $B$ decays", {\it cdf} note 9015, URL http://www.cdf.fnal.gov, Oct 16, 2007.
\bibitem{hfag}
Heavy Flavour Averaging Group (HFAG), "Results for the PDG 2009 web update",
http://www.slac.stanford.edu/xorg/hfag/.
\bibitem{LO}
E. Franco, M. Lusignoli and A. Pugliese,
Nucl. Phys. {\bf B194}, 403 (1982);
L.L. Chau, Phys. Rep. {\bf 95}, 1 (1983);
M.B. Voloshin, N.G. Uraltsev, V.A. Khoze and M.A. Shifman,
Sov. J. Nucl. Phys. {\bf 46}, 112 (1987);
A. Datta, E.A. Paschos and U. T{\"u}rke,
Phys. Lett. {\bf B196}, 382 (1987);
A. Datta, E.A. Paschos and Y.L. Wu,
Nucl. Phys. {\bf B311}, 35 (1988).

\bibitem{bbd1}
M. Beneke, G. Buchalla and I. Dunietz,
Phys. Rev. {\bf D54}, 4419 (1996).

\bibitem{bbgln1}
M.~Beneke, G.~Buchalla, C.~Greub, A.~Lenz and U.~Nierste,
Phys.\ Lett.\ B {\bf 459} (1999) 631
[arXiv:hep-ph/9808385].
\bibitem{rome03}
M.~Ciuchini, E.~Franco, V.~Lubicz, F.~Mescia and C.~Tarantino,
JHEP {\bf 0308} (2003) 031
[arXiv:hep-ph/0308029].

\bibitem{bbln}
M.~Beneke, G.~Buchalla, A.~Lenz and U.~Nierste,
Phys.\ Lett.\ B {\bf 576} (2003) 173
[arXiv:hep-ph/0307344].


\bibitem{soni_gers}
T.~Gershon and A.~Soni,
  J.\ Phys.\ G {\bf 33}, 479 (2007)
  [arXiv:hep-ph/0607230].

\bibitem{Du:1995ez}
  D.~S.~Du and M.~Z.~Yang,
  Phys.\ Rev.\  D {\bf 54}, 882 (1996)
  [arXiv:hep-ph/9510267].


\bibitem{Ali:1998sf}
  A.~Ali and G.~Hiller,
  Eur.\ Phys.\ J.\  C {\bf 8}, 619 (1999)
  [arXiv:hep-ph/9812267].

\bibitem{alok}
  A.~K.~Alok, A.~Dighe and S.~Ray,
  Phys.\ Rev.\  D {\bf 79}, 034017 (2009)
  [arXiv:0811.1186 [hep-ph]].

\bibitem{Aubert:2004it}
  B.~Aubert {\it et al.}  [BABAR Collaboration],
  Phys.\ Rev.\ Lett.\  {\bf 93}, 081802 (2004)
  [arXiv:hep-ex/0404006].

\bibitem{Iwasaki:2005sy}
  M.~Iwasaki {\it et al.}  [Belle Collaboration],
  Phys.\ Rev.\  D {\bf 72}, 092005 (2005)
  [arXiv:hep-ex/0503044].

\bibitem{Ghinculov:2003bx}
  A.~Ghinculov, T.~Hurth, G.~Isidori and Y.~P.~Yao,
  Eur.\ Phys.\ J.\  C {\bf 33}, S288 (2004);
  [arXiv:hep-ph/0310187].

\bibitem{Ali:2002jg}
  A.~Ali, E.~Lunghi, C.~Greub and G.~Hiller,
  Phys.\ Rev.\  D {\bf 66}, 034002 (2002);
  [arXiv:hep-ph/0112300].

\bibitem{Ali:2002ik}
  A.~Ali,
  arXiv:hep-ph/0210183.


\bibitem{Ali:1991is}
  A.~Ali, T.~Mannel and T.~Morozumi,
  Phys.\ Lett.\  B {\bf 273}, 505 (1991).

\bibitem{Huber:2007vv}
  T.~Huber, T.~Hurth and E.~Lunghi,
  Nucl.\ Phys.\  B {\bf 802}, 40 (2008)
  [arXiv:0712.3009 [hep-ph]].


\bibitem{babar-03}  
B.~Aubert {\it et al.}  [BABAR Collaboration], 
  Phys.\ Rev.\ Lett.\  {\bf 91}, 221802 (2003); 
  [arXiv:hep-ex/0308042]. 
 
\bibitem{babar-06} 
 B.~Aubert {\it et al.}  [BABAR Collaboration], 
  Phys.\ Rev.\  D {\bf 73}, 092001 (2006); 
  [arXiv:hep-ex/0604007]. 
 
\bibitem{belle-03} 
 A.~Ishikawa {\it et al.}  [Belle Collaboration], 
  Phys.\ Rev.\ Lett.\  {\bf 91}, 261601 (2003); 
  [arXiv:hep-ex/0308044]. 

\bibitem{lunghi} 
  E.~Lunghi, 
  arXiv:hep-ph/0210379. 

\bibitem{ali-00} 
  A.~Ali, P.~Ball, L.~T.~Handoko and G.~Hiller, 
  Phys.\ Rev.\  D {\bf 61}, 074024 (2000); 
  [arXiv:hep-ph/9910221]. 
  
\bibitem{buras03}
 A.~J.~Buras,
  Phys.\ Lett.\  B {\bf 566}, 115 (2003)
  [arXiv:hep-ph/0303060],
 
M.~Blanke, A.~J.~Buras, D.~Guadagnoli and C.~Tarantino,
 JHEP {\bf 0610}, 003 (2006)
 [arXiv:hep-ph/0604057].


 \bibitem{Grossman:1996qj}
  Y.~Grossman, Z.~Ligeti and E.~Nardi,
  Phys.\ Rev.\  D {\bf 55}, 2768 (1997)
  [arXiv:hep-ph/9607473].


\bibitem{Lenzi:2007nq}
  M.~Lenzi,
  arXiv:0710.5056 [hep-ex].

\bibitem{Smizanska:2008qm}
  M.~Smizanska  [ATLAS Collaboration and CMS Collaboration],
  arXiv:0810.3618 [hep-ex].

\bibitem{latticebag}
 E.~Gamiz, C.~T.~H.~Davies, G.~P.~Lepage, J.~Shigemitsu and M.~Wingate
                  [HPQCD Collaboration],
  Phys.\ Rev.\  D {\bf 80}, 014503 (2009)
  [arXiv:0902.1815 [hep-lat]].

\bibitem{grossman}
  Y.~Grossman, Z.~Ligeti and E.~Nardi,
  Nucl.\ Phys.\  B {\bf 465} (1996) 369
  [Erratum-ibid.\  B {\bf 480} (1996) 753];
  [arXiv:hep-ph/9510378].

\bibitem{burasewerth}
C.~Bobeth, M.~Bona, A.~J.~Buras, T.~Ewerth, M.~Pierini, L.~Silvestrini and A.~Weiler,
  Nucl.\ Phys.\  B {\bf 726}, 252 (2005);
  [arXiv:hep-ph/0505110].

\bibitem{expbsnu}
 R.~Barate {\it et al.}  [ALEPH Collaboration],
  Eur.\ Phys.\ J.\  C {\bf 19}, 213 (2001);
  [arXiv:hep-ex/0010022].




\bibitem{uliburas}
A.~J.~Buras, M.~Gorbahn, U.~Haisch and U.~Nierste,
  Phys.\ Rev.\ Lett.\  {\bf 95}, 261805 (2005);
  [arXiv:hep-ph/0508165];
 A.~J.~Buras, M.~Gorbahn, U.~Haisch and U.~Nierste,
  JHEP {\bf 0611}, 002 (2006),
  [arXiv:hep-ph/0603079];

J.~Brod and M.~Gorbahn,
  Phys.\ Rev.\  D {\bf 78}, 034006 (2008);
  [arXiv:0805.4119 [hep-ph].

\bibitem{Grossman:1996ke}
  Y.~Grossman and M.~P.~Worah,
  Phys.\ Lett.\  B {\bf 395}, 241 (1997);
  [arXiv:hep-ph/9612269];
Y.~Grossman, G.~Isidori and M.~P.~Worah,
  Phys.\ Rev.\  D {\bf 58}, 057504 (1998)
  [arXiv:hep-ph/9708305].

\bibitem{London:1997zk}
  D.~London and A.~Soni,
  Phys.\ Lett.\  B {\bf 407}, 61 (1997)
  [arXiv:hep-ph/9704277].




\bibitem{Beneke:2005pu}
  M.~Beneke,
  Phys.\ Lett.\  B {\bf 620}, 143 (2005)
  [arXiv:hep-ph/0505075].

\bibitem{Cheng:2006dk}
H.~Y.~Cheng, C.~K.~Chua and A.~Soni,
  Phys.\ Rev.\  D {\bf 71}, 014030 (2005)
  [arXiv:hep-ph/0409317];
 H.~Y.~Cheng, C.~K.~Chua and A.~Soni,
  Phys.\ Rev.\  D {\bf 72}, 014006 (2005)
  [arXiv:hep-ph/0502235].

\bibitem{Buchalla:2005us}
  G.~Buchalla, G.~Hiller, Y.~Nir and G.~Raz,
  JHEP {\bf 0509}, 074 (2005)
  [arXiv:hep-ph/0503151].



\bibitem{rf}
  R.~Fleischer, S.~Jager, D.~Pirjol and J.~Zupan,
  Phys.\ Rev.\  D {\bf 78}, 111501 (2008)
  [arXiv:0806.2900 [hep-ph]].


\bibitem{mg1} 
 M.~Gronau and J.~L.~Rosner,
  Phys.\ Lett.\  B {\bf 666}, 467 (2008)
  [arXiv:0807.3080 [hep-ph]].

\bibitem{beak}
S.~Baek, C.~W.~Chiang, M.~Gronau, D.~London and J.~L.~Rosner,
  Phys.\ Lett.\  B {\bf 678}, 97 (2009)
  [arXiv:0905.1495 [hep-ph]].

\bibitem{belle} 
I.~Adachi {\it et al.}  [Belle Collaboration],
  arXiv:0809.4366 [hep-ex].

\bibitem{babar}
  B.~Aubert {\it et al.}  [BABAR Collaboration],
  arXiv:0809.1174 [hep-ex].

\bibitem{lim} T. Inami and C. S. Lim, Prog. Theor. Phys. {\bf 65},
297 (1981); {\it ibid} {\bf 65}, 1772E (1981).
\bibitem{lisanda}
H.~n.~Li, S.~Mishima and A.~I.~Sanda,
  Phys.\ Rev.\  D {\bf 72}, 114005 (2005)
  [arXiv:hep-ph/0508041];
 H.~n.~Li and S.~Mishima,
  Phys.\ Rev.\  D {\bf 73}, 114014 (2006)
  [arXiv:hep-ph/0602214].

\bibitem{shaouly}
S.~Bar-Shalom, G.~Eilam and A.~Soni,
  arXiv:1001.0569 [hep-ph].


\end{thebibliography}
\end{document}